\documentclass[12pt]{article}
\usepackage{amsfonts,amssymb,epsfig,amsmath,mathtools,xcolor}
\usepackage{color}

\usepackage[enableskew]{youngtab}
\usepackage{ytableau}
\usepackage{mathrsfs}
\usepackage{graphicx}
\usepackage{subcaption}
\usepackage{hyperref}

\renewcommand{\baselinestretch}{1.2}

\newcommand{\tr}{\textrm{tr}}
\newcommand{\nn}{\nonumber}
\newcommand{\cO}{\mathcal{O}}
\newcommand{\cT}{\mathcal{T}}
\newcommand{\ep}{\epsilon}
\def\ba#1\ea{\begin{align}#1\end{align}}

\begin{document}

\makeatletter \@addtoreset{equation}{section} \makeatother
\renewcommand{\theequation}{\thesection.\arabic{equation}}
\renewcommand{\thefootnote}{\alph{footnote}}

\begin{titlepage}

\begin{center}
\hfill {\tt SNUTP24-004}\\

\vspace{2cm}

{\Large\bf Brane-fused black hole operators}

\vspace{1.5cm}

\renewcommand{\thefootnote}{\alph{footnote}}

{\large Robert de Mello Koch$^1$, Minkyoo Kim$^2$, Seok Kim$^3$, Jehyun Lee$^3$, 
Siyul Lee$^4$}

\vspace{0.5cm}

\textit{$^1$School of Science, Huzhou University, Huzhou 313000, China.\\
\& School of Physics and Mandelstam Institute for Theoretical Physics,\\
University of the Witwatersrand, Wits, 2050, South Africa.}

\vspace{0.2cm}

\textit{$^2$Center for Quantum Spacetime (CQUeST), Sogang University, Seoul 04107, Korea.}

\vspace{0.2cm}

\textit{$^3$Department of Physics and Astronomy \& Center for
Theoretical Physics,\\
Seoul National University, Seoul 08826, Korea.}

\vspace{0.2cm}

\textit{$^4$Instituut voor Theoretische Fysica, KU Leuven,
Celestijnenlaan 200D, 3001 Leuven, Belgium}

\vspace{0.5cm}

E-mails: {\tt robert@zjhu.edu.cn, mimkim80@gmail.com, seokkimseok@gmail,com,\\
ljs9125@snu.ac.kr, siyul.lee@kuleuven.be}

\end{center}

\vspace{0.1cm}

\begin{abstract}

We construct infinitely many new $\frac{1}{16}$-BPS cohomologies of the 4d maximal super-Yang-Mills 
theory and interpret them as a black hole wrapped by dual giant graviton hairs.
Since the black hole inside a dual giant feels the RR 5-form flux reduced by one unit, 
its microstate should essentially be an $SU(N\!-\!1)$ cohomology. 
However, due to the fortuitous nature of the black hole microstates, 
promoting an $SU(N\!-\!1)$ 
black hole state to $SU(N)$ generally fails to yield a cohomology. We show at $N=3$ 
that suitable fusion products with the dual giants yield cohomologies.
The core black hole size is probed by the minimal size of the dual giant which can wrap it. 
We also discuss two types of large black hole hairs: large conformal 
descendants of gravitons and large dual giants. We prove that any $SU(N)$ black hole 
cohomology admits infinitely many hairs of the first type.

\end{abstract}

\end{titlepage}

\renewcommand{\thefootnote}{\arabic{footnote}}

\setcounter{footnote}{0}

\renewcommand{\baselinestretch}{1}

\tableofcontents

\renewcommand{\baselinestretch}{1.2}

\section{Introduction}

Recently, there have been developments in constructing cohomologies for the
$\frac{1}{16}$-BPS local operators of the $\mathcal{N}=4$ Yang-Mills theory 
\cite{Chang:2022mjp,Choi:2022caq,Choi:2023znd,Chang:2023zqk,Budzik:2023vtr,Budzik:2023xbr,
Chang:2023ywj,Choi:2023vdm,Chang:2024zqi}.
A goal of this program is to better understand the microstates of BPS black holes
\cite{Gutowski:2004ez,Gutowski:2004yv,Chong:2005hr,Kunduri:2006ek}
from the CFT dual.
While the correct microstate structure will be visible in the strongly coupled QFT,
the studies have been focused on the 1-loop BPS states expecting non-renormalization
of their spectrum \cite{minwalla,Chang:2022mjp}. Also, while the semi-classical gravity 
emerges at large $N$, progress has been made in the `quantum' regimes 
at low $N$. Somewhat surprisingly, certain spectral aspects of the black holes 
reminiscent of the BPS black hole hairs were observed even at small $N$ 
\cite{Choi:2023znd}.

In this paper, we construct the representatives of new black hole cohomologies in the
maximal super-Yang-Mills theory with the gauge group $SU(3)$
and provide their `gravity dual' interpretation. 
Our interpretation is related to the novel AdS black holes  
\cite{Choi:2024xnv} (see also \cite{Henriksson:2019zph}) 
with the dual giant graviton hairs \cite{McGreevy:2000cw,Grisaru:2000zn,Hashimoto:2000zp}.
As we explain below, $N=3$ is the minimal value which admits 
such a hair structure in the local microstate operators.
Although this work and \cite{Choi:2024xnv} discuss the opposite extremes,
$N=3$ vs. $\infty$, their qualitative similarities lead us to suspect that
similar cohomologies exist at all $N$ and interpolate the two.

The $\frac{1}{16}$-BPS operators $O$ are annihilated by a pair of mutually 
Hermitian conjugate supercharges $Q$ and $Q^\dag$: $[Q,O\}=0$, $[Q^\dag,O\}=0$. 
Since these supercharges are nilpotent, the BPS operators are in 1-to-1 correspondence
to the cohomology classes of $Q$. The cohomology classes are defined by $Q$-closed operators 
with the identification of two such operators differing by a $Q$-exact term:  
$O\sim O+[Q,\Lambda\}$. So a representative of a cohomology class 
may differ from the BPS operator by a $Q$-exact term: it only 
gives the information on the spectrum of BPS operators. 
In this paper, we consider the cohomologies with the supercharge $Q$ 
of the classical interacting theory, which are in 1-to-1 map to the BPS 
states of the 1-loop Hamiltonian.

One should distinguish the two types of cohomologies: the graviton and the black hole types. 
We follow the recent characterizations \cite{Chang:2024zqi}
for the $\mathcal{N}=4$ Yang-Mills theory.\footnote{The criteria here
apply only to the $\mathcal{N}=4$ theory with the $SU(N)$ gauge group, without 
baryons. More intrinsically, the graviton and baryon states 
can be constructed by quantizing the (super-)moduli space.} 
Consider an infinite sequence $\{O_N\}$ of gauge invariant operators
defined by fixing the `shape' but taking the fields appearing in the operator
to be $N\times N$ traceless matrices. (The operator shape is fixed by specifying 
the fields and how they enter the multi-trace structure,
like ${\rm tr}(XY){\rm tr}(Z\psi_1)$, ${\rm tr}(Z^2f)$.)
With the operator shape fixed, the classical scaling dimension as well as the 
spins and the R-charges are the same for all operators in the sequence.
If $O_N$ is $Q$-closed, so are $O_{N^\prime}$ with $N^\prime < N$. This is
because the $N\times N$ matrix equation $[Q,O_N\}=0$ with the classical supercharge $Q$ 
reduces to the $N^\prime\times N^\prime$ matrix equation $[Q,O_{N^\prime}\}=0$ 
if one consistently turns off all the fields except the $N^\prime\times N^\prime$ 
block-diagonal elements.
So if any operator in the sequence $\{O_N\}$ is $Q$-closed, the operators  
are $Q$-closed for all $N\leq N_\ast$ for some integer $N_\ast$. 
If $N_\ast=\infty$, i.e. if all the operators in the sequence are $Q$-closed, 
we say they represent graviton cohomologies. If a finite $N_\ast$ exists,
we say they represent the black hole cohomologies. The name graviton for 
the former type is clear: increasing $N$ with the fixed shape, and thus fixed energy $E$,  
$O_N$ will eventually describe BPS states at $E\ll N$ and should belong to 
the low energy supergraviton spectrum. $O_N$ in this sequence at finite $N$ 
(or $E\sim N$) represent generalizations of the supergravitons, 
the giant graviton states. On the other hand, calling the latter black holes is 
a somewhat rough notion at this moment. 
The latter type will certainly contain black hole states, while it is not clear whether 
there may be operators which neither deserve to be called gravitons nor 
black holes.\footnote{Although not in the microstate picture, 
conical defects contribute to the Euclidean gravitational path integral in AdS$_3$, 
which are neither gravitons nor black holes \cite{Benjamin:2020mfz}. We thank Masamichi Miyaji 
for this comment.} However, we shall use these intuitive terminologies in this paper.
In \cite{Chang:2024zqi}, the former/latter types are called the monotone/fortuitous cohomologies, 
respectively.

The graviton cohomologies are $Q$-closed without using trace relations. 
Construction of the graviton cohomologies 
reduces to that of the \textit{single trace} cohomologies, 
which are completely understood (e.g. appendix C.3 of \cite{Kinney:2005ej}). 
The general graviton cohomologies can be represented 
by multiplying these single-trace cohomologies.\footnote{These multi-trace representatives 
are overcomplete. This is because, although they are $Q$-closed without 
using trace relations, some linear combinations can become $Q$-exact due to trace relations.}

On the other hand, the black hole cohomologies are $Q$-closed by trace relations,
and there is no systematic way to generate all black hole cohomologies.
Instead, we shall use the `ansatz' devised in \cite{Choi:2023vdm} to construct
new $Q$-closed operators of the black hole type, using a class of known trace relations.
One should also show that the constructed $Q$-closed operator is not $Q$-exact.
There is no known general method for this other than to list all the $Q$-exact operators
with given quantum numbers and show that none of their linear combinations
reproduces the operator under consideration. 
It is in general a very difficult computation.
However, in some non-generic cases one can identify a particular term in the operator
which obstructs the $Q$-exactness, in which case it becomes easier to show that the operator
does not belong to the trivial cohomology class.
The new cohomologies constructed in this paper will naturally have such obstruction terms, 
as we explain below and in section \ref{sec:dghair}.

Black holes in AdS may be dressed by the hairs outside the event horizon. 
Recently, two types of hairs were found for the charged rotating AdS black holes. 
For simplicity, consider the black holes in $AdS_5\times S^5$ carrying two equal 
angular momenta $J\equiv J_1=J_2$ and three equal electric charges 
$R\equiv R_1=R_2=R_3$. If $J$ is too large, the black hole admits 
particle hairs that carry large angular momenta \cite{Kim:2023sig,SUSY-GG,temporary}.
If $R$ is too large, electrically charged hairs are admitted in the form 
of the large dual giant gravitons wrapping the core black hole \cite{Choi:2024xnv}. Both types of 
hairy black holes may reach the BPS limit $E=3R+2J$, implying the existence of hairy 
BPS black holes and their cohomologies.

Since one can distinguish (to certain extent) the graviton/black hole type cohomologies 
at finite $N$ as reviewed above, we can also try to identify the hairy cohomologies at finite $N$. 
The hairs of the first kind with large $J$ have been noticed even for $N=2$
\cite{Choi:2023znd}. It was proposed that such states belong to the cohomologies 
admitting product representatives. That is, the product of a black hole cohomology and 
a graviton cohomology represents another cohomology. This product often becomes 
$Q$-exact if the graviton part does not carry large enough $J$, while it represents
a non-trivial new cohomology when the graviton carries large enough $J$.
See section \ref{sec:genN} for general discussions on such hairy cohomologies with large $J$.

In this paper, we study the cohomologies for BPS black holes wrapped 
by the dual giant graviton hairs.
The physical aspects of these hairy microstates are quite different from those of the
previous paragraph, e.g. they defy the product representatives as we explain now.

The dual giant graviton is a D3-brane extended in the global AdS$_5$.
It backreacts to the background by reducing the RR flux by one unit 
inside it \cite{Hashimoto:2000zp}. So any object inside 
the dual giant is essentially that of the $SU(N\!-\!1)$ gauge theory. 
If a black hole is wrapped by a dual giant graviton, its microstate
should be morally that of the $SU(N\!-\!1)$ theory.
Of course the whole system including the dual giant hair should be described
covariantly in the $SU(N)$ theory, but there should be a sense in which
the shape of the $SU(N\!-\!1)$ core cohomology can be identified.
However, due to the fortuitous nature of the black hole cohomology, $SU(N\!-\!1)$ black hole 
cohomologies fail to be the $SU(N)$ cohomologies if $N_\ast=N-1$
(which will be the case in our examples).
That is, although the operator $O_{N-1}$ satisfies $[Q,O_{N-1}\}=0$
thanks to $SU(N\!-\!1)$ trace relations,
the operator $O_N$ of the same shape may not, i.e. $[Q,O_N\}\neq 0$.
So the hairy cohomologies involving the dual giants do not admit simple product representatives.
Instead, the gauge orientations of the dual giant part and the black hole part
should be entangled for the entire operator to be $Q$-closed, forming 
a \textit{fusion product}.

$N=3$ is the minimal value for which the $SU(N\!-\!1)$ theory is non-trivial.
We shall find $Q$-closed operators which are fusion products of the black hole $O_3$
and the dual giant graviton $G_3$ in the $SU(3)$ theory.
$O_3$ is the $SU(3)$ uplift of an $SU(2)$ black hole cohomology $O_2$.
Unlike $O_2$, $O_3$ is not $Q$-closed in general.
$G_3$ is a (dual giant) graviton cohomology of the $SU(3)$ theory. The whole operator 
is made $Q$-closed by suitably entangling the gauge orientations of the two parts.
To show that these operators are not $Q$-exact, we consider the terms which 
contain only the $SU(2)\times U(1)$ block-diagonal elements of the fields. 
Among those block-diagonal terms, we identify 
a term of the form $O_2 G_1$, a product of the $SU(2)$ black hole cohomology $O_2$ 
and the $U(1)$ graviton cohomology $G_1$. Since both $O_2$ and $G_1$ represent non-trivial 
cohomologies and also since the action of $Q$ does not mix the $SU(2)$ and $U(1)$ blocks, 
this term is not $Q$-exact. This obstructs the full operator from being
$Q$-exact. This proof also hints towards the bulk nature of the operator, 
the $SU(N-1)$ core inside the D-brane and the $U(1)$ part on the brane.

We will show that these obstruction terms exist only when $G_3$ is larger than a 
critical size. We interpret this as the brane probing the size of the core black hole, 
i.e. the brane should be larger than the black hole to wrap it.

The fact that the cohomology remains non-trivial in the $SU(N-1)\times U(1)$ 
sector implies that the corresponding microstate is not a typical $SU(N)$ black 
hole state, but essentially that of an $SU(N\!-\!1)$ theory supplemented by the $U(1)$ brane 
degrees of freedom. Typical black hole cohomologies 
would not have such obstruction terms within the block-diagonal elements.

The rest of this paper is organized as follows.
In section \ref{sec:dghair}, building on the ansatz of \cite{Choi:2023vdm}, 
we construct $Q$-closed operators in which the dual giant graviton part $G_3$
involves $n$ scalars. For $n=1,2$, we show that
no such operators contain obstruction terms against the $Q$-exactness
whereas for all $n\geq3$, we find non-trivial cohomologies.
In section \ref{sec:bothhair}, we find more general hairy black hole cohomologies
using this method. In section \ref{sec:genN}, we make general discussions on the two types 
of hairy cohomologies at arbitrary $N$, in particular proving the existence of infinitely many hairy cohomologies of the product form. We conclude in section \ref{sec:conclusion} with remarks.

\section{Black hole cohomologies with dual giant hairs}\label{sec:dghair}

In this section, we construct novel black hole cohomologies in the
4d $\mathcal{N}=4$ Yang-Mills theory with gauge group $SU(3)$.
They transform in the irreps $[n\geq3,0]$ of the $SU(3)_R\subset SU(4)_R$ 
R-symmetry, and we interpret these as the black hole wrapped 
by the dual giant hair.
We start by clarifying notations before we present our strategy and
results in subsequent subsections.

The $\mathcal{N}=4$ Yang-Mills theory has $16$ Poincar\'e supercharges 
$Q^i_\alpha$, $\overline{Q}_{i\dot\alpha}$,
where $i=1,\cdots, 4$ is the fundamental index for the $SO(6)\sim SU(4)_R$ R-symmetry group
and $\alpha=\pm$, $\dot\alpha=\dot{\pm}$ are doublet indices for the $SO(4)$ rotation.
It also has $16$ conformal supercharges, which are Hermitian
conjugate to the Poincar\'e supercharges in the radial quantization.
We are interested in local operators
annihilated by $Q\equiv Q^4_-$ and its conjugate $Q^\dag$, thus $\frac{1}{16}$-BPS.
Their scaling dimensions $E$ saturate the BPS bound
\begin{equation}
  E\geq R_1+R_2+R_3+J_1+J_2\ ,
\end{equation}
where $R_m$ are the three Cartans of the $SU(4)_R$ and 
$J_i$ are the two Cartans of the $SO(4)$ rotation.

\begin{table}[t]
\begin{center}
\begin{tabular}{| c || c | c | c | c | c || c |}
\hline
Field & $R_{m}$ & $R_{m+1}$ & $R_{m+2}$ & $J_1+J_2$ & $J_1-J_2$ & $j$ \\
\hline
$\phi^m$ & 1 & 0 & 0 & 0 & 0 & 2 \\
$\psi_m$ & $-\frac12$ & $\frac12$ & $\frac12$ & 1 & 0 & 4 \\
$f$ & \hspace{0.4cm}0\hspace{0.4cm} & \hspace{0.4cm}0\hspace{0.4cm} & \hspace{0.4cm}0\hspace{0.4cm} & 2 & 0 & ~6~ \\
\hline
\end{tabular}
\end{center}
\caption{Quantum numbers of the seven BPS elementary fields in the BMN subsector.
$m=1,2,3$ and the subscripts for $R$ are written modulo 3.}\label{tab:fields}
\end{table}

We shall consider a 1d truncation of this problem,
which was referred to as the BMN subsector in \cite{Choi:2023znd}.
(See also \cite{Chang:2022mjp,Choi:2022caq}, which considered the same truncation 
without explicitly relating it to the BMN matrix theory, and also \cite{Chang:2024lkw} 
for further studies.)
With this truncation and at the 1-loop level, BPS operators are constructed
using seven elementary fields: 3 components $\phi^m$ of the scalar,
3 components $\psi_m$ of the fermion, and one $f$ of the field strength,
as summarized in Table \ref{tab:fields}.
$m=1,2,3$ is the (anti-)fundamental index for $SU(3)_R\subset SU(4)_R$ 
that commutes with $Q$ and $R_1+R_2+R_3$.
The charges of an operator are simply the sum of those of the constituent elementary fields.
In Table \ref{tab:fields}, we also listed the values of the charge
\begin{equation}
    j \equiv 2(R_1+R_2+R_3)+3(J_1+J_2)~.
\end{equation}
%
%This will be our measure of the size of the operators, and we 
We will often call it the order, originating from the practice of series expanding
the index as $\mathcal{I}(t) = \sum_j a_j t^j$ with $j=0,1,2,\cdots$.

As explained in the introduction, the BPS operators are in one-to-one correspondence
with cohomology classes with respect to the supercharge $Q$.
We shall consider the cohomologies at the 1-loop level,
i.e. in the interacting classical field theory.
$Q$ acts on the elementary fields as
\begin{equation}\label{Qaction}
  Q\phi^m=0~, \qquad
  Q\psi_m= -\frac{i}{2} \epsilon_{mnp}[\phi^n,\phi^p]~, \qquad
  Qf = -i [\phi^m,\psi_m]~.
\end{equation}

\begin{table}[t]
\begin{center}
\begin{tabular}{c||c|c|c|c|c|c||c|c|c|c}
	\hline
    $j$ & $F_0$ & $F_1$  & $F_2$ & $F_3$ & $F_4$ & $F_{\rm exc}$ & $B_1$ & $B_2$ 
    & $B_3$ & $B_{\rm exc}$ \\
	\hline 
    $24$ & $[0,0]$ &&&&&&&&& \\
    $26$ & \hphantom{$[15,0]$}&\hphantom{$[15,0]$}&\hphantom{$[15,0]$}&
    \hphantom{$[15,0]$}&\hphantom{$[15,0]$}&\hphantom{$[15,0]$}&\hphantom{$[15,0]$}&
    \hphantom{$[15,0]$}&\hphantom{$[15,0]$}&\hphantom{$[15,0]$}\\
    $28$ &&&&&&&&&& \\
    $30$ & $[0,0]$ & $[3,0]$ &&&&&&&& \\
    $32$ && $[4,0]$ &&&&&&&& \\
    $34$ && $[5,0]$ &&&&& $[3,1]$ &&& \\
    $36$ & $[0,0]$ & $[6,0]$ &&&&& $[4,1]$ &&& $[3,0]$ \\
    $38$ && $[7,0]$ &&&& $[1,0]$ & $[5,1]$ &&& \\
    $40$ && $[8,0]$ & $ [5,0]$ && $[3,1]$ && $[6,1]$ &&& \\
    $42$ & $[0,0]$ & $[9,0]$ & $[6,0]$ && $[4,1]$ && $[7,1]$ &&& $[1,1]$ \\
    $44$ && $[10,0]$ & $[7,0]$ && $[5,1]$ && $[8,1]$ & $[5,1]$ && \\
    $46$ && $[11,0]$ & $[8,0]$ && $[6,1]$ & $[2,0]$ & $[9,1]$ & 
    $[6,1]$ && $[5,0]$ \\
    $48$ & $[0,0]$ & $[12,0]$ & $[9,0]$ && $[7,1]$ & $[3,0]$ & $[10,1]$ & $[7,1]$ && $[4,1]$ \\
    $50$ && $[13,0]$ & $[10,0]$ & $[7,0]$ & $[8,1]$ && $[11,1]$ & $[8,1]$ && $[4,0]$ \\
    $52$ & & $[14,0]$ & $[11,0]$ & $[8,0]$ & $[9,1]$ & 
    $[2,0]$ & $[12,1]$ & $[9,1]$ && $[3,1]$\\ 
    $54$ & & $[15,0]$ & $[12,0]$ & $[9,0]$ & $[10,1]$ &
    $[4,1]$ & $[13,1]$ & $[10,1]$ & $[7,1]$ & \\
    \hline
\end{tabular}
\end{center}
\caption{$SU(3)_R$ Dynkin labels of fermionic/bosonic black hole cohomologies
({\bf v2}: as well as non-Coulomb graviton cohomologies)
in the $SU(3)$ theory, after factoring out the $SU(1|3)$ descendants
and the conjectured graviton hairs.}\label{tower}
\end{table}

The goal of this section is to find representatives of new cohomologies in the $SU(3)$ gauge theory. 
Our constructions will be guided by the index of the $SU(3)$ black hole cohomologies computed 
in \cite{Choi:2023vdm}. The black hole cohomology index takes the form of 
\begin{equation}\label{index-schematic}
  \mathcal{I} = \mathcal{I}_{s.c.d.}\cdot\mathcal{I}_{w_2} \cdot \sum_{i}(-1)^{F_i}\chi({\bf R}_i)~t^{j_i}~.
\end{equation}
The last factor stands for `core' black hole primary cohomologies:
they are grouped by irreducible representations ${\bf R}_i$ of $SU(3)_R$,
algebraically represented by its character $\chi({\bf R}_i)$,
and graded according to their order $j$ and their boson/fermion statistics.
The first factor $\mathcal{I}_{s.c.d.}$ is the character of the $SU(1|3)$ multiplet which factors out 
the contributions from the superconformal descendants.
Moreover, it was assumed empirically that $w_2$ type gravitons (which will be defined shortly)
can multiply the core black hole cohomologies to yield new black hole cohomologies, 
whose index is the second factor $\mathcal{I}_{w_2}$: see \cite{Choi:2023vdm} for the details.
Table \ref{tower} presents the black hole cohomology index (\ref{index-schematic})
by listing the $SU(3)_R$ irreps ${\bf R}_i$ for core black hole cohomologies
for each order $j$, and for bosons (B) and fermions (F) separately.\footnote{{\bf v2}:
It was noticed in \cite{Kim:2026} that Table \ref{tower} also includes non-Coulomb gravitons.}
We grouped the states into series of towers along the columns of the table, 
and left out states that do not fit in any of the series
to the exceptional columns $F_{\rm exc}$ and $B_{\rm exc}$.
Whereas the classification into towers and the factoring out of $\mathcal{I}_{w_2}$
have all been based on observation of patterns,
results of this paper will justify and give physical interpretations to 
some of these structures.

In this section we discuss the tower $F_1$.
For every $n \geq 3$, the index predicts
%(from the $F_1$ tower)
fermionic cohomologies at the order 
\begin{equation}\label{charge-F1}
  j\equiv 2(R_1+R_2+R_3)+3(J_1+J_2)=24+2n~,
\end{equation}
in the $[n,0]$ representation of $SU(3)_R$.
We would like to interpret these as the core $SU(2)$ black hole cohomology with $j=24$
wrapped by a dual giant graviton hair made of $n$ scalar fields.
This is a viable proposal at least according to the quantum numbers.
The smallest black hole cohomology in the $SU(2)$ theory has $j=24$,
more concretely $R_1=R_2=R_3=\frac{3}{2}$, $J_1=J_2=\frac{5}{2}$
and it is a singlet under $SU(3)_R$ \cite{Chang:2022mjp}.
The dual giant hair is made of $n$ symmetrized scalars,
therefore transforms in the $[n,0]$ representation of $SU(3)_R$
and carries quantum numbers $R_1+R_2+R_3=n$, $J_1=J_2=0$ and $j=2n$.
All quantum numbers of the $SU(2)$ black hole cohomology and the dual giant hair
add up to those of the tower $F_1$ in Table \ref{tower}.

The `core' black hole cohomology in the $SU(2)$ theory can be represented by \cite{Choi:2023znd}
\begin{equation}\label{SU2O0}
  \cO^2_{24} = \epsilon^{pqr}{v^m}_q{v^n}_r{\rm tr}(\psi_{(m}\psi_n\psi_{p)})\ ,
\end{equation}
where ${v^m}_n={\rm tr}(\phi^m\psi_n)-\frac{1}{3}\delta^m_n{\rm tr}(\phi^p\psi_p)$
is a single-trace graviton operator.
Throughout this paper, we denote black hole cohomologies in the $SU(N)$ theory at order $j$ as $\cO^N_j$.
Note that $\cO^2_{24}$ contains 2 scalars and 5 fermions, and no $f$.
In the $SU(2)$ theory, it is $Q$-closed by $SU(2)$ trace relations.
The right hand side of \eqref{SU2O0} also defines an operator in the $SU(3)$
gauge theory (which we still denote by $\cO^2_{24}$ although it is an $SU(3)$ operator)
in terms of the fields $\phi^m$, $\psi_m$ that are now $3 \times 3$ matrices.
However, this operator is not $Q$-closed because the relevant $SU(2)$ trace relations
do not hold in the $SU(3)$ theory. \cite{Budzik:2023vtr} computed 
the anomalous dimensions of operators in this sector at $N>2$, 
indeed finding that all of them are non-BPS.

Instead, the hairy black hole cohomology in the $SU(3)$ theory that we want to construct
contains $n$ scalars in addition to $\cO^2_{24}$, so it contains $n+2$ scalars and $5$ fermions.
We also want it to be in the $[n,0]$ representation of $SU(3)_R$.
At first sight, one may simply try multiplying $\cO^2_{24}$ by some gauge-invariant 
graviton operator that involves $n$ symmetrized scalars.
However, such a product operator is not $Q$-closed.
Since $Q$ annihilates scalars, it only acts on the $\cO^2_{24}$ part of the product.
So $Q$ acting on the product will be the same graviton operator
times $Q \cO^2_{24}$, which would typically be non-zero in the $SU(3)$ theory.
Therefore we need a more elaborate way to dissolve the $n$ symmetrized scalars in the
black hole operator $\cO^2_{24}$.
Roughly speaking, we want to insert $n$ scalar matrices $\phi^{(a_1},\phi^{a_2},\cdots,\phi^{a_n)}$
suitably inside the multi-trace structure of $\cO^2_{24}$ to form a fusion product. 
%, and then further add suitable backreaction terms as needed in order to make it $Q$-closed.

In the rest of this section, we implement this idea in a systematic framework to construct
black hole cohomologies that correspond to $n\geq 3$,
accounting for the tower $F_1$. We also show that it fails to detect any cohomology 
at $n=1,2$, as anticipated by the index.

\subsection{General methods for the construction}\label{sec:strategy}

In this subsection, we outline our method for constructing the hairy black hole cohomologies.
We will first review the ansatz devised in \cite{Choi:2023vdm} to construct $Q$-closed
non-graviton operators.
Then we will explain how some of the operators can be proved to be not $Q$-exact.
This proof is inspired by the physical picture of the black hole wrapped by a dual giant,
and applies for some operators that admit such an interpretation.

\subsubsection{The ansatz for $Q$-closed operators}\label{sec:ansatz}

Recall the classification of the graviton versus the black hole cohomologies:
an operator is of black hole type if it is $Q$-closed only by trace relations,
which do not hold for arbitrarily large $N$.
The ansatz devised in \cite{Choi:2023vdm} leverages this property to construct
a $Q$-closed operator from trace relations between graviton operators,
which have been obtained as a byproduct of computing the finite $N$ graviton cohomology index.
This ansatz does not construct all black hole type $Q$-closed operators,
but it proves to be successful in constructing a useful class of such operators.
In fact, the representative (\ref{SU2O0}) of the smallest $SU(2)$ black hole cohomology was found along this strategy. 
%as well as the smallest $SU(3)$ black hole cohomology \cite{Choi:2023vdm}.

Let us collectively denote the single trace graviton operators by $\{ g_I \}$.
This set has been classified in \cite{Kinney:2005ej}:
the symmetrized scalars $u_n \equiv {\rm tr}(\phi^{(i_1}\cdots\phi^{i_n)})$
and their $PSU(1,2|3)$ descendants. At finite $N$, only a subset of those with $n\leq N$
need to be used \cite{Choi:2023znd}.
In the BMN subsector of the $SU(3)$ theory, the complete list is given by
\begin{equation}\label{BMN-gen}
    \begin{aligned}
        {u_2}^{ij} \equiv & \; \textrm{tr}  \left(\phi^{(i} \phi^{j)}\right)\ , &
        {u_3}^{ijk} \equiv & \; \textrm{tr} \left(\phi^{(i} \phi^{j} \phi^{k)}\right)\ , \\
        {{v_2}^{i}}_j \equiv & \; \textrm{tr} \left( \phi^i \psi_{j}\right) - {\textstyle \frac{1}{3} }
        \delta^i_j\textrm{tr}\left(\phi^a \psi_{a}\right)\ , &
        {{v_3}^{ij}}_k \equiv & \; \textrm{tr} \left(\phi^{(i} \phi^{j)} \psi_{k} \right) 
        - {\textstyle \frac{1}{4}} \delta^{i}_{k}  
        \textrm{tr} \left( \phi^{(j} \phi^{a)} \psi_{a} \right) 
        - {\textstyle \frac{1}{4}} \delta^{j}_{k}  \textrm{tr} \left( \phi^{(i} \phi^{a)} \psi_{a} \right)\ , \\
        {w_2}^i \equiv & \; \textrm{tr} \left(f \phi^i + {\textstyle \frac{1}{2}} \epsilon^{ia_1a_2} \psi_{a_1} \psi_{a_2}\right)\ , &
        {w_3}^{ij} \equiv & \; \textrm{tr}\left(f \phi^{(i} \phi^{j)} +\epsilon^{a_1a_2(i}\phi^{j)} \psi_{a_1} \psi_{a_2}\right)\ .
    \end{aligned}
\end{equation}
We will often omit the subscripts $2$ and $3$ when it can be inferred from the number of indices.

The space of all graviton type operators is generated by the single trace gravitons \eqref{BMN-gen}
as their Fock space, i.e. all multi-graviton operators are polynomials of $g_I$'s.
However, these polynomials are redundant in that some polynomials are $Q$-exact,
thus belong to trivial cohomologies.
A $Q$-exact polynomial represents a cohomological relation between the monomials in $g_I$'s.
All the relations can be obtained from a finite number of fundamental relations
which we write as $R_a(g_I)=Qr_a$, where $R_a(g_I)$ is a $Q$-exact polynomial in $g_I$
and $r_a$ is the corresponding local gauge-invariant operator.
Each $r_a$ is, unlike $R_a$, not a graviton type operator.
There is an ambiguity in $r_a$ of adding any $Q$-closed operator, 
but we make convenient choices of $r_a$ as listed in Appendix A.
%All our choices are such that $r_a$ become zero under 
%the restriction of the elementary fields into diagonal traceless matrices.
%Here note that the graviton cohomologies originate from the symmetrized scalars, 
%so they are faithfully represented by the eigenvalues of the elementary fields
%\cite{Choi:2023vdm,Choi:2023znd}.
%All $r_a$'s being zero under this restriction illustrates that they are 
%not graviton type.

The exhaustive list of the fundamental relations $R_a$ between $u_2$, $u_3$, $v_2$ and $v_3$
(while excluding $w_2$ and $w_3$ because we do not need them in this work)
is presented in \eqref{tr-rel}.
We also present in \eqref{tr-r} $r_a$'s corresponding to some of the $R_a$'s that are used throughout this paper.

Each fundamental relation $R_a(g_I)$ can be multiplied by any polynomials of gravitons,
say $f_a(g_I)$, resulting in non-fundamental relations at higher orders.
Certain linear combinations of the non-fundamental relations may vanish identically, 
yielding relations of relations:
\begin{equation}\label{relation-of-relations}
  \sum_{a} f_a(g_I) R_a(g_I) =0~.
\end{equation}
%
%For examples of $s_{a,b}$ and relations between them, see Appendix \ref{app:trrel}.
Whenever there is a relation of relations \eqref{relation-of-relations}, one can consider the operator 
\begin{equation}\label{ansatz-general}
  \sum_{a} (-1)^{|f_a|} \, f_a(g_I) r_a~,
\end{equation}
where $|f_a|$ is the Grassmann parity of the graviton polynomial $f_a$ and
$R_a$'s have been replaced with their corresponding $r_a$'s such that $Qr_a = R_a$.
Because all gravitons are annihilated by $Q$, $Q$ acting on \eqref{ansatz-general} is precisely
\eqref{relation-of-relations}, which equals zero.
So (\ref{ansatz-general}) is an operator that is $Q$-closed by trace relations,
and it is not a graviton type operator.
Therefore, it serves as an ansatz for the black hole cohomologies.

\subsubsection{A hairy proof of non-$Q$-exactness}\label{sec:strategy-exact}

To construct a nontrivial cohomology, it remains to prove that the black hole type
$Q$-closed operator obtained using the ansatz is not $Q$-exact.
In other words, we need to exclude $Q$-exact operators among those obtained by the ansatz,
or combinations thereof.
In general, this involves extremely heavy computations because there is no
generally applicable method other than to rule out by brute force
all $Q$-exact operators in the target charge sector.
This is a huge linear problem that has been the computational bottleneck against
finding more black hole cohomologies in \cite{Choi:2023vdm}.
However, for some operators that we present in this work,
it can be proved relatively easily that they cannot be $Q$-exact.

As mentioned in the Introduction, the $SU(3)$ black hole operator with
dual giant hair that we seek after should contain a core $SU(2)$ black hole in some sense.
With this insight, consider restricting gauge orientations of all fields to be included in
the $SU(2) \times U(1)$ subgroup of $SU(3)$.
That is, we write all elementary fields in the basis of Gell-Mann matrices
and turn off $4,\cdots,7^{th}$ components.
Equivalently, we write all fields in the $2+1$ block diagonal form,
turning off the off-block-diagonal components.
In addition, we also turn off the $8^{th}$ component, or the $1\times 1$ block, for the fermions and for $f$.
That is,
\begin{eqnarray}\label{SU2U1-phipsi}
\phi^m &=& {\phi^m}_1 T^1 +{\phi^m}_2 T^2 +{\phi^m}_3 T^3 +\sqrt{3}\hat{\phi}^m T^8~, \nn\\
\psi_m &=& {\psi_m}_1 T^1 +{\psi_m}_2 T^2 +{\psi_m}_3 T^3~, \nn\\
f &=& f_1 T^1 + f_2 T^2 + f_3 T^3~,
\end{eqnarray}
or in matrix form,
\begin{equation}
    \phi^m = \begin{pmatrix} \phi^m|_{SU(2)} & 0 \\ 0 & 0 \end{pmatrix} +
    \hat\phi^m \begin{pmatrix} 1 & 0 & 0 \\ 0 & 1 & 0 \\ 0 & 0 & -2 \end{pmatrix}~, \quad
    \psi_m = \begin{pmatrix} \psi_m|_{SU(2)} & 0 \\ 0 & 0 \end{pmatrix}~, \quad
    f = \begin{pmatrix} f|_{SU(2)} & 0 \\ 0 & 0 \end{pmatrix}~.
\end{equation}
Note that we denote the $U(1)$ component with a hat,
and we will often abuse notation by omitting the $|_{SU(2)}$ sign.
It should be clear that if a field appears in the same expression as the hatted letters,
it indicates the $SU(2)$ block.
Although they are not needed for the present work,
one may turn on the $U(1)$ components $\hat\psi_m$ and $\hat{f}$ if needed.

Let $\cO^3 [R]$ be a gauge-invariant operator in the $SU(3)$ theory
in an irrep $R$ of the $SU(3)_R$ global symmetry.
Restricting each field in the $SU(2) \times U(1)$ block diagonal form \eqref{SU2U1-phipsi},
$\cO^3 [R]$ is expanded into many terms that contain different numbers of
the $SU(2)$ components and the $U(1)$ components.
It can be understood as the $SU(2)$ gauge-invariant operators with the abelian $U(1)$ components
acting as numerical coefficients.
For instance, (note that ${\rm tr}[\phi^{k}]_{SU(2)} = 0$)
\begin{equation}
 {\rm tr}[\phi^{(i}\phi^j\phi^{k)}]
 \xrightarrow{~SU(2)\times U(1)~} {\rm tr}[\phi^{(i}\phi^j\phi^{k)}]_{SU(2)}
 + 3\hat\phi^{(i} {\rm tr}[\phi^j\phi^{k)}]_{SU(2)}
 - 6\hat\phi^{(i} \hat\phi^{j} \hat\phi^{k)}~.
\end{equation}
Meanwhile, since $SU(3)_R$ is a global symmetry that is orthogonal to the gauge groups,
such an expansion in terms of products between $U(1)$ and $SU(2)$ operators
can be organized according to representations of each part under $SU(3)_R$.
Schematically, 
\begin{equation}\label{SU2U1-intoRs}
  \cO^3 [R] \xrightarrow{~SU(2)\times U(1)~} \sum_{R_1 \otimes R_2 \supset R} \hat\cO [R_1] \cO^{2} [R_2]~,
\end{equation}
where the summation is over all irreps $R_1$ and $R_2$ of $SU(3)_R$
such that irrep decomposition of $R_1 \otimes R_2$ contains $R$.
On the right hand side, we imply the $R$ part of the Clebsch-Gordan decomposition of the
tensor product $R_1 \otimes R_2$.
$\cO^{2} [R_2]$ is a gauge-invariant operator in the $SU(2)$ theory,
i.e. a multi-trace operator consisting of the $SU(2)$ fields,
while $\hat\cO [R_1]$ is simply a polynomial in $\hat\phi$
that multiplies $\cO^{2} [R_2]$ like a numerical coefficient.

%As we apply the proof for a specific case of $R=[3,0]$ in the next subsection,
%we present a complementary viewpoint on \eqref{SU2U1-intoRs}.
%This will hopefully provide more intuition for this abstract equation.

Our goal is to prove that $\cO^3 [R]$ is not $Q$-exact using results for the simpler operators $\cO^{2} [R_2]$.
To do so, we first need to ensure that the $Q$-exactness property is preserved under the
$SU(2) \times U(1)$ restriction.
Suppose that $\cO^3 [R]$ is $Q$-exact, so that $\cO^3 [R] = Q {\scriptstyle\mathcal{O}}^3 [R]$.
What we need is that the action of $Q$ and the $SU(2)\times U(1)$ restriction commute with each other:
\begin{center}
$Q$ ($SU(2)\times U(1)$ restriction of ${\scriptstyle\mathcal{O}}^3 [R]$) $=$
$SU(2)\times U(1)$ restriction of $\cO^3 [R]$.
\end{center}
To show this, recall the rule \eqref{Qaction} for the action of $Q$.
First, take the $SU(2)\times U(1)$ restriction \emph{after} the action of $Q$
by substituting each field on the right hand sides of \eqref{Qaction} as in \eqref{SU2U1-phipsi}.
Each equation in \eqref{Qaction} becomes
\begin{eqnarray}\label{Qactionthenrestrict}
    Q\Psi &=& \begin{pmatrix} Q \Psi |_{SU(2)} & 0 \\ 0 & 0 \end{pmatrix}~,
\end{eqnarray}
where $\Psi$ collectively denotes any fields,
and $Q \Psi |_{SU(2)}$ indicates the action of $Q$ \eqref{Qaction} understood
as all fields being $2\times 2$ matrices.
Note that the $U(1)$ components do not survive the commutators.
On the other hand, now take the $SU(2)\times U(1)$ restriction \emph{before} the action of $Q$
by making the restriction \eqref{SU2U1-phipsi} on the left hand sides of \eqref{Qaction}.
Then they become
\begin{eqnarray}\label{restrictthenQaction}
    Q\Psi &=& \begin{pmatrix} Q \Psi |_{SU(2)} & 0 \\ 0 & Q \Psi |_{U(1)} = 0 \end{pmatrix}~.
\end{eqnarray}
Agreement between \eqref{Qactionthenrestrict} and \eqref{restrictthenQaction}
completes the desired proof.%\footnote{The field restriction is basically 
%a consistent truncation. Let
%$L$ and $H$ be the components that we wish to keep and turn off, respectively
%(the $SU(2)\times U(1)$ block-diagonals and the off-diagonals). 
%The full $Q$-action takes the form of $QH\sim LH$, $QL\sim L^2+H^2$.
%The former equation consistently trivializes to $Q0\sim 0$ upon truncation. 
%The latter reduces to $QL\sim L^2$, which is the $Q$ action
%of the $SU(2)\times U(1)$ theory. One can also take $\hat\psi,\hat{f}$ to belong to 
%$H$ instead of $L$, i.e. turn them off, which is also consistent.}

Applying the logic of the previous paragraph, if $\cO^3 [R]$ is $Q$-exact,
then the RHS of \eqref{SU2U1-intoRs} which is the $SU(2)\times U(1)$ restriction of $\cO^3 [R]$,
must also be $Q$-exact.
Meanwhile, the global symmetry $SU(3)_R$ by definition commutes with $Q$,
so $Q$ acting on a definite irrep of $SU(3)_R$ does not change the representation.
So if the RHS of \eqref{SU2U1-intoRs} is $Q$-exact,
every term in the sum must be $Q$-exact separately.
Finally, note that $Q$ does not act on the $U(1)$ components.
Thus, the $SU(2)$ operator $\cO^{2} [R_2]$ for each term in the sum must be $Q$-exact.
In short,
\begin{eqnarray}\label{separatelyQexact}
  Q {\scriptstyle\mathcal{O}}^3 [R]
  &\xrightarrow{~SU(2)\times U(1)~}& \sum_{R_1 \otimes R_2 \supset R} \hat\cO [R_1] \cO^{2} [R_2] \nn\\
  \Rightarrow \qquad {\scriptstyle\mathcal{O}}^3 [R]
  &\xrightarrow{~SU(2)\times U(1)~}& \sum_{R_1 \otimes R_2 \supset R} \hat\cO [R_1]  {\scriptstyle\mathcal{O}}^{2} [R_2]~,
  ~~\text{where}~~ \cO^{2} [R_2] = Q{\scriptstyle\mathcal{O}}^{2} [R_2]~. \quad
\end{eqnarray}
As a result, for $\cO^3 [R]$ to be $Q$-exact,
every $SU(2)$ gauge-invariant operators $\cO^2 [R_2]$
that appears non-trivially in the sum \eqref{SU2U1-intoRs} must separately be $Q$-exact.
Equivalently, if any of $\cO^2 [R_2]$ in the sum is not $Q$-exact,
then the $SU(3)$ operator $\cO^3 [R]$ cannot be $Q$-exact either.

In the rest of this paper, we will use this method
to prove that some $Q$-closed operators that we construct are not $Q$-exact.
In fact, we will find that the smallest $SU(2)$ black hole cohomology $\cO^2_{24}$ of 
(\ref{SU2O0}) serves as $\cO^2 [R_2]$ that obstructs the $Q$-exactness of the $SU(3)$ operators.
This also allows the interpretation of the $SU(3)$ operators as the core $SU(2)$ black hole
microstate wrapped by dual giant hairs, because its $SU(2)\times U(1)$ restriction
\eqref{SU2U1-intoRs} crystallizes in the $SU(2)$ black hole cohomology.

\subsection{A tower of dual giant hairs}\label{sec:dg3}

In this subsection, we construct new black hole cohomologies $\cO^3_{24+2n}$ ($n=3,4,\cdots$)
whose quantum numbers correspond to those in the infinite $F_1$ tower of Table \ref{tower}.
They have charge $j=24+2n$, transform under the representation $[n,0]$ of $SU(3)_R$, and are fermions.

We start with the first cohomology: $n=3$.
Following the method outlined in \ref{sec:strategy}, we first construct
$Q$-closed ans\"atze \eqref{ansatz-general} for the black hole cohomology.
In a hope to find the picture of the core $SU(2)$ black hole (2 scalars, 5 fermions)
wrapped by the dual giant hair (3 scalars), we look for operators that
contain $5$ scalars and $5$ fermions.
Since $Q$ converts one fermion into two scalars, we need to find
relations \eqref{relation-of-relations} between the trace relations
that contain 7 scalars and 4 fermions.
Furthermore, since we look for operators that transform under the representation $[n,0]$,
we consider a set of the relations \eqref{relation-of-relations} that transform
as an $[n,0]$ tensor.
The relations with different field contents or different tensor structures
cannot have linear relations between them,
so we can consistently restrict as such.
%$s_{a,b}$'s with different field contents or not containing $[3,0]$
%cannot participate in the linear relations between $s_{a,b}$'s that we are 
%after, so we can consistently restrict as such.
%\siyul{Our $s_{a,b}$ are already the $[3,0]$ part of the Clebsch-Gordan decomposition.
%I understand that the form of \eqref{relation-of-relations} tells otherwise.
%I will think about what to do.}

In \eqref{tr-rel}, we listed all fundamental trace relations $R_a(g_I)$
of gravitons in the $SU(3)$ theory that contain the letters $\phi^m$ and $\psi_m$ only.
Each $R_a$ has definite numbers of scalars and fermions, which restricts the
graviton polynomial $f_a(g_I)$ that may multiply $R_a$ to enter the desired relation of relations.
As an illustration, let us take the example of $R_{14}^{(0,2)}[1,3]$, 
which contains 3 scalars and 2 fermions
and transforms under the $[1,3]$ representation of $SU(3)_R$.
We need to add 4 scalars and 2 fermions by means of $f_a(g_I)$.
There are two options: $u_2v_2v_2$ and $v_3v_3$.

$R_a$ as well as the gravitons $u_{2,3}$ and $v_{2,3}$
all belong to definite irreps of $SU(3)_R$.
%Recall that $s_{a,b}$ is the part of the Clebsch-Gordan decomposition of $R_a$ and the gravitons
%under the target representation, which is $[3,0]$ in this case.
We need to contract the tensor indices of $u_2v_2v_2$ or $v_3v_3$ and $R_{14}^{(0,2)}$
appropriately to produce a tensor in the $[3,0]$ representation.
A group theory argument can be used to determine the number of independent contractions. 
%, which becomes the size of the basis for the polynomial $f_a(g_I)$ that can multiply $R_a(g_I)$.
%how many independent linear combinations of $s_{a,b}$
%can be produced by multiplying and contracting $u_2v_2v_2$ or $v_3v_3$ with $R_{14}^{(0,2)}$.
First consider $u_2v_2v_2 R_{14}^{(0,2)}$.
Note that two identical fermions $v_2$ in the $[1,1]$ irrep are multiplied.
So only the anti-symmetric part survives in their tensor product:
\begin{equation}
    ( [1,1] \otimes [1,1] )_A = [3,0] + [0,3] + [1,1]~.
\end{equation}
Then $7$ independent $[3,0]$ tensors are obtained by decomposing the product 
$u_2v_2v_2 R_{14}^{(0,2)}$:
\begin{equation}
    [2,0] \otimes ( [1,1] \otimes [1,1] )_A \otimes [1,3] = \cdots + (7 \times [3,0]) + \cdots~.
\end{equation}
Similarly, there are 4 independent $[3,0]$ tensors in the product $v_3v_3 R_{14}^{(0,2)}$:
\begin{equation}
    ( [2,1] \otimes [2,1] )_A \otimes [1,3] = \cdots + (4 \times [3,0]) + \cdots~.
\end{equation}

In the previous paragraph, we illustrated that there are 7 and 4 $[3,0]$ 
tensors in the products $u_2v_2v_2 R_{14}^{(0,2)}$ and $v_3v_3 R_{14}^{(0,2)}$, respectively.
%Therefore, the non-fundamental trace relations $f_a(g_I)R_a(g_I)$
%with the desired field content and the $SU(3)_R$ representation
%are spanned by $7+4=11$ tensors.
Repeating the argument above for every fundamental trace relation $R_a$,
we find that there are $38$ non-fundamental trace relations 
in the irrep $[3,0]$ that span the desired trace relations.
%whose relation will correspond to the relation of relations \eqref{relation-of-relations}.
We list the $38$ basis tensors in \eqref{relation-30},
labeling them generically as $s_a$ with $a=1,\cdots,38$.

Before knowing that the fundamental trace relations $R$'s themselves are polynomials 
of gravitons, all $38$ trace relations $s_a$ are independent. However, as polynomials of 
gravitons, one can find linear relations \eqref{relation-of-relations} between the 38 relations.
As each $s_a$ is a $[3,0]$ tensor with 3 symmetrized fundamental indices,
a relation requires that the same linear combination between $s_a$'s vanishes
for all 10 independent components of $[3,0]$.
We find that there are 13 relations, and we present one of them (chosen with hindsight):
\begin{eqnarray}\label{t30ansatz1}
  0 \!&\!=\!&\! 24 s_1 - \frac{16}{15} s_3 + \frac{4}{15} s_4 - \frac{32}{15} s_5 + 16 s_9 
  + 85 s_{10} + 5 s_{11} - 40 s_{12} - \frac53 s_{13} + \frac53 s_{14} - 40 s_{15} \\
  \!&\!&\!+ \frac{125}{7} s_{17} + 31 s_{18} + \frac{24}{7} s_{19} + \frac{543}{7} s_{20}
  + \frac85 s_{22} + \frac{16}{15} s_{23} + \frac{251}{7} s_{30} - s_{31} - 9 s_{32} + \frac{174}{7} s_{33}
  - \frac{17}{2} s_{34}~. \nn
\end{eqnarray}
Every such relation leads to a $Q$-closed operator \eqref{ansatz-general}
by evaluating the same combination with $R_a$ replaced by $r_a$.

Now with the 13 ans\"atze for the black hole cohomology, one should verify if
any of these are not $Q$-exact.
Instead of determining if each ansatz is $Q$-exact or not, which will not be
computationally viable, we apply the strategy of section \ref{sec:strategy-exact}
to see if we can prove that any one of the ans\"atze is not $Q$-exact.

As explained in section \ref{sec:strategy-exact} in a generic context,
the key is to restrict a target ansatz $\cO^3 [3,0]$
into $SU(2)\times U(1)$ block diagonal fields \eqref{SU2U1-phipsi}
and expand as \eqref{SU2U1-intoRs}.
Among the expansion, we choose to focus on terms with $R_1 = [3,0]$.
Then, $R_2$ can only be the irreps whose tensor product with $R_1=[3,0]$ contains $[3,0]$.
The answer is that $R_2=[0,0]$, $[1,1]$, $[2,2]$ and $[3,3]$ are possible. 
Below, we present a rather algebraic route towards this answer in part to make 
the abstract arguments of section \ref{sec:strategy-exact} more tangible.

Suppose that we have an operator $\cO^{ijk}$ in the $[3,0]$ representation of $SU(3)_R$,
and restrict its field contents as in \eqref{SU2U1-phipsi}.
%We think of each component of $\cO^{ijk}$ simply as a polynomial in the field components.
Equivalent to focusing on the part of \eqref{SU2U1-intoRs} with $R_1 = [3,0]$
is to collect terms cubic in $\hat\phi^m$. (Recall that scalars are the only $U(1)$ parts 
that we keep in our restriction.)
As these are bosonic $U(1)$ components, products of three factors of $\hat\phi^m$
can only appear with their indices symmetrized, thus in the $[3,0]$ representation.
Without any loss of generality, we can organize such terms as
\begin{eqnarray}\label{decompose30-1}
\cO^{ijk} &\xrightarrow{~SU(2)\times U(1)~}& \hat\phi^{a} \hat\phi^{b} \hat\phi^{c} \tilde\cT^{(ijk)}_{(abc)}~.
\end{eqnarray}
%
%where $\tilde\cT^{(ijk)}_{(abc)}$ at this point should be understood as 
%a generic set of coefficients.
$\tilde\cT^{(ijk)}_{(abc)}$ is not a tensor with an $SU(3)_R$ irrep because it has non-zero traces,
e.g. $\tilde\cT^{(ijk)}_{(ibc)} \neq 0$.
Instead, it should be understood as a product $[3,0] \otimes [0,3]$.
By subtracting the traces of $\tilde\cT$, the traces of the traces, and its trace,
one can write $\tilde\cT$ in terms of four traceless tensors $\cT$'s:
\begin{eqnarray}\label{30x03}
\tilde\cT^{(ijk)}_{(abc)} &=& \cT^{(ijk)}_{(abc)} + \delta^i_a \cT^{(jk)}_{(bc)}
+ \delta^{(ij)}_{(ab)} \cT^{k}_{c}  + \delta^{(ijk)}_{(abc)} \cT~.
\end{eqnarray}
Now each $\cT$ on the right hand side is traceless, e.g. $\cT^{(ijk)}_{(ibc)} = 0$,
and is symmetric under permutations both within upper and lower indices as explicitly marked.
Thus they transform in irreps $[3,3]$, $[2,2]$, $[1,1]$ and $[0,0]$ of $SU(3)_R$.
\eqref{30x03} is an algebraic way of writing
$[3,0] \otimes [0,3] = [3,3] \oplus [2,2] \oplus [1,1] \oplus [0,0]$.
As a result, \eqref{decompose30-1} becomes
\begin{eqnarray}\label{decompose30-2}
  \cO^{ijk} &\xrightarrow{~SU(2)\times U(1)~}& \hat\phi^i \hat\phi^j \hat\phi^k \cT + \hat\phi^a \hat\phi^{(i} \hat\phi^j \cT^{k)}_a
  +\hat\phi^a \hat\phi^b \hat\phi^{(i} \cT^{jk)}_{ab} + \hat\phi^a\hat\phi^b\hat\phi^c \cT^{ijk}_{abc} \;.
\end{eqnarray}
This is an algebraic realization of \eqref{SU2U1-intoRs} for $R=[3,0]$ and $R_1 = [3,0]$,
for which $R_2$ runs over $[0,0]$, $[1,1]$, $[2,2]$ and $[3,3]$.

Now recall from section \ref{sec:strategy-exact} that if any of the four $SU(2)$ operators 
$\cT$, $\cT^k_c$, $\cT^{jk}_{bc}$, $\cT^{ijk}_{abc}$ in (\ref{decompose30-2}) is not 
$Q$-exact, then the $SU(3)$ operator $\cO^{ijk}$ cannot be $Q$-exact regardless of all other terms.
Using this argument, we expand all 13 $Q$-closed operators as in \eqref{decompose30-2},
and in particular find the corresponding $SU(2)$ operator $\cT$ which is an $SU(3)_R$ singlet.
There exist linear combinations of the $13$ $Q$-closed operators for which 
their corresponding $\cT$ are non-zero multiples of the smallest black hole cohomology in 
the $SU(2)$ theory, namely $\cO^2_{24}$ in \eqref{SU2O0}. One such operator 
is what corresponds to \eqref{t30ansatz1}, (named according to the convention $\cO^{N}_j$)  
\ba\label{30cohomology}
  \hspace*{-.5cm}(\cO^3_{30})^{ijk} =~&
  -24 \ep^{a_1a_2(i}\ep^{j|b_1b_2} {v^{c}}_d{v^{|k)}}_{a_1}{v^{d}}_{b_1} (r^{(0,2)}_{12})_{a_2b_2c}
  + \frac{16}{15} \ep^{a_1a_2(i} {v^{j|}}_{b}{v^{b}}_c{v^{c}}_{a_1} (r^{(0,2)}_{12})^{|k)}_{a_2} \nn\\
  &- \frac{4}{15} \ep^{a_1a_2a_3} {v^b}_{a_1}{v^{(i}}_{a_2}{v^{j}}_{a_3} (r^{(0,2)}_{12})^{k)}_b 
  + \frac{32}{15} \ep^{a_1a_2(i}{v^{j}}_b{v^{k)}}_{a_1}{v^{b}}_c (r^{(0,2)}_{12})^c_{a_2}
  -16 \ep^{a_1a_2(i} {v^{j|}}_d{v^{b}}_{a_1}{v^{d}}_{c} (r^{(0,2)}_{12})^{|k)c}_{a_2b} \nn\\
  &+ 85 \ep^{a_1a_2(i|}u^{bc}{v^{d}}_{a_1}{v^{|j}}_{b} (r^{(0,3)}_{14})^{k)}_{a_2cd} 
  + 5 \ep^{a_1a_2(i|}u^{bc}{v^{d}}_{b}{v^{|j}}_{a_1} (r^{(0,3)}_{14})^{k)}_{a_2cd}  - 40 \ep^{a_1a_2(i}u^{j|b}{v^{c}}_{a_1}{v^{d}}_{b} (r^{(0,3)}_{14})^{|k)}_{a_2cd}\nn\\  
  &- \frac53 \ep^{a_1a_2(i|}u^{cd}{v^{b}}_{a_1}{v^{|j}}_{b} (r^{(0,3)}_{14})^{k)}_{a_2cd} 
  + \frac53 \ep^{a_1a_2(i}u^{j|c}{v^{b}}_{a_1}{v^{d}}_{b} (r^{(0,3)}_{14})^{|k)}_{a_2cd}  - 40 \ep^{a_1a_2(i}u^{j|b}{v^{c}}_{a_1}{v^{|k)}}_{d} (r^{(0,3)}_{14})^{d}_{a_2bc} \nn\\
  &+ \frac{125}{7} \ep^{a_1a_2(i} {v^{j|b}}_{a_1} {v^{cd}}_b (r^{(0,3)}_{14})^{|k)}_{a_2cd} 
  + 31 \ep^{a_1a_2(i} {v^{bc}}_{a_1} {v^{d|j}}_b (r^{(0,3)}_{14})^{k)}_{a_2cd} + \frac{24}{7} \ep^{a_1a_2(i} {v^{jk)}}_{b} {v^{cd}}_{a_1} (r^{(0,3)}_{14})^{b}_{a_2cd} \nn\\
  &+ \frac{543}{7} \ep^{a_1a_2a_3} {v^{b(i}}_{a_1} {v^{j|c}}_{a_2} (r^{(0,3)}_{14})^{|k)}_{a_3bc} 
  + \frac85 \ep^{a_1a_2(i|}{v^{b}}_{a_1}{v^{|jk)}}_{b}(r^{(0,3)}_{16})_{a_2} + \frac{16}{15} \ep^{a_1a_2(i}{v^{j}}_{a_1}{v^{k)b}}_{a_2}(r^{(0,3)}_{16})_{b} \nn\\
  &+ \frac{251}{7} \ep^{a_1a_2a_3}{v^{b_1}}_{a_1}{v^{b_2(i}}_{a_2}(r^{(0,3)}_{16})^{jk)}_{a_3b_1b_2} 
  - \ep^{a_1a_2a_3}{v^{(i|}}_{a_1}{v^{b_1b_2}}_{a_2}(r^{(0,3)}_{16})^{|jk)}_{a_3b_1b_2} - 9 \ep^{a_1a_2(i}{v^{j|}}_{a_1}{v^{bc}}_d (r^{(0,3)}_{16})^{d|k)}_{a_2bc} \nn\\
  &+ \frac{174}{7} \ep^{a_1a_2(i|}{v^{b}}_{a_1}{v^{c|j}}_d (r^{(0,3)}_{16})^{k)d}_{a_2bc} 
  - \frac{17}{2} \ep^{a_1a_2(i|}{v^b}_{a_1}{v^{cd}}_{a_2} (r^{(0,3)}_{16})^{|jk)}_{bcd}~.
\ea
The operators $r$'s are related to the fundamental trace relations $R$ with same labels by $Qr=R$,
whose expressions are listed in \eqref{tr-r}.
Restricting this operator into the $SU(2) \times U(1)$ form, collecting terms cubic in $\hat\phi$
and expanding as in \eqref{decompose30-2},
the corresponding $SU(2)$ operator $\cT$ is
\begin{equation}\label{30cT}
    \cT = -\frac{648}{5} \cO^2_{24}~.
\end{equation}
The fact that $\cO^2_{24}$ is not $Q$-exact completes the proof that \eqref{30cohomology}
is a nontrivial cohomology at the order $j=30$ in the $SU(3)$ gauge theory.

One also finds that the cohomology represented by (\ref{30cohomology}) is not graviton-like.
An abstract argument is that we used trace relations to show that 
(\ref{30cohomology}) is $Q$-closed.
At any rate, if the original $SU(3)$ operator were a graviton,
the $SU(2)$ part $\cT$ after the restriction should also be a graviton,
which is not the case.
\iffalse
Another practical check is to restrict 
all the elementary fields to diagonals. It has been explained that graviton operators 
can be faithfully represented by the eigenvalues of the elementary fields \cite{Choi:2023znd}.
However, as explained in section \ref{sec:strategy} we have chosen the operators $r_a$'s to be all zero upon the diagonal restriction. So the operator $\cO^3_{30}$, which is 
a linear combination of $r_a$'s, is also zero after the restriction.
\fi
This proves that $\mathcal{O}^3_{30}$ is either trivial or
black hole type cohomology, between which we know that the latter is the case.
Below in this paper, similar arguments will be implicitly assumed whenever we claim that 
certain operators are not graviton-like.
\iffalse
Note also that this argument  
applies to each independent $SU(2)$ term in the $SU(2)\times U(1)$ decomposition (\ref{SU2U1-intoRs})
of the non-graviton operator:
if $\cO^{ijk}$ in \eqref{decompose30-2} is a non-graviton cohomology,
then each $\cT$ on the right hand side must be non-graviton as well.
\fi

Let us make a remark on the other 12 combinations of $Q$-closed operators.
For all $Q$-closed operators, the only possible non-$Q$-exact operator among
$\cT$, $\cT^k_c$, $\cT^{jk}_{bc}$ and $\cT^{ijk}_{abc}$ in their decomposition \eqref{decompose30-2}
is $\cT$, and $\cT$ is cohomologous to $\cO^2_{24}$ if not trivial.
This is because all these operators are $Q$-closed $SU(2)$ operators at 
$j=24$ and not of the graviton type. \cite{Chang:2022mjp} has shown that 
$\mathcal{O}^2_{24}$ is the unique cohomology with these conditions.
Therefore, it is always possible to find 12 linear combinations of the 13 operators
whose corresponding $\cT$ operator is trivial.
%It is important that not all multiplicative factors vanish, so there exists at least
%one linear combination whose $\cT$ part is a non-trivial multiple of $\cO^2_{24}$,
%of which \eqref{30cohomology} is one example that we find the simplest to write.
%Of course, there are other combinations of the 14 operators that are not $Q$-exact as well.
While our arguments can prove that an operator is not $Q$-exact,
the $\cT$ part being zero does not prove that it is $Q$-exact.
Thus, we have not proved that \eqref{30cohomology} is the unique $Q$-cohomology class 
within our ansatz. However, supported by the black hole cohomology index summarized in 
Table \ref{tower} we conjecture that \eqref{30cohomology} indeed represents the unique 
cohomology at $j=30$.

We now repeat all steps of the previous subsection more arduously to find a black hole cohomology
at $j=32$, with the $SU(3)_R$ representation $[4,0]$. This accounts for the second entry
in the tower $F_1$ of Table \ref{tower}.
As the logical steps are identical to the previous subsection,
we minimize the details to reach the result quickly.

We look for a cohomology that contains 6 scalars and 5 fermions,
so we first build non-fundamental trace relations that contain 8 scalars and 4 fermions,
and that belongs to the $[4,0]$ representation of $SU(3)_R$.
By group theory argument, we find that there are 106 independent tensors at $[4,0]$
that span these relations. We list all of them in \eqref{relation32}.

Among the relations \eqref{relation32}, $57$ are independent and there are $49$ linear relations,
yielding $49$ $Q$-closed operators.
As our proof based on the $SU(2)\times U(1)$ restriction can only determine if there is 
at least one non-$Q$-exact operator, we only present one operator
that we will shortly prove to be not $Q$-exact:
\allowdisplaybreaks{
\ba\label{32cohomology}
  \hspace*{-0.5cm}(\cO^3_{32})^{ijkl}
  &= - 162576\ep^{a_1a_2(i}\ep^{j|a_3a_4}{v^{|k}}_{a_1}{v^{l)}}_{a_5}{v^{a_5a_6}}_{a_3} (r_{12}^{(0,2)})_{a_2a_4a_6}
  +3240\ep^{a_1a_2(i}\ep^{j|a_3a_4}{v^{|k|}}_{a_1}{v^{a_5}}_{a_2}{v^{|l)a_6}}_{a_3} (r_{12}^{(0,2)})_{a_4a_5a_6} \nn\\
  &-19080\ep^{a_1a_2(i}\ep^{j|a_3a_4}{v^{|k|}}_{a_1}{v^{a_5}}_{a_3}{v^{|l)a_6}}_{a_2} (r_{12}^{(0,2)})_{a_4a_5a_6}
  +183096\ep^{a_1a_2(i}\ep^{j|a_3a_4}{v^{|k|}}_{a_1}{v^{a_5}}_{a_6}{v^{|l)a_6}}_{a_3} (r_{12}^{(0,2)})_{a_2a_4a_5} \nn\\
  &+283500\ep^{a_1a_2(i}\ep^{j|a_3a_4}{v^{a_5}}_{a_1}{v^{a_6}}_{a_2}{v^{|kl)}}_{a_3} (r_{12}^{(0,2)})_{a_4a_5a_6}
  -126000\ep^{a_1a_2(i}{v^{j}}_{a_1}{v^{k}}_{a_2}{v^{l)a_3}}_{a_4} (r_{12}^{(0,2)})^{a_4}_{a_3}\nn\\
  &+16800\ep^{a_1a_2(i}{v^{j}}_{a_1}{v^{k}}_{a_3}{v^{l)a_3}}_{a_4} (r_{12}^{(0,2)})^{a_4}_{a_2}
  -75600\ep^{a_1a_2(i}{v^{j}}_{a_1}{v^{k}}_{a_4}{v^{l)a_3}}_{a_2} (r_{12}^{(0,2)})^{a_4}_{a_3}\nn\\
  &+46200\ep^{a_1a_2(i}{v^{j|}}_{a_1}{v^{a_3}}_{a_4}{v^{|kl)}}_{a_2} (r_{12}^{(0,2)})^{a_4}_{a_3}
  +21000\ep^{a_1a_2(i}{v^{j|}}_{a_1}{v^{a_3}}_{a_4}{v^{|kl)}}_{a_3} (r_{12}^{(0,2)})^{a_4}_{a_2}\nn\\
  &-75600\ep^{a_1a_2(i}{v^{j|}}_{a_4}{v^{a_3}}_{a_1}{v^{|kl)}}_{a_2} (r_{12}^{(0,2)})^{a_4}_{a_3}
  -75600\ep^{a_1a_2(i}{v^{j|}}_{a_1}{v^{a_3}}_{a_4}{v^{|k|a_4}}_{a_2} (r_{12}^{(0,2)})^{|l)}_{a_3}\nn\\
  &+37800\ep^{a_1a_2(i|}{v^{a_3}}_{a_1}{v^{a_4}}_{a_2}{v^{|jk}}_{a_3} (r_{12}^{(0,2)})^{l)}_{a_4}
  +126000\ep^{a_1a_2(i}{v^{j}}_{a_1}{v^{k|}}_{a_2}{v^{a_3a_4}}_{a_5} (r_{12}^{(0,2)})^{|l)a_5}_{a_3a_4}\nn\\
  &-126000\ep^{a_1a_2(i}{v^{j|}}_{a_1}{v^{a_3}}_{a_4}{v^{|kl)}}_{a_5} (r_{12}^{(0,2)})^{a_4a_5}_{a_2a_3}
  +252000\ep^{a_1a_2(i}{v^{j|}}_{a_1}{v^{a_3}}_{a_5}{v^{a_4|k}}_{a_2} (r_{12}^{(0,2)})^{l)a_5}_{a_3a_4}\nn\\
  &-126000\ep^{a_1a_2(i|}{v^{a_3}}_{a_1}{v^{a_4}}_{a_5}{v^{|jk}}_{a_2} (r_{12}^{(0,2)})^{l)a_5}_{a_3a_4}
  -31500\ep^{a_1a_2a_3} {v^{(i}}_{a_1} {v^{j}}_{a_2} {v^{k}}_{a_3} (r_{14}^{(0,2)})^{l)}\nn\\
  &+31500\ep^{a_1a_2(i|}\ep^{a_3a_4|j} {v^{k}}_{a_1} {v^{l)}}_{a_2} {v^{a_5}}_{a_3} (r_{14}^{(0,2)})_{a_4a_5}
  +315000\ep^{a_1a_2(i|} u^{a_3a_4a_5} {v^{|j}}_{a_1} {v^{k}}_{a_2} (r_{14}^{(0,3)})^{l)}_{a_3a_4a_5}\nn\\
  &-1055700\ep^{a_1a_2(i} u^{j|a_3} {v^{|k|}}_{a_1} {v^{a_4a_5}}_{a_2} (r_{14}^{(0,3)})^{|l)}_{a_3a_4a_5}
  +363660\ep^{a_1a_2(i} u^{j|a_3} {v^{|k|}}_{a_1} {v^{a_4a_5}}_{a_3} (r_{14}^{(0,3)})^{|l)}_{a_2a_4a_5}\nn\\
  &-596400\ep^{a_1a_2(i} u^{j|a_3} {v^{|k|}}_{a_4} {v^{a_4a_5}}_{a_1} (r_{14}^{(0,3)})^{|l)}_{a_2a_3a_5}
  +1905300\ep^{a_1a_2(i} u^{j|a_3} {v^{a_4}}_{a_1} {v^{a_5|k}}_{a_2} (r_{14}^{(0,3)})^{l)}_{a_3a_4a_5}\nn\\
  &+890400\ep^{a_1a_2(i} u^{j|a_3} {v^{a_4}}_{a_1} {v^{a_5|k}}_{a_3} (r_{14}^{(0,3)})^{l)}_{a_2a_4a_5}
  +82500\ep^{a_1a_2(i} u^{j|a_3} {v^{a_4}}_{a_5} {v^{a_5|k}}_{a_1} (r_{14}^{(0,3)})^{l)}_{a_2a_3a_4}\nn\\
  &+2288100\ep^{a_1a_2(i|} u^{a_3a_4} {v^{|j}}_{a_1} {v^{k|a_5}}_{a_2} (r_{14}^{(0,3)})^{|l)}_{a_3a_4a_5}
  -247560\ep^{a_1a_2(i|} u^{a_3a_4} {v^{|j}}_{a_1} {v^{k|a_5}}_{a_3} (r_{14}^{(0,3)})^{|l)}_{a_2a_4a_5}\nn\\
  &+1506900\ep^{a_1a_2(i|} u^{a_3a_4} {v^{|j}}_{a_5} {v^{k|a_5}}_{a_1} (r_{14}^{(0,3)})^{|l)}_{a_2a_3a_4}
  -3441300\ep^{a_1a_2(i|} u^{a_3a_4} {v^{a_5}}_{a_1} {v^{|jk}}_{a_2} (r_{14}^{(0,3)})^{l)}_{a_3a_4a_5}\nn\\
  &-4275300\ep^{a_1a_2(i|} u^{a_3a_4} {v^{a_5}}_{a_1} {v^{|jk}}_{a_3} (r_{14}^{(0,3)})^{l)}_{a_2a_4a_5}
  +30360\ep^{a_1a_2(i} u^{jk} {v^{l)}}_{a_1} {v^{a_3a_4}}_{a_5} (r_{14}^{(0,3)})^{a_5}_{a_2a_3a_4}\nn\\
  &-134400\ep^{a_1a_2(i} u^{jk|} {v^{a_3}}_{a_1} {v^{|l)a_4}}_{a_5} (r_{14}^{(0,3)})^{a_5}_{a_2a_3a_4}
  +340680\ep^{a_1a_2(i} u^{j|a_3} {v^{|k}}_{a_1} {v^{l)a_4}}_{a_5} (r_{14}^{(0,3)})^{a_5}_{a_2a_3a_4}\nn\\
  &+1010400\ep^{a_1a_2(i} u^{j|a_3} {v^{a_4}}_{a_1} {v^{|kl)}}_{a_5} (r_{14}^{(0,3)})^{a_5}_{a_2a_3a_4}
  -414240\ep^{a_1a_2(i|} u^{a_3a_4} {v^{|j}}_{a_1} {v^{kl)}}_{a_5} (r_{14}^{(0,3)})^{a_5}_{a_2a_3a_4}\nn\\
  &+48000\ep^{a_1a_2(i}{v^{jk|}}_{a_1}{v^{a_3a_4}}_{a_5} (r_{16}^{(0,3)})^{|l)a_5}_{a_2a_3a_4}
  -58800\ep^{a_1a_2(i}{v^{jk|}}_{a_5}{v^{a_3a_4}}_{a_1} (r_{16}^{(0,3)})^{|l)a_5}_{a_2a_3a_4}\nn\\
  &-369600\ep^{a_1a_2(i}{v^{j|a_3}}_{a_1}{v^{a_4|k}}_{a_5} (r_{16}^{(0,3)})^{l)a_5}_{a_2a_3a_4}
  +33600\ep^{a_1a_2(i}{v^{j|a_3}}_{a_1}{v^{a_4a_5}}_{a_2} (r_{16}^{(0,3)})^{|kl)}_{a_3a_4a_5}\nn\\
  &+12250\ep^{a_1a_2(i}u^{jk|}{v^{a_3}}_{a_1}{v^{a_4}}_{a_5} (r_{16}^{(0,3)})^{|l)a_5}_{a_2a_3a_4}
  -615850\ep^{a_1a_2(i}u^{j|a_3}{v^{|k|}}_{a_1}{v^{a_4}}_{a_5} (r_{16}^{(0,3)})^{|l)a_5}_{a_2a_3a_4}\nn\\
  &+369250\ep^{a_1a_2(i}u^{j|a_3}{v^{|k|}}_{a_5}{v^{a_4}}_{a_1} (r_{16}^{(0,3)})^{|l)a_5}_{a_2a_3a_4}
  +184625\ep^{a_1a_2(i}u^{j|a_3}{v^{a_4}}_{a_1}{v^{a_5}}_{a_2} (r_{16}^{(0,3)})^{|kl)}_{a_3a_4a_5}\nn\\
  &-504800\ep^{a_1a_2(i|}u^{a_3a_4}{v^{|j}}_{a_1}{v^{k}}_{a_5} (r_{16}^{(0,3)})^{l)a_5}_{a_2a_3a_4}
  -1085950\ep^{a_1a_2(i|}u^{a_3a_4}{v^{|j|}}_{a_1}{v^{a_5}}_{a_2} (r_{16}^{(0,3)})^{|kl)}_{a_3a_4a_5}\nn\\
  &+701100\ep^{a_1a_2(i|}u^{a_3a_4}{v^{|j|}}_{a_1}{v^{a_5}}_{a_3} (r_{16}^{(0,3)})^{|kl)}_{a_2a_4a_5}\ .
\ea

By the $SU(2) \times U(1)$ restriction of the fields, this operator can be expanded
as in \eqref{SU2U1-intoRs}. This time we focus on terms with $R_1 = [4,0]$.
In other words, we collect terms that are quartic in the $U(1)$ components $\hat\phi^m$.
The analogous expansion to \eqref{decompose30-2} is
\begin{equation}\label{decompose32-2}
  \cO^{ijkl} \xrightarrow{~SU(2)\times U(1)~} \hat\phi^i \hat\phi^j \hat\phi^k \hat\phi^l \cT
  + \hat\phi^a \hat\phi^{(i} \hat\phi^j \hat\phi^k \cT^{l)}_a
  + \hat\phi^a \hat\phi^b \hat\phi^{(i} \hat\phi^j \cT^{kl)}_{ab}
  + \hat\phi^a\hat\phi^b\hat\phi^c \hat\phi^{(i} \cT^{jkl)}_{abc}
  + \hat\phi^a\hat\phi^b\hat\phi^c\hat\phi^d \cT^{ijkl}_{abcd}~.
\end{equation}
Restricting \eqref{32cohomology} into the $SU(2) \times U(1)$ form,
collecting terms quartic in $\hat\phi$ and expanding as in \eqref{decompose32-2},
the corresponding $SU(2)$ operator $\cT$ that is an $SU(3)_R$ singlet
is a non-trivial multiple of the $SU(2)$ black hole cohomology operator $\cO^2_{24}$ \eqref{SU2O0}:
\iffalse
%
\begin{equation}\label{32cT}
    \cT = -\frac{216}{125} \cO^2_{24}~.
\end{equation}
%
\fi
%
\begin{equation}\label{32cT}
    \cT = -1020600 \, \cO^2_{24}~.
\end{equation}
This completes the proof that \eqref{32cohomology} is a black hole cohomology at the order
$j=32$ in the $SU(3)$ gauge theory, and we conjecture its uniqueness.

We might repeat the entire procedure for $n=5,6, \cdots$.
Instead, we present a constructive method to obtain the infinite tower of cohomologies
for all $n\geq5$, based on two cohomologies $\cO^3_{30}$ and $\cO^3_{32}$ already constructed.
The cohomologies obtained using this method should be contained in the ansatz of section \ref{sec:strategy}.
The key idea is to multiply the two cohomologies by the $u$-type gravitons. 
This method will be used in the next section as well.

In the $SU(3)$ gauge theory, the $u$-type single graviton operators are
\begin{equation}\label{BMN-gen-u}
    {u_2}^{ij} \equiv \textrm{tr}  \left(\phi^{(i} \phi^{j)}\right)~, \qquad
    {u_3}^{ijk} \equiv \textrm{tr} \left(\phi^{(i} \phi^{j} \phi^{k)}\right)~,
\end{equation}
under the $[2,0]$ and $[3,0]$ representations of $SU(3)_R$, respectively.
Let us consider the simplest example, the product between $u_2$ and $\cO^3_{30}$.
The general case will follow in a straightforward manner.
As $[2,0] \otimes [3,0] = [5,0] \oplus [3,1] \oplus [1,2]$, the product is decomposed into three irreps:
\begin{eqnarray}\label{u2hair-30}
(u_2 \cO^3_{30})^{ijklm} &=& {u_2}^{(ij} (\cO^3_{30})^{klm)}~, \nn\\
(u_2 \cO^3_{30})^{ijk}_l &=& \ep_{lab}{u_2}^{a(i} (\cO^3_{30})^{jk)b}~,\nn\\
(u_2 \cO^3_{30})^{i}_{jk} &=& \ep_{jab}\ep_{kcd} {u_2}^{ac} (\cO^3_{30})^{ibd}~.
\end{eqnarray}
Since $Q$ annihilates both $u_2$ and $\cO^3_{30}$, it is clear that all three operators are $Q$-closed.

Now we shall employ the strategy of section \ref{sec:strategy-exact} to determine whether 
we can prove if these products are not $Q$-exact.
For each of the product operators in \eqref{u2hair-30}, we perform the $SU(2)\times U(1)$
restriction as in \eqref{SU2U1-phipsi} and collect, this time, terms at the $\hat\phi^5$ order.
There are two possibilities for such terms: $(2,3)$ or $(0,5)$ of the $\hat\phi$ factors
may originate from $(u_2,\cO^3_{30})$, respectively.
Note from \eqref{BMN-gen-u} that $u_2$ contains only 2 scalars, and that it is impossible
to extract only one factor of $\hat\phi$ from $u_2$ because ${\rm tr}[\phi^{i}]_{SU(2)}=0$,
\begin{eqnarray}
 {\rm tr}[\phi^{(i}\phi^{j)}]
 &\xrightarrow{~SU(2)\times U(1)~}& {\rm tr}[\phi^{(i}\phi^{j)}]_{SU(2)} + 6 \hat\phi^{i} \hat\phi^{j}~.
\end{eqnarray}
On the other hand, terms cubic in $\hat\phi$ under the $SU(2)\times U(1)$ restriction of
$\cO^3_{30}$ have been formally written down in \eqref{decompose30-2},
among which $\cT = -\frac{648}{5} \cO^2_{24}$ \eqref{32cT}.
The other $\cT$'s with the $[1,1]$, $[2,2]$, $[3,3]$ representations are all $Q$-exact
following a two-step logic.
First, they cannot be graviton operators because the operator on the left
is a non-graviton operator, see discussion below \eqref{30cT}.
Second, the only non-graviton cohomology of the $SU(2)$ theory
at the order $t^{24}$ is the $SU(3)_R$-singlet $\cO^2_{24}$ \cite{Chang:2022mjp},
which belongs to the singlet $\cT$.
Terms quintic in $\hat\phi$ under the $SU(2)\times U(1)$ restriction of
$\cO^3_{30}$ can be collected in a form similar to \eqref{decompose30-2}.
However, all $SU(2)$ blocks that appear on the right hand side will be $Q$-exact
for a similar reason: there is no $SU(2)$ black hole cohomology
at the order $t^{20}$ \cite{Chang:2022mjp}.

Collecting both possibilities for the numbers of $\hat\phi$ factors, we have 
\begin{eqnarray}\label{decompose-u230}
    (u_2 \cO^3_{30})^{ijklm} &\xrightarrow{~SU(2)\times U(1)~}&
    \hat\phi^{i}\hat\phi^{j}\hat\phi^{k}\hat\phi^{l}\hat\phi^{m} \cdot 6\cT + (Q\text{-exact})~, \nn\\
    (u_2 \cO^3_{30})^{ijk}_l &\xrightarrow{~SU(2)\times U(1)~}&
    \ep_{lab} \hat\phi^{a}\hat\phi^{i}\hat\phi^{j}\hat\phi^{k}\hat\phi^{b} \cdot 6\cT
    + (Q\text{-exact})~=~ 
    0 + (Q\text{-exact})~, \nn\\
    (u_2 \cO^3_{30})^{i}_{jk} &\xrightarrow{~SU(2)\times U(1)~}&
    \ep_{jab}\ep_{kcd} \hat\phi^{a}\hat\phi^{c}\hat\phi^{i}\hat\phi^{b}\hat\phi^{d} \cdot 6\cT
    + (Q\text{-exact})~=~ 
    0 + (Q\text{-exact})~,\qquad
\end{eqnarray}
where we put all $Q$-exact terms under the rug.
For the latter two, the obstruction terms from $\cT \propto \mathcal{O}^2_{24}$ 
vanish due to the antisymmetry of the $\ep$-tensor.

\eqref{decompose-u230} are specific applications of the general principle \eqref{SU2U1-intoRs}
on the operators with $R=[5,0]$, $[3,1]$ and $[1,2]$.
We are particularly interested in the term on the right hand side with $R_1 = [5,0]$ and $R_2 = [0,0]$,
i.e. quintic in $\hat\phi$ and the $SU(2)$ block being the $SU(3)_R$ singlet,
which is the only option for being not $Q$-exact.
For the first operator with $R=[5,0]$, \eqref{decompose-u230} shows that
the corresponding $SU(2)$ block $\cO^2[0,0]$ is a non-zero multiple of
$\cT = -\frac{648}{5} \cO^2_{24}$, possibly up to addition of $Q$-exact terms.
Therefore, this term obstructs the $[5,0]$ operator $(u_2 \cO^3_{30})^{ijklm}$ from being $Q$-exact.
In contrast, for the latter two operators with $R=[3,1]$ and $[1,2]$,
$R_1 = [5,0]$ and $R_2 = [0,0]$ are not allowed in the summation in \eqref{SU2U1-intoRs}
since $[5,0] \otimes [0,0]$ clearly does not contain $[3,1]$ or $[1,2]$.
This fact is naturally reflected in \eqref{decompose-u230}
as the corresponding terms being inevitably zero by the tensor structure.
Therefore, the $SU(2)$ black hole cohomology $\cO^2_{24}$ does not obstruct the $[3,1]$ operator
$(u_2 \cO^3_{30})^{ijk}_l$ and the $[1,2]$ operator $(u_2 \cO^3_{30})^{i}_{jk}$ from being $Q$-exact.

So we have proved that among the products between $u_2$ and $\cO^3_{30}$,
the totally symmetric part with the representation $[5,0]$ is not $Q$-exact. 
We propose that this represents the third black hole cohomology 
in the tower $F_1$.

Aiming for $n\geq6$, it is straightforward to extend this proof into products between
arbitrary numbers of $u_2$ and $u_3$, and either $\cO^3_{30}$ or $\cO^3_{32}$.
Consider any such product where the total number of scalars from the $u$-type gravitons is $n-k \geq 2$,
and they multiply $\cO^3_{24+2k}$ with either $k=3$ or $4$.
Among the tensor product between arbitrary numbers of $[2,0]$ and $[3,0]$
(representing the $u$-type gravitons) and $[3,0]$ or $[4,0]$ (representing the black hole cohomology),
there will be exactly one piece of $[n,0]$ represented by the totally symmetric product
\begin{eqnarray}\label{u2hair-gen}
(u \cdots u \cO^3_{24+2k})^{i_1 \cdots i_{n}} &=&
u^{(i_1 \cdots} \cdots u^{ \cdots i_{n-k}} (\cO^3_{24+2k})^{i_{n-k+1} \cdots i_{n})}~,
\end{eqnarray}
and all other irreps in the tensor product will follow by partly contracting 
the $SU(3)_R$ indices of $u\cdots u\mathcal{O}^3_{24+2k}$ 
with $\ep$-tensors, as exemplified in \eqref{u2hair-30}.

Then, we perform the $SU(2)\times U(1)$ restriction \eqref{SU2U1-phipsi}
and collect terms at the $\hat\phi^n$ order.
Again, there are two possibilities on the origin of the $\hat\phi$ factors: $(n-k,k)$ or
$(n-k-2,k+2)$ of the $\hat\phi$ factors may originate from $(u\cdots u,\cO^3_{24+2k})$, respectively.
Note that the two parts carry at maximum $n-k$ and $k+2$ factors of scalars respectively,
and it is impossible to extract all but one scalars as the $U(1)$ component from the gravitons,
because for both $u_2$ and $u_3$, this will leave only one traceless field inside an $SU(2)$ trace.
However, the non-$Q$-exact $SU(2)$ block cannot arise from the second possibility $(n-k-2,k+2)$,
because there is no $SU(2)$ black hole cohomology at the order $j=20$ that may appear as the
$SU(2)$ block in the $\hat\phi^{k+2}$ term in the $SU(2)\times U(1)$ restriction of $\cO^3_{24+2k}$.
So $(n-k,k)$ being the only possibility, every scalar in the $u$-type gravitons must contribute
by its $U(1)$ component and $k$ factors of $\hat\phi$ are extracted from $\cO^3_{24+2k}$.
The relevant expansion is
\begin{eqnarray}
    u^{(i_1 \cdots} \cdots u^{ \cdots i_{n-k})} &\xrightarrow{~SU(2)\times U(1)~}&
    \hat\phi^{i_1} \cdots \hat\phi^{i_{n-k}} ~ C  + \cdots~,
\end{eqnarray}
where $C$ is a non-zero integer, and the omitted terms $\cdots$ will at best 
contribute to $Q$-exact terms in the $SU(2)\times U(1)$ restriction of (\ref{u2hair-gen}).
As for $\mathcal{O}^3_{24+2k}$, one uses either \eqref{decompose30-2} or \eqref{decompose32-2}.

Combining the expansions for each part, we have
\begin{eqnarray}\label{decompose-u2gen}
    (u \cdots u \cO^3_{24+2k})^{i_1 \cdots i_n} &\xrightarrow{~SU(2)\times U(1)~}&
    \hat\phi^{i_1} \cdots \hat\phi^{i_n} ~ C \cT + (Q\text{-exact})~,
\end{eqnarray}
where $\cT = -\frac{648}{5} \cO^2_{24}$ \eqref{30cT}
if $k=3$ and $\cT = -\frac{216}{125} \cO^2_{24}$ \eqref{32cT} if $k=4$.
On the other hand, for all other irreps from the product of the $u$-type gravitons
and $\cO^3_{24+2k}$ with at least one lower index,
the corresponding term vanishes because the contractions with $\ep$-tensors
will annihilate the symmetric product $\hat\phi^{i_1} \cdots \hat\phi^{i_n}$.

Looking at \eqref{decompose-u2gen} as a specific case of \eqref{SU2U1-intoRs}
for $R=R_1 = [n,0]$, it shows that the term with $R_2 = [0,0]$ is not $Q$-exact
because it is proportional to the $SU(2)$ black hole cohomology.
Thus, \eqref{u2hair-gen} defines a black hole cohomology of the $SU(3)$ theory
at the order $t^{24+2n}$, under the $[n,0]$ representation of $SU(3)_R$,
where $n-k$ can be any integer $n-k \geq 2$, and $k=3$ or $4$.

For any $n\geq 6$, \eqref{u2hair-gen} defines the cohomology non-uniquely
as the product of $u$-type gravitons and $\cO^3_{30}$ or $\cO^3_{32}$.
For example, the totally symmetric products $(u_3)^{(i_1i_2i_3} (\cO^{3}_{30})^{i_4i_5i_6)}$
and $(u_2)^{(i_1i_2} (\cO^{3}_{32})^{i_3i_4i_5i_6)}$ both define a $[6,0]$ cohomology
at the order $t^{36}$, and for higher orders the apparent multiplicity grows further.
Moreover, had we directly extended the procedure we used for $n=3$ and $4$
into higher $n$ using relations of relations, we could have obtained cohomologies that may apparently look different from \eqref{u2hair-gen}, potentially adding to the multiplicity at every order.
However, the actual multiplicity will depend on how many combinations of them are $Q$-exact.
It is beyond the scope of the present work to fully answer this question.
As the $SU(3)_R$-singlet $SU(2)$ block $\cT$ is always a multiple of $\cO^2_{24}$,
our method is only able to determine at each order
if there is  at least one black hole cohomology.
So \eqref{u2hair-gen} constructs at least one $SU(3)$ black hole cohomology
at the order $t^{24+2n}$ and under the $SU(3)_R$ representation $[n,0]$ for all $n \geq 5$,
in addition to the cases $n=3$ and $4$ found earlier in this subsection.
They would constitute an infinite series of $SU(3)$ black hole cohomologies
whose overall charges $j$ and the $SU(3)_R$ representations match with the tower $F_1$
detected by the index, as shown in Table \ref{tower}.

Inspired by the gravitational picture of the dual giant hairs at large $N$
\cite{Choi:2024xnv,SUSY-GG}, it is tempting to conjecture that 
the multiplicity at each $n$ is exactly $1$. The basic idea on the bulk side is 
that the $u$ type gravitons carrying no spin have their wavefunctions 
peaked at the central region of AdS, which in the presence of the core black hole does 
not exist due to the event horizon. But collectively exciting these 
gravitons as a large enough spherical dual giant, its radial location 
can be outside the event horizon, providing a hair. 
Since large part of the apparent multiplicities in the previous paragraph comes 
from the multigraviton degeneracies in the absence of black holes, we expect 
that the core black hole operator will render most of them $Q$-exact.
Proving that these multiplicities of hairy cohomologies are all $1$,  
one would be showing that this emergent bulk picture is realized even at $N=3$. 
We again emphasize that we already have strong evidence for this from the index, 
as Table \ref{tower} shows no hint of such large multiplicities beyond the tower $F_1$.

If it is indeed true that the hair at given $n$ (within our ansatz) is unique, 
one may still think that the `graviton part' of (\ref{u2hair-gen}) does not really 
look like dual giant gravitons that are commonly known in the literature. 
For simplicity, let us take all the $[n,0]$ indices to be 
$i_1=\cdots =i_n=3$. This case is supposed to represent a half-BPS dual giant graviton 
made of $n$ scalars $\phi^3\equiv Z$ wrapping the core black hole microstate 
$\mathcal{O}^2_{24}$. In \cite{Corley:2001zk}, 
the symmetric Schur polynomials are expected to represent half-BPS dual giant gravitons 
in the absence of core black holes. On the other hand, in (\ref{u2hair-gen}), $n-k$ of $Z$'s 
are grouped into monomials of $u_2={\rm tr}(Z^2)$ and $u_3={\rm tr}(Z^3)$, while $k$ of them 
are fused with $\mathcal{O}^2_{24}$ to form $\mathcal{O}^3_{24+2k}$. 
The latter fused part has to do with the presence of black holes. Apart from this 
difference, symmetric Schur polynomials are highly nontrivial combinations of 
multitrace operators while (\ref{u2hair-gen}) is almost in a definite multitrace basis. 
These features look very different from each other. Here we note 
that the Schur basis of \cite{Corley:2001zk} is relevant as the physical BPS states 
(in the free limit), while (\ref{u2hair-gen}) are merely representatives of cohomologies. 
It may be possible to cleverly add $Q$-exact terms to (\ref{u2hair-gen}) and make 
the graviton part look more like the Schur polynomial for $n\gg 1$.

\subsection{No cohomologies below the threshold of the tower}\label{sec:dg12}

We have found the $SU(3)$ black hole cohomologies at order $j=24+2n$ in the
representation $[n,0]$ for all $n\geq 3$. From the gravity dual (barring the difference 
between $N=3$ and $\infty$), the presence of an infinite tower is natural because the dual 
giant wrapping the core black hole can be indefinitely large with larger electric 
charge $R$. On the opposite end, the dual giant graviton should be larger than a
certain size to wrap a definite core black hole. It would also be natural to imagine so 
for a definite black hole microstate. At this point, note that Table \ref{tower} manifests 
a curious feature, that the cohomologies only start 
to appear at $n=3$. 
%Although this does not prove rigorously 
%the absence of such hairy cohomologies at $n=1,2$ because the index can in principle 
%suffer from cancellations, 
So we are naturally led to ask what our general methods
yield at $n=1$ and $2$. We show in this subsection that indeed the methods of our section 
\ref{sec:strategy} fail to produce new hairy cohomologies 
at $n=1,2$. Our methods either yield no $Q$-closed operators at all ($n=1$), or 
only yield a $Q$-closed operator with no $SU(2)\times U(1)$ term which obstructs the $Q$-exactness
($n=2$). The failure to find cohomologies with too small hairs at $n \leq 2$ 
adds support to our attributing the tower $F_1$ to the black holes with dual giant hair.
We also claim that $n=3$ indicates the minimal size of the dual giant graviton
to wrap the core black hole, roughly equivalent to the size of the core black hole itself.

Let us proceed similarly to section \ref{sec:dg3}, to find a black hole cohomology
for $n=1$, i.e. at order $j=26$, with 3 scalars and 5 fermions, under the representation $[1,0]$.
There are only two ways to multiply the fundamental trace relations \eqref{tr-rel} by gravitons
to achieve the desired field contents and the representation. They are
\begin{eqnarray}
  (s_{26,1})^i&=&\ep^{a_1a_2a_3}{v^{a_4}}_{a_1}{v^{a_5}}_{a_2}(R_{14}^{(0,2)})^i_{a_3a_4a_5}~, \nn\\
  (s_{26,2})^i&=&{v^{i}}_{a}(R_{20}^{(0,3)})^a~, %\nn\\
  %(s_{26,3})^i&=&(R_{26}^{(0,4)})^i~.
\end{eqnarray}
%
%There are so few possibilities partly because it is difficult to fill the desired number 
%of fermions ($=5$)
%without exceeding the number of scalars ($=3$); the graviton operators available for multiplication
%all have at least as many scalars as fermions.
However, the two operators are independent of each other.
%(It is trivial by definition
%of the fundamental trace relation that $s_{26,3}$ is independent from the other two.)
So there is no relation of relations \eqref{relation-of-relations},
and thus no $Q$-closed operator that can be written under our ansatz. 

We turn to the case $n=2$, i.e. at order $j=28$, with 4 scalars and 5 fermions,
under the representation $[2,0]$.
There are 13 tensors that span the non-fundamental trace relations
with the desired field contents in the representation $[2,0]$:
\begin{eqnarray}
  (s_{28,1})^{ij} &=& \ep^{a_1a_2(i|}{v^{a_3}}_{a_5}{v^{a_4a_5}}_{a_1}(R_{14}^{(0,2)})^{|j)}_{a_2a_3a_4}~, \nn\\
  (s_{28,2})^{ij} &=& \ep^{a_1a_2a_3}{v^{a_4}}_{a_1}{v^{a_5(i}}_{a_2}(R_{14}^{(0,2)})^{j)}_{a_3a_4a_5}~, \nn\\
  (s_{28,3})^{ij} &=& \ep^{a_1a_2(i|}{v^{a_3}}_{a_5}{v^{|j)a_4}}_{a_1}(R_{14}^{(0,2)})^{a_5}_{a_2a_3a_4}~, \nn\\
  (s_{28,4})^{ij} &=& \ep^{a_1a_2(i}{v^{j)}}_{a_1}{v^{a_3a_4}}_{a_5} (R_{14}^{(0,2)})^{a_5}_{a_2a_3a_4}~, \nn\\
  (s_{28,5})^{ij} &=& \ep^{a_1a_2(i}{v^{j)}}_{a_5}{v^{a_3a_4}}_{a_1} (R_{14}^{(0,2)})^{a_5}_{a_2a_3a_4}~, \nn\\
  (s_{28,6})^{ij} &=& \ep^{a_1a_2(i}{v^{j)}}_{a_3}{v^{a_3}}_{a_1}(R_{16}^{(0,2)})_{a_2}~, \nn\\
  (s_{28,7})^{ij} &=& \ep^{a_1a_2(i}{v^{j)}}_{a_1}{v^{a_3}}_{a_2}(R_{16}^{(0,2)})_{a_3}~, \nn\\
  (s_{28,8})^{ij} &=& \ep^{a_1a_2(i|}{v^{a_3}}_{a_4}{v^{a_4}}_{a_1}(R_{16}^{(0,2)})^{|j)}_{a_2a_3}~, \nn\\
  (s_{28,9})^{ij} &=& \ep^{a_1a_2a_3}{v^{a_4}}_{a_1}{v^{(i}}_{a_2}(R_{16}^{(0,2)})^{j)}_{a_3a_4}~, \nn\\
  (s_{28,10})^{ij} &=& \ep^{a_1a_2(i}{v^{j)}}_{a_1}{v^{a_3}}_{a_4}(R_{16}^{(0,2)})^{a_4}_{a_2a_3}~, \nn\\
  (s_{28,11})^{ij} &=& \ep^{a_1a_2a_3}{v^{a_4}}_{a_1}{v^{a_5}}_{a_2}(R_{16}^{(0,2)})^{ij}_{a_3a_4a_5}~, \nn\\
  (s_{28,12})^{ij} &=& {v^{ij}}_a (R_{20}^{(0,3)})^a~, \nn\\
  (s_{28,13})^{ij} &=& {v^{(i}}_{a} (R_{22}^{(0,3)})^{j)a}~.
\end{eqnarray}
There is one linear combination of these that vanishes,
\begin{eqnarray}\label{28srel}
  -30 s_{28,1} + 30 s_{28,3} + \frac{15}{4} s_{28,4} - 15 s_{28,5}
  - \frac{7}{20} s_{28,6} + \frac{21}{80} s_{28,7} && \nn\\
  + \frac{49}{36} s_{28,8} - \frac{7}{12} s_{28,9} + \frac{7}{24} s_{28,10} + s_{28,11} &=& 0~,
\end{eqnarray}
which provides a $Q$-closed operator via \eqref{ansatz-general}.

By the $SU(2) \times U(1)$ restriction, we expand this operator as in \eqref{SU2U1-intoRs},
focusing on terms with $R_1 = [2,0]$.
The analogous expansion to \eqref{decompose30-2} is
\begin{eqnarray}\label{decompose28-2}
  \cO^{ij} &\xrightarrow{~SU(2)\times U(1)~}& \hat\phi^i \hat\phi^j \cT
  + \hat\phi^a \hat\phi^{(i} \cT^{j)}_a
  + \hat\phi^a\hat\phi^b \cT^{ij}_{ab}~.
\end{eqnarray}
Extracting the singlet part $\cT$ from the $Q$-closed operator, we get
\begin{equation}\label{28cT}
    \cT = 0~.
\end{equation}
Therefore, the only $Q$-closed operator that is constructed by our ansatz
cannot be proved to be not $Q$-exact with the method of section \ref{sec:strategy-exact}.
Although this does not prove that it is $Q$-exact, we find it suggestive that 
both our method and the index (Table \ref{tower}) predict no non-trivial 
cohomologies in this charge sector.
We emphasize that \eqref{28cT} is not as trivial as it seems.
If one makes $SU(2)\times U(1)$ restrictions of individual $s_{28,a}$'s, 
many of the corresponding $\cT$ parts are nonzero and proportional to 
the $SU(2)$ cohomology $\mathcal{O}^2_{24}$. However, they all intricately cancel 
in the prescribed linear combination (\ref{28srel}). So our method fails to 
produce new cohomologies nontrivially.

\section{More hairy black hole cohomologies}\label{sec:bothhair}

%We have established the infinite series of $SU(3)$ black hole cohomologies
%of the $F_1$ tower of Table \ref{tower}.
%After explicitly constructing two cohomologies with dual giant hair,
%we used products with the $u$-type gravitons to build the infinite tower.
In this section, we consider the hairy black hole cohomologies 
obtained by multiplying the $w$-type gravitons to the cohomologies in the $F_1$ 
tower of Table \ref{tower}.
We show that products with the $w_2$ gravitons always lead to new cohomologies,
partially justifying hairs speculated in \cite{Choi:2023vdm}.
Then we comment on possibilities of new cohomologies
by further dressing with the $w_3$ graviton hair. 
%Again our method of showing non-$Q$-exactness is the 
%$SU(2)\times U(1)$ restriction, using our knowledge of the $SU(2)$ black hole cohomologies. 
%With the $w$ type graviton hairs, a few things are known in the $SU(2)$ 
%theory. Furthermore, in the BMN sector of the $SU(2)$ theory, there is a conjecture 
%\cite{Choi:2023znd} on the full set of black hole cohomologies. 
%Since this conjecture is quite compelling to us, we shall often assume it in this 
%section to construct new $SU(3)$ cohomologies. 

We start by summarizing the results from the previous section.
For every integer $n\geq3$, there exists a black hole cohomology
$(\cO^3_{24+2n})^{i_1 \cdots i_n}$ at order $24+2n$ in the $[n,0]$ representation of $SU(3)_R$.
For each $n$, upon the $SU(2) \times U(1)$ restriction of the fields
and decomposing as in \eqref{decompose30-2},
the singlet part $\cT$ that corresponds to $\cO^2[R_2]$ in \eqref{SU2U1-intoRs} with
$R_1 = [n,0]$ and $R_2=[0,0]$ is a non-zero multiple of the $SU(2)$ black hole cohomology
$\cO^2_{24}$, and other $\cT$'s with non-trivial representations under $SU(3)_R$ are $Q$-exact.
In short,
\begin{eqnarray}\label{wbase}
    (\cO^3_{24+2n})^{i_1 \cdots i_n} &\xrightarrow{~SU(2)\times U(1)~}&
    \hat\phi^{i_1} \cdots \hat\phi^{i_n} ~ C_n\cO^2_{24} + (Q\text{-exact})~.
    \qquad (n\geq 3,~ C_n \neq 0)
\end{eqnarray}

First consider the product $w_2\cO^3_{24+2n}$,
where we recall that ${w_2}^i \equiv \textrm{tr}
\left(f \phi^i + \frac12 \ep^{ia_1a_2} \psi_{a_1}\psi_{a_2}\right)$.
%$w_2$ and $\cO^3_{24+2n}$ transform under the representations $[1,0]$ and 
%$[n,0]$ of $SU(3)_R$, respectively. 
The product decomposes into two irreps, 
\begin{equation}\label{10xn0}
    [1,0] \otimes [n,0] = [n+1,0] \oplus [n-1,1]~.
\end{equation}
The explicit forms of the operators are
\begin{eqnarray}\label{w2-n0-products}
  (w_2\cO^3_{24+2n})^{i_1\cdots i_{n+1}}
  &=& w^{(i_1}(\cO^3_{24+2n})^{i_2\cdots i_{n+1})}~, \nn\\
  (w_2\cO^3_{24+2n})^{i_1\cdots i_{n-1}}_{j}
  &=& \ep_{jab} w^{a}(\cO^3_{24+2n})^{bi_1\cdots i_{n-1}}~.
\end{eqnarray}
%
%Note that for the second operator, contraction of any of $i_1 \cdots i_{n-1}$ with $j$
%vanishes because of the antisymmetry of the $\ep$-tensor.
%This traceless condition is necessary to argue that the second operator,
%with $n-1$ symmetrized upper indices and one lower index, transforms under the irrep $[n-1,1]$.
%This property is already anticipated by the absence of $[n-2,0]$ in \eqref{10xn0},
%because if the trace, had it not vanished, will contribute an $[n-2,0]$ tensor to the product.
Both operators in \eqref{w2-n0-products} are automatically $Q$-closed,
because both $w_2$ and $\cO^3_{24+2n}$ separately are.

Now let us try to determine if they are not $Q$-exact, using the method of section \ref{sec:strategy-exact}.
We perform the $SU(2)\times U(1)$ restriction of the fields \eqref{SU2U1-phipsi}
and collect terms that contain $n$ factors of the $U(1)$ components $\hat{\phi}$.
No factor of $\hat\phi$ can be extracted from ${w_2}^i \equiv \textrm{tr}
\left(f \phi^i + \frac{1}{2} \epsilon^{ia_1a_2} \psi_{a_1} \psi_{a_2}\right)$,
because if one extracts $\hat\phi$ from the only $\phi$ inside the trace,
it will leave the traceless field $f$ alone in the $SU(2)$ trace.
So the $w_2$ graviton must be entirely $SU(2)$,
and all $n$ factors of $\hat\phi$ must come from $\cO^3_{24+2n}$.
The latter is summarized in \eqref{wbase}, so we can easily write the ${\hat\phi}^n$ terms
in the $SU(2)\times U(1)$ restriction of the two operators in \eqref{w2-n0-products}:
\begin{eqnarray}\label{w2hair-SU2U1}
  (w_2\cO^3_{24+2n})^{i_1\cdots i_{n+1}} &\xrightarrow{~SU(2)\times U(1)~}&
  \hat\phi^{(i_2} \cdots \hat\phi^{i_{n+1}} ~ C_n w^{i_1)} \cO^2_{24} + (Q\text{-exact})~, \nn\\
  (w_2\cO^3_{24+2n})^{i_1\cdots i_{n-1}}_{j} &\xrightarrow{~SU(2)\times U(1)~}&
  \ep_{jab} \hat\phi^{b} \hat\phi^{i_1} \cdots \hat\phi^{i_{n-1}} ~ C_n w^{a} \cO^2_{24} + (Q\text{-exact})~.
\end{eqnarray}
\eqref{w2hair-SU2U1} are specific applications of the general principle \eqref{SU2U1-intoRs}
on the operators with $R=[n+1,0]$ and $[n-1,1]$.
We focus on the term with $R_1 = [n,0]$ and $R_2 = [1,0]$ on the right hand side.
$R_1 \otimes R_2 = [n,0] \otimes [1,0]$ contains both $[n+1,0]$ and $[n-1,1]$,
so this term is allowed for both operators.
The advantage of this choice of $R_1$ and $R_2$ will become clear shortly.
\eqref{w2hair-SU2U1} shows that for both operators, the corresponding $SU(2)$ block
$\cO^2[1,0]$ is a non-zero multiple of $w^i \cO^2_{24}$.
Meanwhile, the $SU(2)$ operator $w^i \cO^2_{24}$ has been proved to be a black hole cohomology
in \cite{Choi:2023znd}, so it is not $Q$-exact.
Therefore, the terms proportional to $w^i \cO^2_{24}$ obstruct the operators on the left hand sides
from being $Q$-exact, proving that the entire product between $w_2$ and $\cO^3_{24+2n}$,
which consists of irreps $[n+1,0]$ and $[n-1,1]$, will be the new black hole cohomologies.

This argument can be repeated for products with arbitrary number of $w_2$ gravitons.
Consider the product between $k$ $w_2$ gravitons and $\cO^3_{24+2n}$.
The representation of the product would be
$[1,0]^{\otimes k}_S \otimes [n,0] = [k,0] \otimes [n,0]$.
$[1,0]^{\otimes k}_S$ is the symmetric product of $k$ $[1,0]$'s
because $w_2$ are identical bosons, which is $[k,0]$.
We write the product operator schematically as
\begin{eqnarray}\label{w2-gen}
    (w_2)^k \cO^3_{24+2n}~,
\end{eqnarray}
neglecting the tensor structure.
In other words, \eqref{w2-gen} collectively denotes all possible irreps that arise
from the tensor product $[k,0] \otimes [n,0]$.

We then perform the $SU(2)\times U(1)$ restriction and collect terms in $\hat\phi^n$.
Because no factor of $\hat\phi$ can be extracted from the $w_2$ gravitons,
we can use \eqref{wbase} directly to write the $SU(2)\times U(1)$ restriction of \eqref{w2-gen}:
\begin{eqnarray}\label{w2-gen-decompose}
    (w_2)^k \cO^3_{24+2n} &\xrightarrow{~SU(2)\times U(1)~}&
    {\hat\phi}^n ~C_n (w_2)^k \cO^2_{24} + (Q\text{-exact})~.
\end{eqnarray}
It is again written schematically neglecting the tensor structure,
but the tensor structure on the left between $[k,0]$ of $(w_2)^k$
and $[n,0]$ of $\cO^3_{24+2n}$ is identical to that on the right between
$[k,0]$ of $(w_2)^k$ and $[n,0]$ of ${\hat\phi}^n$.

\eqref{w2-gen-decompose} is a specific application of the general principle \eqref{SU2U1-intoRs}
on the operator $(w_2)^k \cO^3_{24+2n}$
with $SU(3)_R$ irrep $R$ that is contained in $[k,0] \otimes [n,0]$.
We are focusing on the term with $R_1 = [n,0]$ and $R_2 = [k,0]$ on the right hand side,
which is clearly allowed because $R$ is contained in $R_1\otimes R_2$ by definition.
\eqref{w2-gen-decompose} shows that the $SU(2)$ block $\cO^2[R_2]$ for this term
is a non-zero multiple of $(w_2)^k \cO^2_{24}$.
Meanwhile, all $SU(2)$ operators $(w_2)^k \cO^2_{24}$ have been conjectured
to be black hole cohomologies, i.e. not $Q$-exact, in \cite{Choi:2023znd}.
While it was shown only for $k=1$, it was conjectured for all $k>1$ based on the index over
black hole cohomologies which exhibits a geometric series that can be interpreted as
being freely generated by the $w_2$ gravitons.
It therefore acts as the obstruction term against the $Q$-exactness
of all product operators $(w_2)^k \cO^3_{24+2n}$.

This partially justifies the scenario of \cite{Choi:2023vdm} that
the $SU(3)$ black hole cohomologies would be freely dressed by the $w_2$ gravitons hairs.
This scenario was established rather empirically, based on 
i) the observation in \cite{Choi:2023znd} that the $w_2$ graviton 
can dress the black hole cohomologies to yield new black hole cohomologies in 
the $SU(2)$ theory, and ii) that the $SU(3)$ index simplifies significantly 
when the geometric series with respect to the $w_2$ gravitons is factored out.
Table \ref{tower} is the result of the factorization.
Here we have shown that the black hole cohomologies in the $F_1$ tower
can be dressed by any number of the $w_2$ graviton hair.

Finally, we comment on the $w_3$ graviton hair.
Products with the $w_3$ gravitons will only allow us to speculate, as opposed to guarantee,
new hairy black hole cohomologies.

Consider the product between a single $w_3$ graviton and $\cO^3_{24+2n}$.
They transform in the irreps $[2,0]$ and $[n,0]$ of $SU(3)_R$, respectively.
The product decomposes into three irreps, 
\begin{equation}\label{20xn0}
    [2,0] \otimes [n,0] = [n+2,0] \oplus [n,1]  \oplus [n-2,2]~.
\end{equation}
The explicit forms of operators are
\begin{eqnarray}\label{w3-n0-products}
  (w_3\cO^3_{24+2n})^{i_1\cdots i_{n+2}}
  &=& w^{(i_1i_2}(\cO^3_{24+2n})^{i_3\cdots i_{n+2})}~, \nn\\
  (w_3\cO^3_{24+2n})^{i_1\cdots i_n}_{j}
  &=& \ep_{ja_1a_2} w^{a_1(i_1}(\cO^3_{24+2n})^{i_2\cdots i_{n})a_2}~, \nn\\
  (w_3\cO^3_{24+2n})^{i_1\cdots i_{n-2}}_{jk}
  &=& \ep_{a_1a_2(j}\ep_{k)b_1b_2} w^{a_1b_1}(\cO^3_{24+2n})^{a_2b_2i_1\cdots i_{n-2}}~.
\end{eqnarray}
%
%Similar remarks to that below \eqref{w2-n0-products} regarding the traceless condition 
%for irreps follow.
All three operators are automatically $Q$-closed,
because both $w_3$ and $\cO^3_{24+2n}$ separately are.

Now we study if they are not $Q$-exact, using the method of section \ref{sec:strategy-exact}.
We perform the $SU(2)\times U(1)$ restriction of the fields \eqref{SU2U1-phipsi}
and collect the terms that contain $n+1$ factors of $\hat{\phi}$.
The one added factor of $\hat\phi$ is because one and only one factor of $\hat\phi$
can be extracted from $w_3 = \textrm{tr}\left(f \phi^{(i} \phi^{j)}
+\epsilon^{a_1a_2(i}\phi^{j)} \psi_{a_1} \psi_{a_2}\right)$.
$w_3$ contains at most two scalars, but it is impossible to extract
two scalar $U(1)$ components from $\tr[f\phi^{(i}\phi^{j)}]$
because it will leave the $SU(2)$ block of the traceless $f$ alone inside the $SU(2)$ trace.
It is also impossible to extract no $\hat\phi$ from $w_3$, because the entire $w_3$
graviton operator will then be restricted to the $SU(2)$ block, which is zero.
Thus, terms with $n+1$ factors of $\hat\phi$ can only be obtained by extracting one $\hat\phi$
from $w_3$ and $n$ from $\cO^3_{24+2n}$.
Under the $SU(2)\times U(1)$ restriction of $w_3$, terms that contain one factor of $\hat\phi$ is 
\begin{eqnarray}\label{w3phi1}
    w_3^{ij} \to 2\hat\phi^{(i}w_2^{j)}~.
\end{eqnarray}
%
%where we recall that $w_2$ is another single graviton operator given in \eqref{BMN-gen}.
On the other hand, the ${\hat\phi}^n$ terms in the $SU(2)\times U(1)$ restriction of $\cO^3_{24+2n}$
was already given in \eqref{wbase}.
Combining the two, the ${\hat\phi}^{n+1}$ terms from the three operators
in \eqref{w3-n0-products} are
\begin{eqnarray}\label{w3hair-SU2U1}
  w^{(i_1i_2}(\cO^3_{24+2n})^{i_3\cdots i_{n+2})} &\xrightarrow{~SU(2)\times U(1)~}&
  2C_n\hat{\phi}^{(i_1\cdots i_{n+1}}~ w^{i_{n+2})}\cO^2_{24} + (Q\text{-exact})~, \nn\\
  \ep_{ja_1a_2} w^{a_1(i_1}(\cO^3_{24+2n})^{i_2\cdots i_{n})a_2} &\xrightarrow{~SU(2)\times U(1)~}&
  -C_n\ep_{ja_1a_2} \hat{\phi}^{(i_1\cdots i_n)a_1}~ w^{a_2}\cO^2_{24} + (Q\text{-exact})~, \nn\\
  \ep_{a_1a_2(j}\ep_{k)b_1b_2} w^{a_1b_1}(\cO^3_{24+2n})^{a_2b_2i_1\cdots i_{n-2}}
  &\xrightarrow{~SU(2)\times U(1)~}& (Q\text{-exact})~.
\end{eqnarray}

\eqref{w3hair-SU2U1} are specific applications of the general principle \eqref{SU2U1-intoRs}
on the operators with $R = [n+2,0]$, $[n,1]$ and $[n-2,2]$,
and we are focusing on the term with $R_1 = [n+1,0]$ and $R_2 = [1,0]$ on the right hand side.
For the first two operators, the $SU(2)$ block $\cO^2[R_2] \propto w_2 \cO^2_{24}$
obstructs the operators on the left hand sides from being $Q$-exact.
In contrast, this term cannot exist for the third operator,
because $[n+1,0] \otimes [1,0]$ does not contain $[n-2,2]$.
So there is no obstruction against the $Q$-exactness of the third operator.

This argument for a single factor of $w_3$ can be generalized to an arbitrary number of them,
analogously to the factors of $w_2$.
Consider the product between $k$ $w_3$ gravitons and $\cO^3_{24+2n}$.
The $SU(3)_R$ representation of the product will be $[2,0]^{\otimes k}_S \otimes [n,0]$,
where $[2,0]^{\otimes k}_S$ is the symmetric product of $k$ $[2,0]$'s.
We write the product operator schematically as
\begin{eqnarray}\label{w3-gen}
    (w_3)^k \cO^3_{24+2n}~,
\end{eqnarray}
neglecting the tensor structure.
In other words, \eqref{w3-gen} collectively denotes all possible irreps that arise
from the tensor product $[2,0]^{\otimes k}_S \otimes [n,0]$.

We then perform the $SU(2)\times U(1)$ restriction and collect terms at 
the $\hat\phi^{n+k}$ order,
anticipating each $w_3$ to contribute one factor of $\hat\phi$.
We aim to find the $SU(2)$ operator $(w_2)^k \cO^2_{24}$ as the obstruction term 
against the $Q$-exactness. Combining \eqref{w3phi1} and \eqref{wbase},
\begin{eqnarray}\label{w3-gen-decompose}
    (w_3)^k \cO^3_{24+2n} &\xrightarrow{~SU(2)\times U(1)~}&
    {\hat\phi}^{n+k}~ C'_n (w_2)^k \cO^2_{24} + (Q\text{-exact})~,
\end{eqnarray}
where $C'_n$ is the product of $C_n$ and the numerical factor that arises
from the tensor structure, such as $2$ and $-1$ for the first two of \eqref{w3hair-SU2U1}.
It is again written schematically neglecting the tensor structure,
but it is not as trivial as the case for $w_2$, so let us comment on it.
Each $[2,0]_{w_3}$ on the LHS is decomposed on the RHS to a symmetric product
$[1,0]_{\hat\phi} \otimes_S [1,0]_{w_2}$, see \eqref{w3phi1}.
Of the product between $(w_3)^k$ and $\cO^3_{24+2n}$ on the LHS,
it is only the part where the $k$ factors of $[1,0]_{\hat\phi}$ and $[n,0]_{\cO^3_{24+2n}}$
are multiplied symmetrically into $[n+k,0]$
that appears as $\hat\phi^{n+k}$ on the RHS.
%Since $\hat\phi$ is a $U(1)$ boson, no antisymmetry between the $n+k$ indices is allowed.
Then, since the $k$ factors of $[1,0]_{\hat\phi}$ are symmetrized,
the $k$ factors of $[1,0]_{w_2}$ must be symmetrized among themselves as well.
This is what appears as $(w_2)^k \cO^2_{24}$ on the RHS.
To summarize, among the operator $(w_3)^k \cO^3_{24+2n}$ on the LHS that collectively denotes
many irreps that appear in $[2,0]^{\otimes k}_S \otimes [n,0]
= ([1,0]_{\hat\phi} \otimes_S [1,0]_{w_2})^{\otimes k}_S \otimes [n,0]$,
only the part where $k$ factors of $[1,0]_{\hat\phi}$ and $[n,0]_{\cO^3_{24+2n}}$ are
symmetrized into $[n+k,0]$ and $k$ factors of $[1,0]_{w_2}$ are symmetrized into $[k,0]$
admit the term proportional to $(w_2)^k \cO^2_{24}$ as displayed on the RHS.

Provided that such a term on the RHS is present,
this obstructs the operator on the LHS from being $Q$-exact.
However, it is too quick to declare that we have found new cohomologies.
Recall the product operator $(w_2)^k \cO^3_{24+2(n+k)}$ and the proof for its
non-$Q$-exactness presented earlier in this section, but with the value of $n$ shifted.
Its representation is $[1,0]^{\otimes k}_S \otimes [n+k,0] = [k,0] \otimes [n+k,0]$.
Reflecting on the previous paragraph, this is precisely the representation
of the subset of $(w_3)^k \cO^3_{24+2n}$ that admits the obstruction term,
and the obstruction terms for the two operators are identical.

As we have commented towards the end of section \ref{sec:dg3} on $\cO^3_{24+2n}$ with $n\geq6$,
our method guarantees only one nontrivial cohomology if many $Q$-closed operators share 
the common obstruction term. So we cannot conclude that the black hole cohomology with the
$w_3$ graviton hair $(w_3)^k \cO^3_{24+2n}$ yields any new black hole cohomology
in addition to those with the $w_2$ hair $(w_2)^k \cO^3_{24+2(n+k)}$.
On the other hand, we cannot conclude that they are not cohomologous either. 
With pending definiteness, we briefly comment on possible scenarios.

Taking the most conservative viewpoint that we only take into account the guaranteed
black hole cohomologies,\footnote{This was our viewpoint of section
\ref{sec:dg3}, where we advocated a single $F_1$ tower of dual giant hairs
despite possible multiplicities for $n\geq 6$. In this case, we were also strongly guided 
by the gravity results of \cite{Choi:2024xnv,SUSY-GG}.}
the $w_3$ hair does not add any new cohomology.
%The $w_2$ hair on the dual giant hair cohomologies in the tower $F_1$ justifies
%the factorization that has resulted in Table \ref{tower},
%but no other entry in Table \ref{tower} is addressed.

Now we explore more progressive viewpoints.
First consider the dual giant hair cohomologies in the $F_1$ tower dressed by
one factor of the $w_3$ graviton hair.
Among the product $w_3 \cO^3_{24+2n}$ that transforms under
$[2,0] \otimes [n,0] = [n+2,0] \oplus [n,1] \oplus [n-2,2]$,
those that have the $w_2 \cO^2_{24}$ obstruction term correspond to
$[1,0] \otimes [n+1,0] = [n+2,0] \oplus [n,1]$.
Note that $n\geq3$, and that $w_3 \cO^3_{24+2n}$ has the order $j=34+2n$.
Thus, if $w_3 \cO^3_{24+2n}$ belongs to the different cohomology class from
$w_2 \cO^3_{24+2(n+1)}$, then they will nicely account for the $F_2$ and $F_4$ towers
in Table \ref{tower}.
%Of course, it is possible that only one of the towers, or even only some entries in the towers,
%are the new cohomologies.
Second and finally, consider the dual giant hair cohomologies in the $F_1$ tower dressed by
two factors of the $w_3$ graviton hair.
Among the product $(w_3)^2 \cO^3_{24+2n}$,
those that have the $(w_2)^2 \cO^2_{24}$ obstruction term correspond to
$[2,0]\otimes [n+2,0] = [n+4,0] \oplus [n+2,1] \oplus [n,2]$.
If they are not cohomologous to the corresponding parts of
$(w_2)^2 \cO^3_{24+2(n+2)}$, they will account for towers  
that start with $[7,0]$, $[5,1]$ and $[3,2]$ at $j=50$.
From Table \ref{tower}, we note that the first such tower nicely accounts for the
series $F_3$, while the other two have not been detected by the index.
%We repeat that it is possible that only some of the towers, or even only some entries 
%in the towers, are the new cohomologies.
We feel that some of the product cohomologies discussed in this paragraph (if not all) 
have chances to be new cohomologies, independent of the the $w_2$ hairs, 
accounting for some towers in Table \ref{tower}. Since 
there are no independent obstruction terms against the $Q$-exactness in the $SU(2)\times U(1)$ 
restriction, it will require a different method to prove or disprove this.

\section{General comments on two types of hairs}\label{sec:genN}

In section 2, we studied novel hairy black hole cohomologies which are 
fusion products of core black hole microstates and the dual giant graviton hairs.
On the other hand, more traditional hairs given by the point 
particles in AdS and their quantum wave functions 
have been discussed in \cite{Choi:2023znd,Choi:2023vdm} and in section \ref{sec:bothhair}.
Their cohomologies were suggested to admit product representatives of the form 
$(\textrm{black hole})\times(\textrm{graviton})$. Concrete examples of both hairy cohomologies 
are available only at small $N$.
In this section, we argue the existence of the product hairy cohomologies at general $N$, 
and comment on related works in the literature.

The product cohomologies were suggested to represent the microstates 
of hairy black holes that have fast rotating point particle hairs 
outside the horizon.\footnote{More specifically, they can be interpreted as the BPS limits 
of the `gray galaxies' \cite{Kim:2023sig,SUSY-GG,temporary}.}
To be definite, consider the cohomology represented by the product operator
\begin{equation}\label{product-hair}
  \cO \cdot \partial_{1}^{j_1}\partial_{2}^{j_2}{\rm tr}(X^2)\ ,
\end{equation}
where $\cO$ is a fixed black hole cohomology, $\partial_{1,2}$ are the two holomorphic derivatives
on $\mathbb{R}^4$ that commute with $Q$, and $\partial_{1}^{j_1}\partial_{2}^{j_2}{\rm tr}(X^2)$ 
is a conformal descendant of the primary graviton operator ${\rm tr}(X^2)$ at 
the origin of $\mathbb{R}^4$, where $X\equiv \phi^1$ 
in the notation of this paper.
\eqref{product-hair} is $Q$-closed because both factors separately are.
The goal of this section is to discuss the (non)-$Q$-exactness of \eqref{product-hair} for general $N$.
The discussions in this section will hold for more general
graviton operators replacing ${\rm tr}(X^2)$.
Also, most of the arguments (before the gravity dual analysis) will hold for any $\cO$
that represents a nontrivial cohomology, either of graviton or of black hole type.

We first show a modest claim, that it is impossible for (\ref{product-hair}) 
to be $Q$-exact for all $j_{1,2}\geq 0$.
To prove this statement, let us assume otherwise, that they are all $Q$-exact.
One can equivalently state that the operator
\begin{equation}\label{product-hair-local}
  \cO \cdot {\rm tr}(X(z)X(z))
\end{equation}
is $Q$-exact for any $z=(z_1,z_2)\equiv z_\alpha$,
%nonzero value of $z=(z_1,z_2)\equiv z^{\dot\alpha}\neq (0,0)$,
where $X(z)$ is the holomorphic field operator defined in terms of the letters at the origin by
\begin{equation}
  X(z)\equiv \sum_{n=0}^\infty\frac{1}{n!}(z_1 D_1 +z_2 D_2)^nX(0)
  =\sum_{n=0}^\infty \frac{1}{n!}(z_{\alpha}D_{\alpha})^nX(0)\ .
\end{equation} 
(\ref{product-hair-local}) for a nonzero $z_\alpha$ is Taylor expanded into (\ref{product-hair}).

In terms of holomorphic fields, the $Q$-transformation takes the following 
local form \cite{Grant:2008sk,Chang:2013fba},
\ba\label{Qaction-local}
  Q\phi^m(z)&=[\lambda(z),\phi^m(z)]~, &
  Q\psi_m(z)&=\epsilon_{mnp}[\phi^n(z),\phi^p(z)]+\{\lambda(z),\psi_m(z)\}~, \nn\\
  Qf(z)&=[\phi^m(z),\psi_m(z)]+[\lambda(z),f(z)]~, &
  Q\lambda(z)&=\{\lambda(z),\lambda(z)\}~,
\ea
where we have included the gaugino $\lambda$,
\ba
  \phi^m(z)&=\sum_{n=0}^\infty\frac{1}{n!}(z_{\alpha}D_{\alpha})^n\phi^m(0)~, &
  \psi_m(z)&=\sum_{n=0}^\infty\frac{1}{n!}(z_{\alpha}D_{\alpha})^n\psi_m(0)~, \nn\\
  f(z)&=\sum_{n=0}^\infty\frac{1}{n!}(z_{\alpha}D_{\alpha})^nf(0)~, & 
  \lambda(z)&=\sum_{n=0}^\infty\frac{1}{(n+1)!}(z_{\alpha}D_{\alpha})^n
  (z_\beta \lambda_\beta(0))~.
\ea
In the holomorphic field basis, the supercharge $Q$ acts `locally' at given $z$,
i.e. it transforms a local operator at $z$ into a local operator at the same $z$.

Now we come back to the assumption (to be disproved) that
(\ref{product-hair-local}) for any $z$ is $Q$-exact, say $Q\Lambda$.
Since $\cO$ and the graviton part ${\rm tr}(X(z)X(z))$ are not $Q$-exact by themselves
but only via a trace relation,
There must be a term in $\Lambda$ on which $Q$ acts and produces a letter $X(z)$
and another letter at the same $z$:
note from \eqref{Qaction-local} that $Q$ always produces two letters at the same location.
Then by a trace relation, $X(z)$ enters the graviton part while the other enters $\cO$.
So $\cO$ must contain a local field at $z$.
%So if (\ref{product-hair-local}) is $Q$-exact, say $Q\Lambda$,
%by trace relations, it should be so by having a letter $X(z)$ in the graviton part 
%and another letter in $O$ generated by acting $Q$ on $\Lambda$. But since 
%$X(z)$ is local, this is possible only if every term of $O$ contains a factor of 
%local field at the same location $z$. However, note that (\ref{product-hair-local}) 
%should be $Q$-exact for all nonzero $z$ for if to be $Q$-exact. So for this to be possible, 
%every term in $O$ should contain a factor of local field at precisely that $z$. 
%It is easy to convince oneself that this is impossible.
However, the local field at $z\neq 0$ is a sum of infinitely many letters at the origin,
with indefinitely many derivatives.
As $\cO$ is an operator of fixed shape defined at the origin,
it cannot have a term with infinitely many letters,
and thus cannot contain the local field at $z \neq 0$.
This proves that (\ref{product-hair-local}) cannot be $Q$-exact,
and it follows that (\ref{product-hair}) cannot be $Q$-exact for all $j_{1,2} \geq 0$.
Therefore, \textit{some} conformal descendant graviton hair should be allowed, 
admitting the product representative (\ref{product-hair}).

We can slightly generalize the argument to show that, for a given black hole cohomology $\cO$,
an infinite number of the products with conformal descendant gravitons \eqref{product-hair}
must be not $Q$-exact.
Let us assume the opposite, that all but a finite number $n\geq1$ of products \eqref{product-hair} are $Q$-exact.
Then, among the Taylor expansion at the origin of the operator \eqref{product-hair-local},
up to $n$ of the expansion terms will be not $Q$-exact.
%so \eqref{product-hair-local} will be also not $Q$-exact.
However, consider a more general class of operators
\begin{equation}\label{product-hair-local2}
  \cO \cdot \left[{\rm tr}(X(z)^2)+\sum_{i=1}^na_i{\rm tr}(X(z_i)^2)\right]~,
\end{equation}
for some $z$ and $z_i$ ($i=1,\cdots,n$) that are all different.
Now consider the Taylor expansion of \eqref{product-hair-local2} at the origin.
In general, it will be possible to tune $n$ coefficients $a_i$'s
so that the coefficients for the $n$ non-$Q$-exact expansion terms vanish.
For these suitably chosen values of $a_i$'s for given $z$ and $z_i$,
\eqref{product-hair-local2} must be $Q$-exact by trace relations.
However, repeating the previous argument, 
$\cO$ must contain local fields at some of $z$ and $z_i$'s,
which is impossible for the black hole cohomology with fixed shape defined at the origin.
%Generalizing the argument of the previous paragraph, this operator
%cannot be $Q$-exact by using trace relations. Tuning 
%the values of $a_i$'s as functions of $z$ and $z_i$'s, one can have $n$ different 
%operators among (\ref{product-hair}) to cancel in the square bracket of (\ref{product-hair-local2}). 
So we conclude that, even after excluding any finite subset of operators
in (\ref{product-hair}), there still exist non-$Q$-exact operators in the remaining set. 
A consequence is that infinitely many operators of the form 
(\ref{product-hair}) are not $Q$-exact.\footnote{The argument of this paragraph 
does not extend to the case with $n=\infty$. For instance, one can construct a local 
operator at $z=0$ 
by an infinite linear combination $\Psi(0)=\oint\frac{dz}{z}\Psi(z)$. If the graviton 
part consists of local fields at $z=0$ only, there are known examples in which 
(\ref{product-hair-local2}) can be $Q$-exact: see appendix D of \cite{Choi:2023znd}.}

This is still weaker than what we conjecture to be true. 
We expect that, if (\ref{product-hair}) 
for some values $(j_1,j_2)=(j_1^{(0)},j_2^{(0)})$ of $j_{1,2}$ is not $Q$-exact,
then the same is true for all $j_1 \geq j_1^{(0)}$, $j_2\geq j_2^{(0)}$.
Namely, we conjecture that (\ref{product-hair}) is nontrivial beyond certain lower bound. 
We leave either its proof or disproof as a future work.

The conjecture of the previous paragraph is partly motivated by \cite{Choi:2023znd},
which studied the index for the cohomologies at $N=2$ and 
also the perturbative black hole hair at $N=\infty$.
Here we discuss the latter studies in more detail.
From the gravity dual perspective, our conjecture is natural because the size of the
core black hole is fixed by $\cO$ and the wavefunction of the graviton will be farther away 
from the black hole as we increase $j_{1,2}$,
making it easier to be a hair outside the horizon.
For simplicity, consider the AdS$_5$ BPS black hole with macroscopic angular momenta $J_1=J_2\equiv J$ 
and charges $R_1=R_2=R_3\equiv R$ \cite{Gutowski:2004ez}, whose microstate is $\cO$. 
Such BPS black hole solutions are labeled by one parameter, which we call $q$.
The charges carried by this black hole are given by
\begin{equation}
  R=\frac{N^2}{2}\left(\frac{q}{\ell^2}+\frac{q^2}{2\ell^4}\right)\ \ ,\ \  
  J=\frac{N^2}{2}\left(\frac{3q^2}{2\ell^4}+\frac{q^3}{\ell^6}\right)\ ,
\end{equation}
where $\ell$ is the AdS radius, satisfying a charge relation 
\begin{equation}\label{charge-relation}
  J(R)=-\frac{N^2}{2}-3R+\frac{(N^2+4R)^{\frac{3}{2}}}{2N}\ .
\end{equation}
It was noted \cite{Choi:2023znd} that the lower bound on $j_1,j_2$ for the hair 
to exist is related to the point particle 
trajectory (or the peak of its quantum wavefunction) 
staying outside the event horizon. 
The black hole hair for the bulk field dual to 
${\rm tr}(X^2)$
is shown to exist for $j_1+j_2>\frac{4q}{\ell^2}$ 
\cite{Choi:2023znd}\footnote{\cite{Choi:2023znd} also studied the linearized BPS equation 
for a scalar  
$\phi$ in AdS$_5$ dual to the operator ${\rm tr}(X^2)$, valid when its amplitude is small. 
Such solutions were found when $j_1+j_2>\frac{4q}{\ell^2}$, while otherwise the solution for 
$\phi$ diverges at the horizon and invalidates the linearized approximation assumed. 
Even for the finite solutions at $j_1+j_2>\frac{4q}{\ell^2}$, 
there is an issue of higher derivative singularities: 
we conjecture that they are acceptable singularities in string theory,
partly because similar results are obtained from the probe particle analysis \cite{Choi:2023znd}.
We thank Shiraz Minwalla for raising the issue of singularities to us.}.
For later use, we note that in the large charge regime $q\gg \ell^2$,
the bound for the BPS hair to exist can be written as
\begin{equation}\label{product-hair-bound}
  j\equiv j_1+j_2>2\left(\frac{2J}{N^2}\right)^{\frac{1}{3}}
\end{equation}
in terms of the macroscopic charge.

Now we reconsider the bound \eqref{product-hair-bound} by combining the QFT and gravity insights. 
If the black hole microstate $\cO$ carries angular momentum $J$, it may come either 
from derivatives or from the intrinsic spin carried by the fields.
At very large charges $R\gg N^2$ and $J\gg N^2$,
the charge relation (\ref{charge-relation}) for the large black holes is approximated by
\begin{equation}\label{charge-relation-large}
  J(R)\sim\frac{4R^{\frac{3}{2}}}{N}\ .
\end{equation}
In this `Cardy' limit, one expects that the angular momentum $J$ come almost entirely from the derivatives. 
This is because in the Cardy limit, the QFT dual derivation of the free energy yields divergence
due to unsuppressed numbers of derivatives 
\cite{Choi:2018hmj,ArabiArdehali:2019tdm,Honda:2019cio,Kim:2019yrz,Cabo-Bizet:2019osg}.
So the numbers $J_1$, $J_2$ of derivatives carried by the black hole microstate operator $\cO$
are approximately given by
\begin{equation}
  J_1\approx J_2=J\sim\frac{4R^{\frac{3}{2}}}{N}\ .
\end{equation}
At least for the order of magnitude estimate, the number of letters can be taken to be
proportional to $R$ unless $\cO$ is dominantly made by the field strength $f$,
since all other elementary fields carry $O(1)$ R-charges.
Therefore, distributing the $J$ derivatives to $\sim R$ letters, 
each letter carries roughly the following number of derivatives,
\begin{equation}
  \frac{J}{R}\sim \frac{J^{\frac{1}{3}}}{N^{\frac{2}{3}}}~,
\end{equation}
for a typical large black hole microstate.
So the bound (\ref{product-hair-bound}) for the angular momentum of the BPS hair
obtained from the gravity side can be interpreted as
the letters in the graviton $\partial^j{\rm tr}(X^2)$ carrying more derivatives
than the typical letters in the black hole microstate $\cO$ do.
In other words, the parton momentum distributions 
inside the graviton and the black hole should be separated to admit the  
nontrivial product hairs. It will be interesting to 
make a field theoretic analysis of the allowed product cohomologies 
focusing on the parton momentum distributions.

Related to the product cohomologies with a large number of derivatives, 
we also note that the anomalous dimensions of such product operators  
have been studied in the literature: see \cite{Alday:2007mf,Komargodski:2012ek}. 
Namely, the operator product expansion of $O_1(0)O_2(x)$ has been discussed  
(in the Lorentzian signature), from which one can study the anomalous dimensions of operators
of the form $O_1(0)\partial^j O_2(0)$. For the composite operator with smallest dimension,
the anomalous dimension at large $j$ is suppressed as 
\begin{equation}\label{product-small-anomalous}
  E-E_{\rm BPS}=cj^{-\tau}+\cdots\ ,
\end{equation}
with a certain number $c$ and a negative exponent $-\tau$. Although this does not 
say which operators are exactly BPS (even for the operators with vanishing leading coefficient 
$c$), perhaps one can combine the intuitions of these papers and our findings 
in this section to prove stronger statements.

On the other hand, for the new types of hairy cohomologies explored in this paper which defy 
the product representatives, we have little to say at general $N$ at this moment. 
However, like the product operators at large $j$ with small anomalous dimensions quoted 
in the previous paragraph, one may be able to show from QFT that the anomalous dimensions 
are small or zero when the dual giant size $n$ (as in the $[n,0]$
representation in section \ref{sec:dghair}) is sufficiently large. 
From the gravity perspective, there is a common feature between the product operators at
large $j$ and the dual giant hairy operators at large $n$.
Since we consider the core black hole operator with definite size, it occupies a definite finite 
region around the center of AdS. For the product hair, large $j$ indicates that the hair is very far away from 
the black hole, implying that the mutual back-reaction is parametrically small. This would be the 
gravity dual interpretation of (\ref{product-small-anomalous}).
As for the dual giant hair, large R-charge indicates that the dual giant radius is parametrically large,
again implying that the mutual back-reaction is small.
In fact, this was the key idea for constructing the approximate 
hairy black hole solutions in \cite{Kim:2023sig,Choi:2024xnv,SUSY-GG}.

As for the black hole microstates with large dual giant hairs, 
we consider them as the BPS states on $S^3\times \mathbb{R}$. 
Exciting one large dual giant graviton would 
approximately correspond to exciting the large classical solution in the $U(1)$ Cartan of $SU(N)$. 
This is effectively like going deep into the Coulomb branch so that the gauge symmetry is broken to
$U(1)\times SU(N\!-\!1)$. In this description, morally, the $1\times (N-1)$ and the $(N-1)\times 1$ 
off-diagonal modes would be made very heavy by the $U(1)$ Coulomb branch VEV, 
acquiring mass proportional to $\sqrt{n}$.
Exciting the large $U(1)$ part to realize the large dual giant hair and then exciting the BPS 
black hole state in the $SU(N-1)$ part, we may systematically study the quantum back-reactions 
by the heavy off-diagonal modes. This may perhaps shed more light on the $SU(2)\times U(1)$ 
terms identified in section \ref{sec:dghair} versus the rest,
where the former terms obstructed the full $SU(3)$ operator from being $Q$-exact. 
For concrete studies in this setup, it would be helpful to use the large charge 
effective field theory on the moduli space \cite{Cuomo:2024fuy} (see also \cite{Ivanovskiy:2024vel}).
Perhaps we may be able to derive a large charge asymptotic formula for the small anomalous dimensions 
of the hairy black hole state, analogous to (\ref{product-small-anomalous}). 
The $SU(N-1)$ core black hole part may be studied from the classical BPS solutions using 
the setup explored in \cite{Grant:2008sk,Baiguera:2022pll}.

\section{Conclusion and remarks}\label{sec:conclusion}

We constructed infinitely many new $\frac{1}{16}$-BPS cohomologies in the
$SU(3)$ maximal super-Yang-Mills theory and interpreted them as the finite $N$ versions 
of the black hole microstates containing dual giant graviton hairs. 
Our constructions explain the tower of black hole states $F_1$ appearing in the $SU(3)$
index, Table \ref{tower}. We also partly justified the product hairs 
conjectured in \cite{Choi:2023vdm}, by showing that the product between the gravitons $w_2$ 
and the cohomologies in the $F_1$ tower are not $Q$-exact.
Based on our methods,
it may be possible to construct the cohomologies in the fermionic towers $F_2$, $F_3$, $F_3$ 
and the bosonic towers $B_1$, $B_2$, $B_3$. We illustrated the possibilities for
the fermionic towers $F_{2,3,4}$ in section \ref{sec:bothhair}.
As for the fermionic cohomologies in the tower $F_0$, 
we may need different ideas, perhaps generalizing the ansatz of \cite{Choi:2023znd}.
We also discussed the general structures of hairy black hole cohomologies of two types, 
the point particle hair and the dual giant graviton hair. In particular, we proved
the existence of infinitely many cohomologies with rotating point particle hairs at general $N$.

We used a novel method to prove that the $Q$-closed operators 
constructed are not $Q$-exact. We showed that the terms in the operators made of 
the $SU(N-1)\times U(1)$ block diagonal components
%, which close under the action of $Q$, 
cannot be $Q$-exact, using our knowledge on the $SU(N-1)$ cohomologies. This setup 
is motivated from the fact that the objects inside a dual giant graviton essentially looks like those 
of the $SU(N-1)$ theory. So this proof also clarifies the bulk interpretation
as
%that the constructed cohomologies represent
the black hole microstate wrapped by a dual giant graviton.
Although this is quite intuitive, it would be more desirable to find an $SU(N)$ covariant
characterization of the brane wrapping the core black hole.
Since the two parts of the operator have their gauge orientations 
entangled to each other, the characterization would hopefully be purely group theoretic. 
We expect it will help us visualize the emergent 
bulk physics, say diagnosing the relative radial locations of the event horizon 
and the dual giant. We hope to come back to this question in the near future.

In this paper we studied how a black hole microstate can be 
dressed by hairs, and discussed in the bulk how the size of the black hole 
can be probed by the hairs.
However, this is perhaps a secondary question in the cohomology 
program for the BPS black hole microstates.
A more central and challenging question would be the
structures of the `core' black hole operators.
In the $SU(2)$ theory, an infinite series of the core black hole operators was
found explicitly \cite{Choi:2023znd}, but the well-organized nature of these operators
is unlike what we would expect from the `typical' black hole microstates.
%In the $SU(3)$ theory, one black hole cohomology without an apparent hair structure was found in \cite{Choi:2023vdm}.
In order to study the typical black hole states more extensively,
one needs a broader framework than the ansatz used in this paper.
We note that the large $N$ and large charge index
of these cohomologies were studied in the literature (see
\cite{Cabo-Bizet:2018ehj,Choi:2018hmj,Benini:2018ywd} and the references therein),
which reproduces the Bekenstein-Hawking entropy of the BPS black holes.
These results ensure that there should be a large number of black hole cohomologies at weak coupling.

We remark one way to extend the ansatz of \cite{Choi:2023vdm}.
Let us denote the graviton operators by $C_0=\{g_I\}$, the level-$0$ cohomologies.
As we reviewed in section \ref{sec:ansatz}, the ansatz for the $Q$-closed non-graviton
operators \eqref{ansatz-general} is based on the relations of relations \ref{relation-of-relations}
between the gravitons, $C_0$.
The ansatz leads to a subset of the non-graviton cohomologies,
which we denote by $C_1$, the level-$1$ cohomologies.
Then, one can find cohomologous relations ($Q$-exact polynomials) between $C_0\cup C_1$,
not just between the gravitons $C_0$.
Such new trace relations $R_a^{(1)}=Qr_a^{(1)}$ at level $1$ will extend the ansatz
for the $Q$-closed operators and possibly yield new cohomologies, that we denote by $C_2$.
This procedure can be repeated recursively to construct level-$n$ cohomologies $C_n$ for all $n\geq 0$. 
It will be interesting to know if all cohomologies can be constructed from
$C_0\cup C_1\cup C_2\cup \cdots$, or if not, how much can be covered by this recursive construction.
The level $n$ may also have a physical interpretation,
perhaps the complexity of the black hole microstates.
There may be simpler models than the $\mathcal{N}=4$ Yang-Mills theory
where these questions can be answered more concretely. (For instance, see \cite{SYK} 
for a recent study of fortuitous cohomologies in the supersymmetric SYK models.)

Another promising approach to uncovering new cohomologies associated with trace relations
involves utilizing the restricted Schur polynomial basis \cite{Bhattacharyya:2008rb,Bhattacharyya:2008xy}
(see also \cite{Brown:2007xh,Brown:2008ij} for closely related constructions
and \cite{deMelloKoch:2024sdf} for a pedagogical review).
The restricted Schur polynomials exploit the permutation symmetry that exchanges identical fields,
to construct the basis.
Each element of this basis is labeled by a collection of Young diagrams which label
irreducible representations of the permutation group which swaps fields in the composite operator.
A key advantage of this basis is its systematic treatment of trace relations.
Specifically, the complete set of trace relations are captured by the statement that
operators labelled by Young diagrams with more than $N$ rows vanish.
It is interesting to see if this approach is able to construct new $Q$-closed operators.

\vskip 0.5cm

\hspace*{-0.8cm} {\bf\large Acknowledgements}
\vskip 0.2cm

\hspace*{-0.75cm} 
We thank Yiming Chen, Jaehyeok Choi, Sunjin Choi, Seunggyu Kim, Eunwoo Lee, Shiraz Minwalla,
Masamichi Miyaji and Jaewon Song for helpful discussions.
This work is supported in part by a start up research fund of Huzhou University, 
a Zhejiang Province talent award, a Changjiang Scholar award (RdMK), 
NRF grants 2023R1A2C200536011, 2022R1I1A1A01069032, 
2020R1A6A1A03047877 (CQUeST) (MK),
NRF grant 2021R1A2C2012350 (SK, JL)
and FWO project G094523N (SL).

%%%%%APPENDIX
\appendix
\section{Trace relations of $SU(3)$ gravitons}\label{app:trrel}

In this appendix, we present the expressions for the graviton trace relations
that are referred to throughout the body of this paper.

\subsection{The fundamental trace relations}

We list the fundamental trace relations between the $SU(3)$ gravitons that only involve
the scalars $\phi^m$ and the fermions $\psi_m$.
These relations are referred to as $R_a(g_I)$ in section \ref{sec:ansatz},
representing polynomials of single-trace gravitons $g_I$ that are $Q$-exact.
These are fundamental in the sense that they generate all such polynomials.
In short, they generate all cohomologous relations between the $SU(3)$ multi-graviton
operators that only consist of scalars and fermions.

The single trace graviton operators only involving scalars and fermions are given by
\begin{equation}
    \begin{aligned}
        {u_2}^{ij} \equiv & \; \textrm{tr}  \left(\phi^{(i} \phi^{j)}\right)\ , &
        {u_3}^{ijk} \equiv & \; \textrm{tr} \left(\phi^{(i} \phi^{j} \phi^{k)}\right)\ , \\
        {{v_2}^{i}}_j \equiv & \; \textrm{tr} \left( \phi^i \psi_{j}\right) - \frac13
        \delta^i_j\textrm{tr}\left(\phi^a \psi_{a}\right)\ , &
        {{v_3}^{ij}}_k \equiv & \; \textrm{tr} \left(\phi^{(i} \phi^{j)} \psi_{k} \right) 
        - \frac{1}{4} \delta^{i}_{k}  
        \textrm{tr} \left( \phi^{(j} \phi^{a)} \psi_{a} \right) 
        - \frac{1}{4} \delta^{j}_{k}  \textrm{tr} \left( \phi^{(i} \phi^{a)} \psi_{a} \right)\ ,
    \end{aligned}
\end{equation}
The complete list of fundamental trace relations of the gravitons listed above are \cite{Choi:2023vdm}\footnote{{\bf v2}: three of these, namely
$(R_{24}^{(0,3)})$, $(R_{26}^{(0,4)})^i$ and $(R_{30}^{(0,5)})$,
vanish on the Coulomb branch (each gauge-adjoint letter restricted to diagonal)
but are not $Q$-exact \cite{Kim:2026}.
These should not be used for the ansatz.},
{\allowdisplaybreaks
    \begin{align}\label{tr-rel}
        &t^{10} [1,2](u_2u_3)&&:~(R_{10}^{(0,0)})^i_{jk} &=&~ \epsilon_{a_1 a_2 (j} \epsilon_{k) b_1 b_2} u^{a_1 b_1} u^{i a_2 b_2} \nonumber\\
        &t^{12} [0,0](u_2u_2u_2) &&:~ (R_{12}^{(0,0)}) &=&~ \epsilon_{a_1 a_2 a_3} \epsilon_{b_1 b_2 b_3} u^{a_1 b_1}u^{a_2 b_2}u^{a_3 b_3} \nonumber\\
        &t^{12} [2,2](u_2u_2u_2,u_3u_3) &&:~(R_{12}^{(0,0)})^{ij}_{kl} &=&~ \epsilon_{a_1 a_2 (k} \epsilon_{l) b_1 b_2} \left( u^{a_1 b_1} u^{a_2 b_2} u^{ij} + 6 u^{a_1 b_1 (i} u^{j) a_2 b_2} \right)  \nonumber\\
        &t^{12} [0,3](u_2v_3)&&:~ (R_{12}^{(0,1)})_{ijk} &=&~ \epsilon_{(i|a_1a_2}\epsilon_{|j|b_1b_2} u^{a_1b_1} {v^{a_2 b_2}}_{|k)} \nonumber\\
        &t^{12} [1,1](u_2v_3, u_3v_2) &&:~ (R_{12}^{(0,1)})^i_j &=&~  \epsilon_{j a_1 a_2} \left( 4 u^{a_1 b} {v^{i a_2}}_{b} + 3u^{i a_1b} {v^{a_2}}_{b}\right) \nonumber\\
        &t^{12} [2,2](u_2v_3, u_3v_2) &&:~ (R_{12}^{(0,1)})^{ij}_{kl} &=&~\epsilon_{a_1 a_2 (k}\left(u^{a_1(i}{v^{j)a_2}}_{l)}+
        u^{ij a_1}{v^{a_2}}_{l)}\right) \nonumber\\
        &t^{14} [1,0](u_2u_2v_2) &&:~ (R_{14}^{(0,1)})^i &=&~ \epsilon_{a_1a_2a_3} u^{i a_1} u^{b a_2} {v^{a_3}}_{b}  \nonumber\\
        &t^{14} [0,2](u_2u_2v_2, u_3v_3) &&:~ (R_{14}^{(0,1)})_{ij} &=&~ \epsilon_{a_1a_2 (i|}\left(\epsilon_{b_1b_2b_3} u^{a_1b_1} u^{a_2b_2}{v^{b_3}}_{|j)} -2 \epsilon_{|j) b_1b_2 } u^{a_1 b_1 c} {v^{a_2b_2}}_c \right) \nonumber\\
        &t^{14} [2,1](u_2u_2v_2, u_3v_3) &&:~ (R_{14}^{(0,1)})^{ij}_k &=&~\epsilon_{k a_1a_2} \left(3 u^{(a_1 b} u^{ij)} {v^{a_2}}_{b} + 4u^{a_1b} u^{a_2 (i} {v^{j)}}_b +24 u^{a_1 b (i} {v^{j) a_2}}_b \right)  \nonumber\\
        &t^{14} [1,3](u_2u_2v_2, u_3v_3) &&:~ (R_{14}^{(0,1)})^i_{jkl} &=&~\epsilon_{(j|a_1 a_2} \epsilon_{|k| b_1 b_2} \left(u^{a_1 b_1} u^{a_2 b_2} {v^i}_{|l)} + 6 u^{ia_1b_1} {v^{a_2b_2}}_{|l)}\right) \nonumber\\
        &t^{14} [3,2] (u_2u_2v_2, u_3v_3) &&:~ (R_{14}^{(0,1)})^{ijk}_{lm} &=&~\epsilon_{a_1a_2(l} \left( u^{(a_1 i} u^{jk)} {v^{a_2}}_{m)} + 6 u^{a_1 (ij} {v^{k)a_2}}_{m)}\right)  \nonumber\\
        &t^{14} [1,3](v_2v_3) &&:~ (R_{14}^{(0,2)})^{i}_{jkl} &=&~\epsilon_{a_1 a_2 (j} {v^{a_1}}_{k} {v^{i a_2}}_{l)} \nonumber\\
        &t^{16}[0,1](u_2v_2v_2, v_3v_3)&&:~(R_{16}^{(0,2)})_i &=&~\epsilon_{i a_1 a_2} \left(12u^{bc}{v^{a_1}}_b{v^{a_2}}_c +13u^{a_1 b}{v^{a_2}}_c{v^c}_b + 12{{v}^{a_1b}}_c{v^{a_2c}}_b \right) \nonumber\\
        &t^{16}[1,2] (u_2v_2v_2, v_3v_3)&&:~ (R_{16}^{(0,2)})^i_{jk} &=&~
        \epsilon_{a_1a_2(j}\!\left(3u^{ib}{v^{a_1}}_{k)}{v^{a_2}}_b \!-\! 
        7u^{ia_1}{v^b}_{k)}{v^{a_2}}_b \!+\! 6u^{a_1b}{v^i}_{k)}{v^{a_2}}_b 
        \!+\! 24{v^{a_1b}}_{k)}{v^{ia_2}}_b \right) \nonumber\\
        &t^{16}[2,3] (u_2v_2v_2,v_3v_3)&&:~(R_{16}^{(0,2)})^{ij}_{klm} &=&~ \epsilon_{a_1a_2(k}\left({u^{a_1(i}{v^{j)}}_{l}v^{a_2}}_{m)}+ 3{v^{a_1(i}}_{l}{v^{j)a_2}}_{m)}\right) \nonumber\\
        &t^{18}[0,0](u_3v_2v_2)&&:~(R_{18}^{(0,2)}) &=&~\epsilon_{a_1a_2a_3} u^{a_1bc} {v^{a_2}}_b {v^{a_3}}_c \nonumber\\
        &t^{20}[1,0](v_2v_2v_3) &&:~(R_{20}^{(0,3)})^i &=&~2{v^a}_c{v^b}_a{v^{ic}}_b - 3{v^i}_a{v^c}_b{v^{ab}}_c \nonumber\\
        &t^{22}[2,0](u_2v_2v_2v_2)&&:~(R_{22}^{(0,3)})^{ij} &=&~u^{ij}{v^a}_b{v^b}_c{v^c}_a -3 u^{a(i}{v^{j)}}_b{v^b}_c{v^c}_a +3 u^{ab}{v^{(i}}_a{v^{j)}}_c{v^c}_b \nonumber\\
        &t^{24}[0,0] (u_2v_2v_2v_3)&&:~(R_{24}^{(0,3)}) &=&~ \epsilon_{a_1a_2a_3}u^{a_1b}{v^{a_2}}_b{v^{a_3c}}_{d}{v^{d}}_c \nonumber\\
        &t^{26}[1,0] (v_2v_2v_2v_3)&&:~(R_{26}^{(0,4)})^i&=&~{v^{i}}_a {v^a}_b{v^d}_c{v^{bc}}_d 
        \nonumber\\
        &t^{30}[0,0] (v_2v_2v_2v_2v_2)&&:~(R_{30}^{(0,5)}) &=&~{v^{a}}_b{v^{b}}_c{v^{c}}_d{v^{d}}_e{v^{e}}_a \nonumber\\
        &t^{30}[3,0] (v_2v_2v_2v_2v_2)&&:~(R_{30}^{(0,5)})^{ijk}&=&~\epsilon^{a_1a_2(i}{v^{j}}_{a_1}{v^{k)}}_{a_2} {v^{b}}_c{v^{c}}_d{v^{d}}_b\ .
\end{align}
}
The superscripts $(n_f,n_\psi)$ indicate the numbers of $f$ and $\psi$
in the term with the largest number of $f$.
Note that we only consider examples with $n_f = 0$.
On the left, we also indicated the types of graviton monomials that appear in the polynomial.

As these graviton trace relations stand for $Q$-exact graviton polynomials,
each of them can be written as $R=Qr$. 
%, where the non-graviton operators $r$ are crucial
%for constructing the ansatz \eqref{ansatz-general} for the $Q$-closed operators.
Below we present the explicit expressions for the relevant $r$'s.
Note that there is an ambiguity in choosing $r$ given $R$.
We take advantage of this ambiguity to write all $r$'s such that they vanish
if all fields are diagonal.
This ensures that $r$'s are non-graviton operators,
see also discussion below \eqref{30cT}.\footnote{{\bf v2:}
the last term in $(r_{14}^{(0,2)})_{ij}$ has changed from v1.
Both expressions give the same $(R_{14}^{(0,1)})_{ij}$ as the difference between
the two is $Q$-closed, and either may enter \eqref{32cohomology}.}
{\allowdisplaybreaks
\begin{align}\label{tr-r}
        &(r_{10}^{(0,1)})^i_{jk} &=&~ -2\, \epsilon_{a_1a_2(j} \tr \left( \phi^{a_1}\phi^{a_2}\phi^{i}\psi_{k)} \right)\ ,  \nonumber\\
        &r_{12}^{(0,1)} &=&~ \epsilon_{a_1a_2a_3} \left[ 6 \tr\left( \psi_b \phi^{a_1} \right) \tr \left( \phi^b \phi^{a_2} \phi^{a_3}\right) - \tr\left( \psi_b \phi^{a_1} \phi^{a_2}\right) \tr \left( \phi^b  \phi^{a_3}\right)  \right] \nonumber\\
        &&& -3\epsilon_{a_1a_2a_3} \left[ \tr\left(\psi_b \phi^b \phi^{a_1} \phi^{a_2} \phi^{a_3}\right)
        \!+\!\tr\left(\psi_b \phi^{a_1} \phi^b \phi^{a_2} \phi^{a_3}\right)
        \!+\!\tr\left(\psi_b \phi^{a_1} \phi^{a_2}\phi^b  \phi^{a_3}\right)
        \!+\!\tr\left(\psi_b\phi^{a_1} \phi^{a_2} \phi^{a_3} \phi^b \right)\right] \ , 
        \nonumber\\
        &(r_{12}^{(0,1)})^{ij}_{kl} &=&~  -2\epsilon_{a_1a_2(k}\left[ \tr\left(\psi_{l)}\phi^{(i}\phi^{j)}\phi^{a_1}\phi^{a_2}\right) +7\tr\left(\psi_{l)}\phi^{(i|}\phi^{a_1}\phi^{|j)}\phi^{a_2}\right) \right]\ , 
        \nonumber\\
        &(r_{12}^{(0,2)})_{ijk} &=&~ \frac{1}{2} \epsilon_{a_1 a_2 (i} \tr \left( \phi^{a_1} \psi_{j}\phi^{a_2} \psi_{k)} \right) \ , \nonumber\\
        &(r_{12}^{(0,2)})^{i}_{j} &=&~  6 \tr \left( \phi^{(i} \phi^{a)} \psi_{(a} \psi_{j)}\right) - 5 \tr\left(\phi^{[i} \psi_{a}\phi^{a]} \psi_{j}\right)\ ,
        \nonumber\\
        &(r_{12}^{(0,2)})^{ij}_{kl} &=&~   \tr \left(\phi^{(i}\phi^{j)} \psi_{(k}\psi_{l)}\right)\ , \nonumber\\
        &(r_{14}^{(0,2)})^i  &=&~3\, \tr \left( \phi^i \psi_{a_1} \phi^{a_1} \phi^{a_2} \psi_{a_2} \right) + 2\, \tr \left( \phi^i \phi^{a_1} \right) \tr\left( \phi^{a_2} \psi_{(a_1} \psi_{a_2)} \right) \nonumber\\ 
        &&&-6\, \tr\left( \phi^i \psi_{a_1} \right) \tr \left( \phi^{[a_1} \phi^{a_2]} \psi_{a_2} \right) -\, \tr \left(\phi^i \psi_{a_1}\psi_{a_2} \right) \tr\left(\phi^{a_1} \phi^{a_2}\right)\ , \nonumber\\
        &(r_{14}^{(0,2)})_{ij}  &=&~   \frac{5}{9}\epsilon_{a_1a_2a_3}\left[2 \tr\left(\psi_{(i}\psi_{j)} \phi^{a_1} \phi^{a_2}\phi^{a_3}\right) + \tr\left(\psi_{(i} \phi^{a_1}\psi_{j)}\phi^{a_2}\phi^{a_3}\right)\right] \nonumber\\
        &&&+\epsilon_{a_1a_2(i}\left[\tr\left(\psi_{j)} \psi_{a_3}\phi^{a_1}\phi^{a_2}\phi^{a_3}\right)+\tr\left(\psi_{j)} \psi_{a_3}\phi^{a_1}\phi^{a_3}\phi^{a_2}\right)+\tr\left(\psi_{j)} \psi_{a_3}\phi^{a_3}\phi^{a_1}\phi^{a_2}\right)\right] \nonumber\\
        &&&+\epsilon_{a_1a_2(i}\left[\tr\left(\psi_{j)} \phi^{a_1}\psi_{a_3}\phi^{a_2}\phi^{a_3}\right)+\tr\left(\psi_{j)} \phi^{a_1}\psi_{a_3}\phi^{a_3}\phi^{a_2}\right)+\tr\left(\psi_{j)} \phi^{a_3}\psi_{a_3}\phi^{a_1}\phi^{a_2}\right)\right] \nonumber\\
        &&&+\epsilon_{a_1a_2(i}\left[\tr\left(\psi_{j)} \phi^{a_1}\phi^{a_2}\psi_{a_3}\phi^{a_3}\right)+\tr\left(\psi_{j)} \phi^{a_1}\phi^{a_3}\psi_{a_3}\phi^{a_2}\right)+\tr\left(\psi_{j)} \phi^{a_3}\phi^{a_1}\psi_{a_3}\phi^{a_2}\right)\right] \nonumber\\
        &&&+\epsilon_{a_1a_2(i}\left[\tr\left(\psi_{j)} \phi^{a_1}\phi^{a_2}\phi^{a_3}\psi_{a_3}\right)+\tr\left(\psi_{j)} \phi^{a_1}\phi^{a_3}\phi^{a_2}\psi_{a_3}\right)+\tr\left(\psi_{j)} \phi^{a_3}\phi^{a_1}\phi^{a_2}\psi_{a_3}\right)\right] \nonumber\\
        &&&-\frac13 \epsilon_{a_1a_2(i} \big[5\tr\left(\psi_{j)}\phi^{a_1}\phi^{a_2}\right)\tr\left(\psi_{a_3}\phi^{a_3}\right)
        +2\tr\left(\psi_{j)}\phi^{(a_1}\phi^{a_3)}\right)\tr\left(\psi_{a_3}\phi^{a_2}\right) \nn\\
        &&&
        \qquad \qquad \quad
        -2 \tr\left(\psi_{j)}\phi^{a_2}\right)\tr\left(\psi_{a_3}\phi^{(a_1}\phi^{a_3)}\right) \big]\ , \nonumber\\
        &(r_{14}^{(0,2)})^{ij}_{k}  &=&~12\tr\left(\phi^{(i} \phi^{a} \phi^{j)} \psi_{(a} \psi_{k)}\right)+12\tr\left(\phi^{(i|} \phi^{a} \phi^{|j)} \psi_{(a} \psi_{k)}\right) +54 \tr\left( \phi^{(i} \phi^{j} \psi_{(a} \phi^{a)} \psi_{k)} \right)-36 \tr\left( \phi^{(i} \phi^{j)} \psi_{(a} \phi^a \psi_{k)} \right) \ ,\nonumber\\
        &(r_{14}^{(0,2)})^i_{jkl}  &=&~ 2\epsilon_{a_1a_2(j}\left[\tr\left(\phi^i \phi^{a_1} \phi^{a_2}\psi_k \psi_{l)} \right) +3\tr\left(\phi^i \phi^{a_1} \psi_k \phi^{a_2} \psi_{l)} \right)-2\tr\left(\phi^i  \psi_k \phi^{a_1} \phi^{a_2} \psi_{l)}\right)\right]\ , 
        \nonumber\\
        &(r_{14}^{(0,2)})^{ijk}_{lm} &=&~ 4 \tr\left(\phi^{(i}\phi^j\phi^{k)}\psi_{(l}\psi_{m)}\right) + 3\tr\left(\phi^{(i}\phi^j\psi_{(l}\phi^{k)}\psi_{m)}\right) \nonumber\\
        &(r_{14}^{(0,3)})^i_{jkl}  &=&~  -\frac{1}{2} \tr \left(\phi^i \psi_{(j}\psi_{k}\psi_{l)} \right)\ , \nonumber\\
        &(r_{16}^{(0,3)})_i &=&~ 
        \frac{39}{4} \tr \left( \psi_i \{\psi_{b_1}\psi_{b_2}, \phi^{b_1}\phi^{b_2}\} \right) \!+\!
        2\tr \left( \psi_i \psi_{b_1} \phi^{b_1}\psi_{b_2}\phi^{b_2} \right) \!-\!
        \frac{61}{4}\tr \left( \psi_i \psi_{b_1} \phi^{b_2}\psi_{b_2}\phi^{b_1} \right) \nn\\
        &&&+\frac{97}{4}\tr \left( \psi_i  \phi^{b_1}\psi_{b_1}\psi_{b_2}\phi^{b_2} \right)  
        -\frac{41}{4} \tr \left( \psi_i  \phi^{b_2}\psi_{b_1}\psi_{b_2}\phi^{b_1} \right)
        -5 \tr \left( \psi_i  \psi_{b_1}\phi^{b_1}\phi^{b_2}\psi_{b_2} \right) \nn\\
        &&& -\frac{25}{2} \tr \left( \psi_i  \psi_{b_1}\phi^{b_2}\phi^{b_1}\psi_{b_2} \right)
        +2\tr \left( \psi_i  \phi^{b_1}\psi_{b_1}\phi^{b_2}\psi_{b_2} \right)
        - \frac{61}{4}\tr \left( \psi_i  \phi^{b_2}\psi_{b_1}\phi^{b_1}\psi_{b_2} \right) \nn\\
        &&& - \frac{11}{4} \tr \left(  \phi^{b_1}\phi^{b_2}\right) \tr \left(\psi_i \psi_{b_1}\psi_{b_2} \right)
        -\frac{27}{2} \tr \left( \psi_{b_1}\psi_{b_2} \right) \tr \left(\psi_i \phi^{b_1}\phi^{b_2} \right)
        + \frac{29}{4} \tr \left(\phi^{b_2} \psi_{b_2} \right) \tr \left(\psi_i [\psi_{b_1},\phi^{b_1}] \right)\ , \nonumber\\
        &(r_{16}^{(0,3)})^i_{jk} &=&~ 2 \tr \left( \psi_{(j} \psi_{k)} \psi_b \phi^b \phi^i \right)
        -4\tr \left( \psi_{(j} \psi_{k)} \psi_b \phi^i \phi^b \right)
        - \tr \left( \psi_{(j|} \psi_b \psi_{|k)} \{\phi^b ,\phi^i\} \right)
        -4 \tr \left( \psi_{(j} \psi_{k)} \phi^{(b}\psi_b  \phi^{i)} \right)  \nonumber\\
        &&&+7\tr \left( \psi_{(j|} \{\psi_b, \phi^{b}\}\psi_{|k)}  \phi^{i} \right)
        -11\tr \left( \psi_{(j|} \{\psi_b, \phi^{i}\}\psi_{|k)}  \phi^{b} \right)
        -4\tr \left( \psi_{(j} \psi_{k)}  \phi^b \phi^i \psi_b\right) \nonumber\\
        &&&+2\tr \left( \psi_{(j} \psi_{k)}  \phi^i \phi^b \psi_b\right)
        +3 \tr \left( \psi_{(j|} \psi_b \right) \tr \left( \psi_{|k)} [\phi^b, \phi^i] \right)
        +6 \tr \left( \psi_{(j} \phi^{[b} \right) \tr \left( \{\psi_{k)},\psi_b\} \phi^{i]} \right) \ , \nn\\
        &(r_{16}^{(0,3)})^{ij}_{klm} &=&~ -\tr(\phi^{(i}\phi^{j)}\psi_{(k}\psi_{l}\psi_{m)}) -2\tr(\phi^{(i}\psi_{(k}\phi^{j)}\psi_{l}\psi_{m)}) \;.
\end{align}}
All these expressions except for $r_{14}^{(0,2)} [3,2]$ and $r_{16}^{(0,3)} [2,3]$
were brought from \cite{Choi:2023vdm}.
Note that we only listed $r$'s for $j\leq 16$.
Although $r$'s with higher $j$ ($j\geq 18$) are included in our ansatz for constructing
some $Q$-closed operators in section \ref{sec:dg3},
the above $r$'s are sufficient to show the existence of the cohomology.

\subsection{The bases for the non-fundamental trace relations}

In this subsection, we list the bases for the non-fundamental graviton trace relations
whose field contents and the $SU(3)_R$ representation
are relevant to constructing the new cohomologies in section \ref{sec:dg3}.

We first list all 38 tensors that span the relations,
that trasform under the $[3,0]$ representation and contain 7 scalars and 4 fermions.
Linear relations between these relations (e.g. \eqref{t30ansatz1})
correspond to the relations of relations \eqref{relation-of-relations},
which lead to the ansatz \eqref{ansatz-general} for the $Q$-closed operator with the
$[3,0]$ representation and $j=30$ as relevant for the $n=3$ case of section \ref{sec:dg3}.
We label them with a simplified subscript $s_{1,\cdots,38}$.
{\allowdisplaybreaks \ba\label{relation-30}
  s_1 &= \ep^{a_1a_2(i}\ep^{j|b_1b_2} {v^{c}}_d{v^{|k)}}_{a_1}{v^{d}}_{b_1} (R^{(0,1)}_{12})_{a_2b_2c} \;, &
  s_2 &= \ep^{a_1a_2(i}\ep^{j|b_1b_2} {v^{|k)}}_d{v^{c}}_{a_1}{v^{d}}_{b_1} (R^{(0,1)}_{12})_{a_2b_2c} \;,\nn\\
  s_3 &= \ep^{a_1a_2(i} {v^{j|}}_{b}{v^{b}}_c{v^{c}}_{a_1} (R^{(0,1)}_{12})^{|k)}_{a_2} \;, &
  s_4 &= \ep^{a_1a_2a_3} {v^b}_{a_1}{v^{(i}}_{a_2}{v^{j}}_{a_3} (R^{(0,1)}_{12})^{k)}_b \;,\nn\\
  s_5 &= \ep^{a_1a_2(i}{v^{j}}_b{v^{k)}}_{a_1}{v^{b}}_c (R^{(0,1)}_{12})^c_{a_2} \;,&
  s_6 &= \ep^{a_1a_2(i|}{v^b}_c{v^c}_d{v^d}_{a_1} (R^{(0,1)}_{12})^{|jk)}_{a_2b} \;,\nn\\
  s_7 &= \ep^{a_1a_2a_3} {v^{b}}_{a_1}{v^{c}}_{a_2}{v^{(i}}_{a_3} (R^{(0,1)}_{12})^{jk)}_{bc} \;,&
  s_8 &= \ep^{a_1a_2(i} {v^{j|}}_{a_1}{v^{b}}_{d}{v^{d}}_{c} (R^{(0,1)}_{12})^{|k)c}_{a_2b} \;,\nn\\
  s_9 &=  \ep^{a_1a_2(i} {v^{j|}}_d{v^{b}}_{a_1}{v^{d}}_{c} (R^{(0,1)}_{12})^{|k)c}_{a_2b} \;,&
  s_{10} &= \ep^{a_1a_2(i|}u^{bc}{v^{d}}_{a_1}{v^{|j}}_{b} (R^{(0,2)}_{14})^{k)}_{a_2cd}\;,\nn\\
  s_{11} &= \ep^{a_1a_2(i|}u^{bc}{v^{d}}_{b}{v^{|j}}_{a_1} (R^{(0,2)}_{14})^{k)}_{a_2cd} \;,&
  s_{12} &= \ep^{a_1a_2(i}u^{j|b}{v^{c}}_{a_1}{v^{d}}_{b} (R^{(0,2)}_{14})^{|k)}_{a_2cd} \;,\nn\\
  s_{13} &= \ep^{a_1a_2(i|}u^{cd}{v^{b}}_{a_1}{v^{|j}}_{b} (R^{(0,2)}_{14})^{k)}_{a_2cd} \;,&
  s_{14} &= \ep^{a_1a_2(i}u^{j|c}{v^{b}}_{a_1}{v^{d}}_{b} (R^{(0,2)}_{14})^{|k)}_{a_2cd} \;,\nn\\
  s_{15} &= \ep^{a_1a_2(i}u^{j|b}{v^{c}}_{a_1}{v^{|k)}}_{d} (R^{(0,2)}_{14})^{d}_{a_2bc}\;,&
  s_{16} &= \ep^{a_1a_2a_3}u^{(ij|}{v^{c}}_{a_1}{v^{d}}_{a_2} (R^{(0,2)}_{14})^{|k)}_{a_3cd} \;,\nn\\
  s_{17} &= \ep^{a_1a_2(i} {v^{j|b}}_{a_1} {v^{cd}}_b (R^{(0,2)}_{14})^{|k)}_{a_2cd} \;,&
  s_{18} &= \ep^{a_1a_2(i} {v^{bc}}_{a_1} {v^{d|j}}_b (R^{(0,2)}_{14})^{k)}_{a_2cd} \;,\nn\\
  s_{19} &= \ep^{a_1a_2(i} {v^{jk)}}_{b} {v^{cd}}_{a_1} (R^{(0,2)}_{14})^{b}_{a_2cd} \;,&
  s_{20} &= \ep^{a_1a_2a_3} {v^{b(i}}_{a_1} {v^{j|c}}_{a_2} (R^{(0,2)}_{14})^{|k)}_{a_3bc}  \;,\nn\\
  s_{21} &= \ep^{a_1a_2(i}{v^{j}}_b{v^{k)b}}_{a_1}(R^{(0,2)}_{16})_{a_2} \;,&
  s_{22} &= \ep^{a_1a_2(i|}{v^{b}}_{a_1}{v^{|jk)}}_{b}(R^{(0,2)}_{16})_{a_2} \;,\nn\\
  s_{23} &= \ep^{a_1a_2(i}{v^{j}}_{a_1}{v^{k)b}}_{a_2}(R^{(0,2)}_{16})_{b} \;,&
  s_{24} &=  \ep^{a_1a_2(i}{v^{j|}}_b{v^{bc}}_{a_1}(R^{(0,2)}_{16})^{|k)}_{a_2c} \;,\nn\\
  s_{25} &= \ep^{a_1a_2(i|}{v^{c}}_b{v^{|j|b}}_{a_1}(R^{(0,2)}_{16})^{|k)}_{a_2c} \;,&
  s_{26} &= \ep^{a_1a_2(i|}{v^{b}}_{a_1}{v^{|j|c}}_{b}(R^{(0,2)}_{16})^{|k)}_{a_2c} \;,\nn\\
  s_{27} &= \ep^{a_1a_2b}{v^{c}}_{a_1}{v^{(ij}}_{a_2}(R^{(0,2)}_{16})^{k)}_{bc} \;,&
  s_{28} &= \ep^{a_1a_2(i}{v^j}_{a_1}{v^{k)b}}_c (R^{(0,2)}_{16})^{c}_{a_2b} \;,\nn\\
  s_{29} &= \ep^{a_1a_2(i|}{v^b}_{a_1}{v^{|jk)}}_c (R^{(0,2)}_{16})^{c}_{a_2b} \;,&
  s_{30} &= \ep^{a_1a_2a_3}{v^{b_1}}_{a_1}{v^{b_2(i}}_{a_2}(R^{(0,2)}_{16})^{jk)}_{a_3b_1b_2} \;,\nn\\
  s_{31} &= \ep^{a_1a_2a_3}{v^{(i|}}_{a_1}{v^{b_1b_2}}_{a_2}(R^{(0,2)}_{16})^{|jk)}_{a_3b_1b_2} \;,&
  s_{32} &= \ep^{a_1a_2(i}{v^{j|}}_{a_1}{v^{bc}}_d (R^{(0,2)}_{16})^{d|k)}_{a_2bc} \nn\\
  s_{33} &= \ep^{a_1a_2(i|}{v^{b}}_{a_1}{v^{c|j}}_d (R^{(0,2)}_{16})^{k)d}_{a_2bc} &
  s_{34} &= \ep^{a_1a_2(i|}{v^b}_{a_1}{v^{cd}}_{a_2} (R^{(0,2)}_{16})^{|jk)}_{bcd}  \nn\\
  s_{35} &= \ep^{ab(i}{v^j}_a{v^{k)}}_b R_{18}^{(0,2)} \;,&
  s_{36} &= u^{a(i}{v^j}_a (R_{20}^{(0,3)})^{k)} \;,\nn\\
  s_{37} &= u^{(ij}{v^{k)}}_a (R_{20}^{(0,3)})^{a} \;,&
  s_{38} &= {v^{(ij}}_a (R_{22}^{(0,3)})^{k)a} \;. %\nn\\
  % s_{39} &= u^{(ij} (R_{26}^{(0,4)})^{k)} \;.
\ea}
Among these 38 trace relations, 25 are independent
and there are 13 relations of relations, among which is \eqref{t30ansatz1}.
They provide $13$ operators which are $Q$-closed by trace relations.
One of these is presented in \eqref{30cohomology}, and we prove subsequently
that it is not $Q$-exact and thus represents a black hole cohomology.

Similarly, we now list all 106 tensors that span the relations,
that trasform under the $[4,0]$ representation and contain 8 scalars and 4 fermions.
Linear relations between these relations correspond to \eqref{relation-of-relations},
which lead to the ansatz \eqref{ansatz-general} for the $Q$-closed operator with the
$[4,0]$ representation and $j=32$ as relevant for the $n=4$ case of section \ref{sec:dg3}.
We label them with a simplified subscript $s_{1,\cdots,106}$.
\allowdisplaybreaks{
\ba\label{relation32}
  %v_2v_2v_3 t^{12}[0,3]:
  s_1&= \ep^{a_1a_2(i}\ep^{j|a_3a_4}{v^{|k}}_{a_1}{v^{l)}}_{a_2}{v^{a_5a_6}}_{a_3} (R_{12}^{(0,1)})_{a_4a_5a_6}~,&
  s_2&= \ep^{a_1a_2(i}\ep^{j|a_3a_4}{v^{|k}}_{a_1}{v^{l)}}_{a_5}{v^{a_5a_6}}_{a_3} (R_{12}^{(0,1)})_{a_2a_4a_6}~,\nn\\
  s_3&= \ep^{a_1a_2(i}\ep^{j|a_3a_4}{v^{|k|}}_{a_1}{v^{a_5}}_{a_2}{v^{|l)a_6}}_{a_3} (R_{12}^{(0,1)})_{a_4a_5a_6}~,&
  s_4&= \ep^{a_1a_2(i}\ep^{j|a_3a_4}{v^{|k|}}_{a_1}{v^{a_5}}_{a_3}{v^{|l)a_6}}_{a_2} (R_{12}^{(0,1)})_{a_4a_5a_6}~,\nn\\
  s_5&= \ep^{a_1a_2(i}\ep^{j|a_3a_4}{v^{|k|}}_{a_1}{v^{a_5}}_{a_6}{v^{|l)a_6}}_{a_3} (R_{12}^{(0,1)})_{a_2a_4a_5}~,&
  s_6&= \ep^{a_1a_2(i}\ep^{j|a_3a_4}{v^{|k|}}_{a_6}{v^{a_5}}_{a_1}{v^{|l)a_6}}_{a_3} (R_{12}^{(0,1)})_{a_2a_4a_5}~,\nn\\
  s_7&= \ep^{a_1a_2(i}\ep^{j|a_3a_4}{v^{a_5}}_{a_1}{v^{a_6}}_{a_2}{v^{|kl)}}_{a_3} (R_{12}^{(0,1)})_{a_4a_5a_6}~,&
  %v_2v_2v_3 t^{12}[1,1]:
  s_8&= \ep^{a_1a_2(i}{v^{j}}_{a_1}{v^{k}}_{a_2}{v^{l)a_3}}_{a_4} (R_{12}^{(0,1)})^{a_4}_{a_3}~,\nn\\
  s_9&= \ep^{a_1a_2(i}{v^{j}}_{a_1}{v^{k}}_{a_3}{v^{l)a_3}}_{a_4} (R_{12}^{(0,1)})^{a_4}_{a_2}~,&
  s_{10}&= \ep^{a_1a_2(i}{v^{j}}_{a_1}{v^{k}}_{a_4}{v^{l)a_3}}_{a_2} (R_{12}^{(0,1)})^{a_4}_{a_3}~,\nn\\
  s_{11}&= \ep^{a_1a_2(i}{v^{j|}}_{a_1}{v^{a_3}}_{a_2}{v^{|kl)}}_{a_4} (R_{12}^{(0,1)})^{a_4}_{a_3}~,&
  s_{12}&= \ep^{a_1a_2(i}{v^{j|}}_{a_1}{v^{a_3}}_{a_4}{v^{|kl)}}_{a_2} (R_{12}^{(0,1)})^{a_4}_{a_3}~,\nn\\
  s_{13}&= \ep^{a_1a_2(i}{v^{j|}}_{a_1}{v^{a_3}}_{a_4}{v^{|kl)}}_{a_3} (R_{12}^{(0,1)})^{a_4}_{a_2}~,&
  s_{14}&= \ep^{a_1a_2(i}{v^{j|}}_{a_4}{v^{a_3}}_{a_1}{v^{|kl)}}_{a_2} (R_{12}^{(0,1)})^{a_4}_{a_3}~,\nn\\
  s_{15}&= \ep^{a_1a_2(i}{v^{j|}}_{a_1}{v^{a_3}}_{a_2}{v^{|k|a_4}}_{a_3} (R_{12}^{(0,1)})^{|l)}_{a_4}~,&
  s_{16}&= \ep^{a_1a_2(i}{v^{j|}}_{a_1}{v^{a_3}}_{a_4}{v^{|k|a_4}}_{a_2} (R_{12}^{(0,1)})^{|l)}_{a_3}~,\nn\\
  s_{17}&= \ep^{a_1a_2(i|}{v^{a_3}}_{a_1}{v^{a_4}}_{a_2}{v^{|jk}}_{a_3} (R_{12}^{(0,1)})^{l)}_{a_4}~,&
  %v_2v_2v_3 t^{12}[2,2]:
  s_{18}&= \ep^{a_1a_2(i}{v^{j}}_{a_1}{v^{k}}_{a_4}{v^{l)a_3}}_{a_5} (R_{12}^{(0,1)})^{a_4a_5}_{a_2a_3}~,\nn\\
  s_{19}&= \ep^{a_1a_2(i}{v^{j}}_{a_1}{v^{k|}}_{a_2}{v^{a_3a_4}}_{a_5} (R_{12}^{(0,1)})^{|l)a_5}_{a_3a_4}~,&
  s_{20}&= \ep^{a_1a_2(i}{v^{j}}_{a_1}{v^{k|}}_{a_5}{v^{a_3a_4}}_{a_2} (R_{12}^{(0,1)})^{|l)a_5}_{a_3a_4}~,\nn\\
  s_{21}&= \ep^{a_1a_2(i}{v^{j|}}_{a_1}{v^{a_3}}_{a_4}{v^{|kl)}}_{a_5} (R_{12}^{(0,1)})^{a_4a_5}_{a_2a_3}~,&
  s_{22}&= \ep^{a_1a_2(i}{v^{j|}}_{a_4}{v^{a_3}}_{a_1}{v^{|kl)}}_{a_5} (R_{12}^{(0,1)})^{a_4a_5}_{a_2a_3}~,\nn\\
  s_{23}&= \ep^{a_1a_2(i}{v^{j|}}_{a_4}{v^{a_3}}_{a_5}{v^{|kl)}}_{a_1} (R_{12}^{(0,1)})^{a_4a_5}_{a_2a_3}~,&
  s_{24}&= \ep^{a_1a_2(i}{v^{j|}}_{a_1}{v^{a_3}}_{a_2}{v^{a_4|k}}_{a_5} (R_{12}^{(0,1)})^{l)a_5}_{a_3a_4}~,\nn\\
  s_{25}&= \ep^{a_1a_2(i}{v^{j|}}_{a_1}{v^{a_3}}_{a_4}{v^{a_4|k}}_{a_5} (R_{12}^{(0,1)})^{l)a_5}_{a_2a_3}~,&
  s_{26}&= \ep^{a_1a_2(i}{v^{j|}}_{a_1}{v^{a_3}}_{a_5}{v^{a_4|k}}_{a_2} (R_{12}^{(0,1)})^{l)a_5}_{a_3a_4}~,\nn\\
  s_{27}&= \ep^{a_1a_2(i}{v^{j|}}_{a_1}{v^{a_3}}_{a_5}{v^{a_4|k}}_{a_3} (R_{12}^{(0,1)})^{l)a_5}_{a_2a_4}~,&
  s_{28}&= \ep^{a_1a_2(i}{v^{j|}}_{a_5}{v^{a_3}}_{a_1}{v^{a_4|k}}_{a_2} (R_{12}^{(0,1)})^{l)a_5}_{a_3a_4}~,\nn\\
  s_{29}&= \ep^{a_1a_2(i}{v^{j|}}_{a_1}{v^{a_3}}_{a_2}{v^{a_4a_5}}_{a_3} (R_{12}^{(0,1)})^{kl)}_{a_4a_5}~,&
  s_{30}&= \ep^{a_1a_2(i}{v^{j|}}_{a_1}{v^{a_3}}_{a_4}{v^{a_4a_5}}_{a_2} (R_{12}^{(0,1)})^{kl)}_{a_3a_5}~,\nn\\
  s_{31}&= \ep^{a_1a_2(i|}{v^{a_3}}_{a_1}{v^{a_4}}_{a_2}{v^{|jk}}_{a_5} (R_{12}^{(0,1)})^{l)a_5}_{a_3a_4}~,&
  s_{32}&= \ep^{a_1a_2(i|}{v^{a_3}}_{a_1}{v^{a_4}}_{a_5}{v^{|jk}}_{a_2} (R_{12}^{(0,1)})^{l)a_5}_{a_3a_4}~,\nn\\
  s_{33}&= \ep^{a_1a_2(i|}{v^{a_3}}_{a_1}{v^{a_4}}_{a_5}{v^{|jk}}_{a_4} (R_{12}^{(0,1)})^{l)a_5}_{a_2a_3}~,&
  s_{34}&= \ep^{a_1a_2(i|}{v^{a_3}}_{a_1}{v^{a_4}}_{a_2}{v^{a_5|j}}_{a_3} (R_{12}^{(0,1)})^{kl)}_{a_4a_5}~,\nn\\
  %v_2v_2v_2 t^{14}[1,0]:
  s_{35}&= \ep^{a_1a_2a_3} {v^{(i}}_{a_1} {v^{j}}_{a_2} {v^{k}}_{a_3} (R_{14}^{(0,1)})^{l)}~,&
  %v_2v_2v_2 t^{14}[0,2]:
  s_{36}&= \ep^{a_1a_2(i|}\ep^{a_3a_4|j} {v^{k}}_{a_1} {v^{l)}}_{a_2} {v^{a_5}}_{a_3} (R_{14}^{(0,1)})_{a_4a_5}~,\nn\\
  %v_2v_2v_2 t^{14}[2,1]:
  s_{37}&= \ep^{a_1a_2(i} {v^{j|}}_{a_1} {v^{a_3}}_{a_2} {v^{a_4}}_{a_3} (R_{14}^{(0,1)})^{|kl)}_{a_4}~,&
  s_{38}&= \ep^{a_1a_2(i} {v^{j|}}_{a_3} {v^{a_3}}_{a_1} {v^{a_4}}_{a_2} (R_{14}^{(0,1)})^{|kl)}_{a_4}~,\nn\\
  s_{39}&= \ep^{a_1a_2(i} {v^{j}}_{a_1} {v^{k|}}_{a_2} {v^{a_3}}_{a_4} (R_{14}^{(0,1)})^{|l)a_4}_{a_3}~,&
  %v_2v_2v_2 t^{14}[1,3]:
  s_{40}&= \ep^{a_1a_2(i|}\ep^{a_3a_4|j} {v^{k|}}_{a_1} {v^{a_5}}_{a_2} {v^{a_6}}_{a_3} (R_{14}^{(0,1)})^{|l)}_{a_4a_5a_6}~,\nn\\
  s_{41}&= \ep^{a_1a_2(i|}\ep^{a_3a_4|j} {v^{k|}}_{a_1} {v^{a_5}}_{a_3} {v^{a_6}}_{a_5} (R_{14}^{(0,1)})^{|l)}_{a_2a_4a_6}~,&
  %v_2v_2v_2 t^{14}[3,2]:
  s_{42}&= \ep^{a_1a_2(i|} {v^{a_3}}_{a_1} {v^{a_4}}_{a_2} {v^{a_5}}_{a_3} (R_{14}^{(0,1)})^{|jkl)}_{a_4a_5}~,\nn\\
  s_{43}&= \ep^{a_1a_2(i|} {v^{a_3}}_{a_1} {v^{a_4}}_{a_3} {v^{a_5}}_{a_4} (R_{14}^{(0,1)})^{|jkl)}_{a_2a_5}~,&
  s_{44}&= \ep^{a_1a_2(i} {v^{j|}}_{a_1} {v^{a_3}}_{a_2} {v^{a_4}}_{a_5} (R_{14}^{(0,1)})^{|kl)a_5}_{a_3a_4}~,\nn\\
  s_{45}&= \ep^{a_1a_2(i} {v^{j|}}_{a_5} {v^{a_3}}_{a_1} {v^{a_4}}_{a_3} (R_{14}^{(0,1)})^{|kl)a_5}_{a_2a_4}~,&
  %u_3v_2v_2 t^{14}[1,3]:
  s_{46}&= \ep^{a_1a_2(i} u^{jk|a_3} {v^{a_4}}_{a_1} {v^{a_5}}_{a_2} (R_{14}^{(0,2)})^{|l)}_{a_3a_4a_5}~,\nn\\
  s_{47}&= \ep^{a_1a_2(i} u^{jk|a_3} {v^{a_4}}_{a_1} {v^{a_5}}_{a_3} (R_{14}^{(0,2)})^{|l)}_{a_2a_4a_5}~,&
  s_{48}&= \ep^{a_1a_2(i} u^{jk|a_3} {v^{a_4}}_{a_1} {v^{a_5}}_{a_4} (R_{14}^{(0,2)})^{|l)}_{a_2a_3a_5}~,\nn\\
  s_{49}&= \ep^{a_1a_2(i} u^{j|a_3a_4} {v^{|k|}}_{a_1} {v^{a_5}}_{a_2} (R_{14}^{(0,2)})^{|l)}_{a_3a_4a_5}~,&
  s_{50}&= \ep^{a_1a_2(i} u^{j|a_3a_4} {v^{|k|}}_{a_1} {v^{a_5}}_{a_3} (R_{14}^{(0,2)})^{|l)}_{a_2a_4a_5}~,\nn\\
  s_{51}&= \ep^{a_1a_2(i} u^{j|a_3a_4} {v^{|k|}}_{a_5} {v^{a_5}}_{a_1} (R_{14}^{(0,2)})^{|l)}_{a_2a_3a_4}~,&
  s_{52}&= \ep^{a_1a_2(i|} u^{a_3a_4a_5} {v^{|j}}_{a_1} {v^{k}}_{a_2} (R_{14}^{(0,2)})^{l)}_{a_3a_4a_5}~,\nn\\
  s_{53}&= \ep^{a_1a_2(i} u^{jk|a_3} {v^{|l)}}_{a_5} {v^{a_4}}_{a_1} (R_{14}^{(0,2)})^{a_5}_{a_2a_3a_4}~,&
  %u_2v_2v_3 t^{14}[1,3]:
  s_{54}&= \ep^{a_1a_2(i} u^{jk|} {v^{a_3}}_{a_1} {v^{a_4a_5}}_{a_2} (R_{14}^{(0,2)})^{|l)}_{a_3a_4a_5}~,\nn\\
  s_{55}&= \ep^{a_1a_2(i} u^{jk|} {v^{a_3}}_{a_1} {v^{a_4a_5}}_{a_3} (R_{14}^{(0,2)})^{|l)}_{a_2a_4a_5}~,& 
  s_{56}&= \ep^{a_1a_2(i} u^{jk|} {v^{a_3}}_{a_4} {v^{a_4a_5}}_{a_1} (R_{14}^{(0,2)})^{|l)}_{a_2a_3a_5}~,\nn\\
  s_{57}&= \ep^{a_1a_2(i} u^{j|a_3} {v^{|k|}}_{a_1} {v^{a_4a_5}}_{a_2} (R_{14}^{(0,2)})^{|l)}_{a_3a_4a_5}~,& 
  s_{58}&= \ep^{a_1a_2(i} u^{j|a_3} {v^{|k|}}_{a_1} {v^{a_4a_5}}_{a_3} (R_{14}^{(0,2)})^{|l)}_{a_2a_4a_5}~,\nn\\
  s_{59}&= \ep^{a_1a_2(i} u^{j|a_3} {v^{|k|}}_{a_4} {v^{a_4a_5}}_{a_1} (R_{14}^{(0,2)})^{|l)}_{a_2a_3a_5}~,& 
  s_{60}&= \ep^{a_1a_2(i} u^{j|a_3} {v^{a_4}}_{a_1} {v^{a_5|k}}_{a_2} (R_{14}^{(0,2)})^{l)}_{a_3a_4a_5}~,\nn\\
  s_{61}&= \ep^{a_1a_2(i} u^{j|a_3} {v^{a_4}}_{a_1} {v^{a_5|k}}_{a_3} (R_{14}^{(0,2)})^{l)}_{a_2a_4a_5}~,& 
  s_{62}&= \ep^{a_1a_2(i} u^{j|a_3} {v^{a_4}}_{a_5} {v^{a_5|k}}_{a_1} (R_{14}^{(0,2)})^{l)}_{a_2a_3a_4}~,\nn\\
  s_{63}&= \ep^{a_1a_2(i|} u^{a_3a_4} {v^{|j}}_{a_1} {v^{k|a_5}}_{a_2} (R_{14}^{(0,2)})^{|l)}_{a_3a_4a_5}~,& 
  s_{64}&= \ep^{a_1a_2(i|} u^{a_3a_4} {v^{|j}}_{a_1} {v^{k|a_5}}_{a_3} (R_{14}^{(0,2)})^{|l)}_{a_2a_4a_5}~,\nn\\
  s_{65}&= \ep^{a_1a_2(i|} u^{a_3a_4} {v^{|j}}_{a_5} {v^{k|a_5}}_{a_1} (R_{14}^{(0,2)})^{|l)}_{a_2a_3a_4}~,& 
  s_{66}&= \ep^{a_1a_2(i|} u^{a_3a_4} {v^{a_5}}_{a_1} {v^{|jk}}_{a_2} (R_{14}^{(0,2)})^{l)}_{a_3a_4a_5}~,\nn\\
  s_{67}&= \ep^{a_1a_2(i|} u^{a_3a_4} {v^{a_5}}_{a_1} {v^{|jk}}_{a_3} (R_{14}^{(0,2)})^{l)}_{a_2a_4a_5}~,& 
  s_{68}&= \ep^{a_1a_2(i} u^{jk} {v^{l)}}_{a_1} {v^{a_3a_4}}_{a_5} (R_{14}^{(0,2)})^{a_5}_{a_2a_3a_4}~,\nn\\
  s_{69}&= \ep^{a_1a_2(i} u^{jk|} {v^{a_3}}_{a_1} {v^{|l)a_4}}_{a_5} (R_{14}^{(0,2)})^{a_5}_{a_2a_3a_4}~,& 
  s_{70}&= \ep^{a_1a_2(i} u^{j|a_3} {v^{|k}}_{a_1} {v^{l)a_4}}_{a_5} (R_{14}^{(0,2)})^{a_5}_{a_2a_3a_4}~,\nn\\
  s_{71}&= \ep^{a_1a_2(i} u^{j|a_3} {v^{a_4}}_{a_1} {v^{|kl)}}_{a_5} (R_{14}^{(0,2)})^{a_5}_{a_2a_3a_4}~,& 
  s_{72}&= \ep^{a_1a_2(i|} u^{a_3a_4} {v^{|j}}_{a_1} {v^{kl)}}_{a_5} (R_{14}^{(0,2)})^{a_5}_{a_2a_3a_4}~,\nn\\
  %v_3v_3 t^{16}[2,3]:
  s_{73}&= \ep^{a_1a_2(i}{v^{jk|}}_{a_1}{v^{a_3a_4}}_{a_5} (R_{16}^{(0,2)})^{|l)a_5}_{a_2a_3a_4}~,& 
  s_{74}&= \ep^{a_1a_2(i}{v^{jk|}}_{a_5}{v^{a_3a_4}}_{a_1} (R_{16}^{(0,2)})^{|l)a_5}_{a_2a_3a_4}~,\nn\\
  s_{75}&= \ep^{a_1a_2(i}{v^{j|a_3}}_{a_1}{v^{a_4|k}}_{a_5} (R_{16}^{(0,2)})^{l)a_5}_{a_2a_3a_4}~,& 
  s_{76}&= \ep^{a_1a_2(i}{v^{j|a_3}}_{a_1}{v^{a_4a_5}}_{a_2} (R_{16}^{(0,2)})^{|kl)}_{a_3a_4a_5}~,\nn\\
  %v_3v_3 t^{16}[1,2]:
  s_{77}&= \ep^{a_1a_2a_3}{v^{(ij}}_{a_1}{v^{k|b}}_{a_2}(R_{16}^{(0,2)})^{|l)}_{a_3b}~,& 
  s_{78}&= \ep^{a_1a_2(i}{v^{jk}}_{a_1}{v^{l)c}}_{b}(R_{16}^{(0,2)})^{b}_{a_2c}~,\nn\\
  s_{79}&= \ep^{a_1a_2(i}{v^{jk|}}_{a_1}{v^{bc}}_{a_2}(R_{16}^{(0,2)})^{|l)}_{bc}~,& 
  s_{80}&= \ep^{a_1a_2(i}{v^{j|b}}_{a_1}{v^{c|k}}_{a_2}(R_{16}^{(0,2)})^{|l)}_{bc}~,\nn\\
  %v_3v_3 t^{16}[0,1]:
  s_{81}&= \ep^{a_1a_2a_3}{v^{(ij}}_{a_1}{v^{kl)}}_{a_2} (R_{16}^{(0,2)})_{a_3}~,& 
  s_{82}&= \ep^{a_1a_2(i}{v^{jk}}_{a_1}{v^{l)b}}_{a_2} (R_{16}^{(0,2)})_{b}~,\nn\\
  %u_2v_2v_2 t^{16}[2,3]:
  s_{83}&= \ep^{a_1a_2(i}u^{jk|}{v^{a_3}}_{a_1}{v^{a_4}}_{a_5} (R_{16}^{(0,2)})^{|l)a_5}_{a_2a_3a_4}~,& 
  s_{84}&= \ep^{a_1a_2(i}u^{j|a_3}{v^{|k|}}_{a_1}{v^{a_4}}_{a_5} (R_{16}^{(0,2)})^{|l)a_5}_{a_2a_3a_4}~,\nn\\
  s_{85}&= \ep^{a_1a_2(i}u^{j|a_3}{v^{|k|}}_{a_5}{v^{a_4}}_{a_1} (R_{16}^{(0,2)})^{|l)a_5}_{a_2a_3a_4}~,& 
  s_{86}&= \ep^{a_1a_2(i}u^{j|a_3}{v^{a_4}}_{a_1}{v^{a_5}}_{a_2} (R_{16}^{(0,2)})^{|kl)}_{a_3a_4a_5}~,\nn\\
  s_{87}&= \ep^{a_1a_2(i|}u^{a_3a_4}{v^{|j}}_{a_1}{v^{k}}_{a_5} (R_{16}^{(0,2)})^{l)a_5}_{a_2a_3a_4}~,& 
  s_{88}&= \ep^{a_1a_2(i|}u^{a_3a_4}{v^{|j|}}_{a_1}{v^{a_5}}_{a_2} (R_{16}^{(0,2)})^{|kl)}_{a_3a_4a_5}~,\nn\\
  s_{89}&= \ep^{a_1a_2(i|}u^{a_3a_4}{v^{|j|}}_{a_1}{v^{a_5}}_{a_3} (R_{16}^{(0,2)})^{|kl)}_{a_2a_4a_5}~,& 
  %u_2v_2v_2 t^{16}[1,2]:
  s_{90}&= \ep^{a_1a_2(i}u^{jk}{v^{l)}}_{a_1}{v^{a_3}}_{a_4} (R_{16}^{(0,2)})^{a_4}_{a_2a_3}~,\nn\\
  s_{91}&= \ep^{a_1a_2(i}u^{jk}{v^{l)}}_{a_4}{v^{a_3}}_{a_1} (R_{16}^{(0,2)})^{a_4}_{a_2a_3}~,&
  s_{92}&= \ep^{a_1a_2(i}u^{jk|}{v^{a_3}}_{a_1}{v^{a_4}}_{a_2} (R_{16}^{(0,2)})^{|l)}_{a_3a_4}~,\nn\\
  s_{93}&= \ep^{a_1a_2(i}u^{j|a_3}{v^{|k}}_{a_1}{v^{l)}}_{a_4} (R_{16}^{(0,2)})^{a_4}_{a_2a_3}~,&
  s_{94}&= \ep^{a_1a_2(i}u^{j|a_3}{v^{|k|}}_{a_1}{v^{a_4}}_{a_2} (R_{16}^{(0,2)})^{|l)}_{a_3a_4}~,\nn\\
  s_{95}&= \ep^{a_1a_2(i}u^{j|a_3}{v^{|k|}}_{a_1}{v^{a_4}}_{a_3} (R_{16}^{(0,2)})^{|l)}_{a_2a_4}~,&
  s_{96}&= \ep^{a_1a_2(i|}u^{a_3a_4}{v^{|j}}_{a_1}{v^{k}}_{a_2} (R_{16}^{(0,2)})^{l)}_{a_3a_4}~,\nn\\
  %u_2v_2v_2 t^{16}[0,1]:
  s_{97}&= \ep^{a_1a_2a_3}u^{(ij}{v^{k}}_{a_1}{v^{l)}}_{a_2} (R_{16}^{(0,2)})_{a_3}~,&
  s_{98}&= \ep^{a_1a_2(i}u^{jk}{v^{l)}}_{a_1}{v^{b}}_{a_2} (R_{16}^{(0,2)})_{b}~,\nn\\
  s_{99}&= \ep^{a_1a_2(i}u^{j|b}{v^{|k}}_{a_1}{v^{l)}}_{a_2} (R_{16}^{(0,2)})_{b}~,&
  %v_2v_3 t^{18}[0,0]:
  s_{100}&= \ep^{a_1a_2(i}{v^{j}}_{a_1}{v^{kl)}}_{a_2} (R_{18}^{(0,2)})~,\nn\\
  %u_2v_3 t^{20}[1,0]:
  s_{101}&= u^{a(i}{v^{jk}}_a (R_{20}^{(0,3)})^{l)}~,&
  s_{102}&= u^{(ij}{v^{kl)}}_a (R_{20}^{(0,3)})^{a}~,\nn\\
  %u_3v_2 t^{20}[1,0]:
  s_{103}&= u^{a(ij}{v^{k}}_a (R_{20}^{(0,3)})^{l)}~,&
  s_{104}&= u^{(ijk}{v^{l)}}_a (R_{20}^{(0,3)})^{a}~,\nn\\
  %u_2v_2 t^{22}[2,0]:
  s_{105}&= u^{a(i}{v^{j}}_a (R_{22}^{(0,3)})^{kl)}~,&
  s_{106}&= u^{(ij}{v^{k}}_a (R_{22}^{(0,3)})^{l)a}~.%\nn\\
  %u_3 t^{26}[1,0]:
  %s_{107}&= u^{(ijk} (R_{26}^{(0,4)})^{l)}~,&
\ea}
Among these 106 graviton trace relations,
57 are independent and there are 49 relations of relations.
They provide $49$ operators which are $Q$-closed by trace relations.
One of these is presented in \eqref{32cohomology}, and we prove subsequently
that it is not $Q$-exact and thus represents a black hole cohomology.

\end{document}